%% file: ws-ijqi_smo_arxiv.tex
\documentclass{ws-ijqi_arxiv}

\pdfoutput=1
\usepackage{amsmath}
\usepackage{amssymb}
\usepackage{graphicx}			
\usepackage{dcolumn}			
\usepackage{bm}						
\usepackage{verbatim}
\usepackage{color}				

\input{setmathcommands_ijqi}
\allowdisplaybreaks[1]

\begin{document}

\markboth{S. Olmschenk et al.}
{Quantum Logic Between Distant Trapped Ions}

\catchline{}{}{}{}{}

\title{Quantum Logic Between Distant Trapped Ions}

\author{S. Olmschenk$^{1,\dagger}$, D. Hayes$^1$, D. N. Matsukevich$^1$, P. Maunz$^1$, D. L. Moehring$^2$, and C. Monroe$^1$}

\address{$^1$Joint Quantum Institute, University of Maryland Department of Physics and National Institute of Standards and Technology, College Park, Maryland 20742, United States of America\\
$^2$Max-Planck-Institut f\"{u}r Quantenoptik, 85748 Garching, Germany\\
$^\dagger$smolms@umd.edu}

\maketitle

\begin{history}
\end{history}

\begin{abstract}
Trapped atomic ions have proven to be one of the most promising candidates for the realization of quantum computation due to their long trapping times, excellent coherence properties, and exquisite control of the internal atomic states.  Integrating ions (quantum memory) with photons (distance link) offers a unique path to large-scale quantum computation and long-distance quantum communication.  In this article, we present a detailed review of the experimental implementation of a heralded photon-mediated quantum gate between remote ions, and the employment of this gate to perform a teleportation protocol between two ions separated by a distance of about one meter.
\end{abstract}


\section{Introduction}
	\label{sec:intro}

Quantum information research has the potential to drastically alter the fields of communication and computation.  Efforts in quantum computation are driven by the prospect of using the features of quantum physics to tackle otherwise intractable computational problems.  It was realized early on that controllable quantum systems may be used to simulate larger quantum systems far more efficiently than is possible using conventional computers,\cite{feynman:qsim} and that individual quantum systems might be used as quantum bits (qubits) for information processing.\cite{deutsch:universalQC}  While both of these instances were crucial points, interest in quantum information increased dramatically in 1994 when Peter Shor unveiled an algorithm that could be implemented on a quantum computer that would enable an exponential speed-up in the factorization of large numbers.\cite{shor:factoring}  Given that current encryption techniques, such as the RSA algorithm, rely on the relative inability of a conventional computer to factor large numbers, Shor's factorization algorithm linked quantum computation to the immediate and important issue of code-breaking.

On the other hand, quantum information research has also yielded a new method of secure communication, where the security of the encrypted information is guaranteed not by a lack of efficient mathematical or computational algorithms, but by the physical quantum properties of the information carriers.  It has been shown that entangled pairs of qubits can be used to securely transfer information, where the presence of an eavesdropper is detected by the measurements made during the protocol.\cite{bennett:bb84,ekert:q_crypto_bell}  The essential component for this protocol is the quantum no-cloning theorem, which proves quantum physics prohibits the cloning (copying) of an unknown quantum state.\cite{wootters:no-clone}  By using a quantum communication protocol with pairs of correlated qubits, the back-action of measurement by an eavesdropper destroys the correlations, announcing their presence before any information is transmitted.

Of course, the same features of quantum mechanics that are used to ensure the security of communication also prevent a simple ``read-and-send'' approach to transmitting quantum information.  Any attempt to read or measure a single quantum superposition state results in a single measurement outcome, and thereby lacks the information needed to reconstruct the probabilities of the quantum superposition.  While an estimation of an unknown quantum state could be obtained by simply measuring a large number of copies, the quantum no-cloning theorem forbids generating identical copies of a single unknown quantum state.  Nevertheless, a quantum state can still be transferred through the process of quantum teleportation.\cite{bennett:teleportation}  In the quantum teleportation protocol, a quantum state initially stored in system A can be recovered at system B without ever having traversed the space between the systems.  The ability to teleport quantum information is an essential ingredient for long-distance quantum communication and may be a vital component to achieve the exponential processing speed-up promised by quantum computation.  The essence of quantum teleportation lies in the non-local correlations, or entanglement, afforded by quantum physics.

Quantum communication and quantum computation both use entanglement as an essential resource.  Entanglement is the quantum correlations between systems that do not have well-defined individual properties.  In addition to being useful for quantum communication and quantum computation, entanglement embodies the counter-intuitive depiction of nature predicted by quantum physics, and allows for explicit experimental tests of quantum theory.\cite{bell:inequal}  Entanglement has been observed in a wide variety of systems, including: photons;\cite{kok:linear_op_qc} atomic ions;\cite{blatt:ions_entangled} superconducting Josephson junctions;\cite{clarke:superconducting_qubits} and neutral atoms in cavities,\cite{kimble:qinternet} confined by an optical lattice,\cite{bloch:ultracold_lattices} and in small ensembles.\cite{kimble:qinternet}  Hybrid systems composed of a single photonic qubit and a matter qubit (ion,\cite{blinov:ion-photon} ensemble,\cite{matsukevich:matter-light,riedmatten:atom-ensemble,chen:atom-ensemble_entangle,sherson:light-ensemble_teleport} or atom\cite{volz:atom-photon,wilk:atom-photon}) have also shown entanglement.  Here, we will focus on photons and atomic ions, both of which have already proven amenable to applications in quantum information.

Atomic ions are one of the most promising systems for quantum information processing due to their long trapping times, excellent coherence properties, and exquisite control of the internal atomic states.  To date, the largest entangled state of individually addressable qubits has been an 8-particle W-state,\cite{haffner:8_ion_w} realized in a system of trapped atomic ions that used the collective motion of the confined atoms to implement the entangling protocol.  The experimental effort is now focused on scaling this system to larger numbers of qubits.  One approach to scaling this system\cite{kielpinski:ion_architecture}
 is to use microfabricated ion traps\cite{stick:microtrap,seidelin:surface_trap} and advanced ion trap arrays.\cite{hensinger:t-trap,blakestad:x-trap}  Since these deterministic gate operations utilize the common modes of motion of the trapped ions, an understanding of motional decoherence is also actively being pursued.\cite{turchette:heating,deslauriers:needle,labaziewicz:heating}

Photons, on the other hand, are a natural choice of qubit for communication purposes, as they can quickly traverse the distance between locations with only small perturbations to the encoded quantum information.  Already, a series of seminal experiments have used photons in quantum communication protocols over distances as large as 144 km.\cite{ursin:144km}  Ultimately, though, the direct communication of quantum information over long-distances is impeded by the attenuation of light in air and optical fibers.  For transmission through optical fibers, even at telecom wavelengths, which experience the least attenuation in fiber, the probability of transmitting a single photon over 1000 km is $< 10^{-20}$.  Current fiber-optic information transfer mitigates the loss in signal amplitude by introducing repeaters along the transmission path to ``boost'' the signal along the way.  Although application of the standard model of a repeater to quantum information is prohibited by the no-cloning theorem, an analogous ``quantum repeater'' has been suggested to enable the transfer of quantum information over arbitrary distances.\cite{briegel:quantum_repeater,duan:dlcz}  In this method, the distance between two points is broken up into a series of shorter segments, with a quantum memory at each connection point, or node.  Entanglement can then be established between pairs of nodes, and subsequent segments connected via entanglement swapping, which is used to extend the entanglement over the entire length of the repeater.  The final step is to use this long-distance entanglement as a resource to transfer quantum information over that distance by the process of quantum teleportation.  

Integrating atomic ions (quantum memory) with photons (distance link) offers a unique path to large-scale quantum computation and long-distance quantum communication.  Combining ions with photons enables long-distance quantum operations between stable quantum memories.  These photon-mediated operations are mostly insensitive to the motional state of the ions, and thus can tolerate motional heating and do not require ground-state cooling.  Moreover, the two-photon scheme presented here is not interferometrically sensitive to the optical path length difference.  Large-scale implementation may also be simplified by avoiding the need for complex trap arrays that allow for shuttling ions, or by enabling a higher-level architecture that links distant trap arrays.  In addition, since these operations are mediated by photons, it may be possible to create hybrid matter quantum systems: interfacing atoms with solid-state qubits, such as quantum dots or NV centers.  Finally, while this photon-mediated operation is inherently probabilistic, it can still be efficiently scaled to enable generation of the large entangled states required for quantum information processing.

In the following sections, we present a detailed review of the experimental implementation of a probabilistic quantum gate, and employment of this gate in a quantum teleportation protocol to transfer a qubit between two atomic ions separated by about one meter.  We begin in Sec.~\ref{sec:ion_trap} with the basics of trapping charged particles.  Next we delve into the specifics of our atomic qubit, the ytterbium ion (\ybion), in Sec.~\ref{sec:ybion}.  Section~\ref{sec:photon_interference} presents the two-photon interference effect.  In Sec.~\ref{sec:entanglement} are the results of the experimental implementation of the heralded quantum gate.  Finally, Sec.~\ref{sec:teleportation} illustrates the use of this gate in the long-distance teleportation protocol.  We conclude with an outlook of the potential practical application of these results to long-distance quantum communication and large-scale quantum computation.

\section{Ion Trap}
	\label{sec:ion_trap}

The radiofrequency (rf) ion trap was invented by Wolfgang Paul, for which he shared the 1989 Nobel prize.\cite{paul:nobel}  The first laser-cooling experiments with atoms were reported independently by Wineland et al.\cite{wineland:cooling} using Mg${}^{+}$, and Neuhauser et al.\cite{neuhauser:cooling} using Ba${}^{+}$.\footnote{The experiment by Wineland et al. used a Penning trap to confine the ions.  Subsequent experiments by Wineland et al. have used the type of rf trap presented here.}  Since its introduction, the rf ion trap has been used for a wealth of applications, including atomic clocks, measurements of fundamental constants, mass and frequency spectroscopy,\cite{fisk:171s12,march:mass_spect,fortier:fund,rosenband:AltoHg} and quantum information science.\cite{blatt:ions_entangled}

According to Earnshaw's Theorem, ``a charged particle cannot be held in a stable equilibrium by electrostatic forces alone.''\cite{griffiths:em}  This result is also embodied in the freespace Maxwell equation $\nabla \cdot \hat{E} = 0$, which in words states that electric field lines entering a region free of charges also need to exit the region; thus, any configuration of static fields will always be anti-trapping in some direction.  Nevertheless, it is still possible to confine charged particles with electromagnetic fields by using either a combination of electric and magnetic fields (Penning trap) or dynamic electric fields (rf or Paul trap).  In the following, we concentrate exclusively on the rf trap.

\begin{figure}
	\centering
	\includegraphics[width=0.5\columnwidth,keepaspectratio]{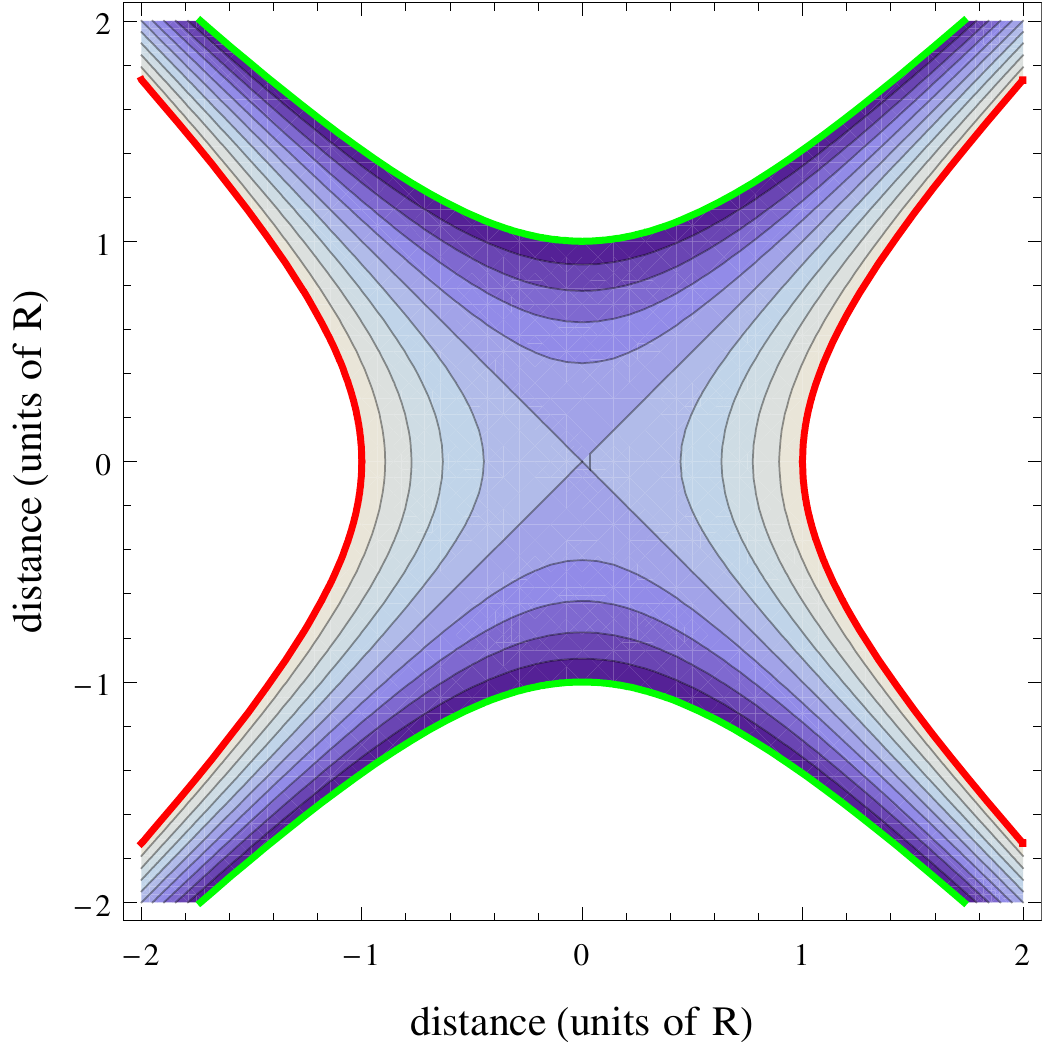}
	\caption{Hyperbolic electrodes.  This two-dimensional view shows hyperbolic electrodes where the \textcolor{green}{green} electrodes are held at ground, and the \textcolor{red}{red} electrodes are at $V_{0} \cos(\Omega_{T} t)$.  The resulting potential is shown as a contour plot between the electrodes (shades of blue) for $t = 0$.}
	\label{fig:hyperbolic_elec}
\end{figure}

We consider dynamic quadrupole potentials to trap charged particles.  As an illustration in two dimensions, consider the hyperbolic electrode configuration shown in Fig.~\ref{fig:hyperbolic_elec}, which has the nearest electrode a distance $R$ from the center of the structure.  Given that the left and right electrodes have potential $V_{0} \cos(\Omega_{T} t)$ applied, while the top and bottom electrodes are held at $0$ V, the potential between the electrodes is given by
\begin{equation}
	\label{eq:hyperbolic_poten}
	\phi_{hyp} = \frac{V_{0}}{2} \cos(\Omega_{T} t) \left( 1 + \frac{x^{2} - y^{2}}{R^{2}} \right) ,
\end{equation}
which satisfies the boundary conditions, and Laplace's equation.  The electric field produced by this potential is
\begin{eqnarray}
	\label{eq:hyperbolic_E-field}
	\hat{E}(x,y,t) & = & - \nabla \phi_{hyp} \nonumber \\
								 & = & - \frac{V_{0}}{R^{2}}(x \hat{x} - y \hat{y}) \cos(\Omega_{T} t) \nonumber \\
								 & = & - \hat{E}_{0}(x,y) \cos(\Omega_{T} t) ,
\end{eqnarray}
where we have written the electric field as a product of its spatial variation $\hat{E}_{0}(x,y)$ and its time variation $\cos(\Omega_{T} t)$.
The force in the $x$-direction on a particle of mass $m$ and charge $e$ is then
\begin{equation}
	\label{eq:hyperbolic_Fx}
	F_{x} = m \ddot{x} = - \frac{e V_{0}}{R^{2}} \cos(\Omega_{T} t) x
\end{equation}
yielding the equation of motion
\begin{equation}
	\label{eq:hyperbolic_x-motion}
	\ddot{x} + \frac{e V_{0}}{m R^{2}} \cos(\Omega_{T} t) x = 0 .
\end{equation}
This is actually just a simplified version of a Mathieu equation.  The general form of the Mathieu equation is:
\begin{equation}
	\label{eq:mathieu}
	\frac{d^2 u}{d \tau^2} + \left( a_{u} + 2 q_{u} \cos(2 \tau) \right) u = 0 .
\end{equation}
Equation~\ref{eq:hyperbolic_x-motion} can be put in this form by making the substitution $2 \tau = \Omega_{T} t$.  By the chain rule, we then have $\frac{d}{dt} = \frac{d}{d\tau} \frac{d\tau}{dt} = \frac{\Omega_{T}}{2} \frac{d}{d\tau}$, so that Eq.~\ref{eq:hyperbolic_x-motion} can be written as
\begin{equation}
	\label{eq:hyperbolic_mathieu}
	\frac{d^2 x}{d \tau^2} + 2 q_{x} \cos(2 \tau) x = 0 ,
\end{equation}
where in the last line we have defined $q_{x} = (2 e V_{0})/(m R^2 \Omega_{T}^2)$.  Note that compared to the general form of Eq.~\ref{eq:mathieu}, here we have $a_{x} = 0$.\footnote{In the present case, we have neglected the third dimension of the problem.  The ion can be confined in the third dimension using static potentials (see below), in which case $a_{x} \neq 0$; a good review is Ref.~\refcite{wineland:nist-jnl}.}

The Floquet Theorem suggests a general solution of the form\cite{mclachlan:mathieu}
\begin{equation}
	\label{eq:floquet_soln}
	x(\tau) = A \sum_{n=-\infty}^{\infty} C_{2n} \cos((2n+\beta) \tau) + \ii B \sum_{n=-\infty}^{\infty} C_{2n} \sin((2n+\beta) \tau) ,
\end{equation}
where $A,B$ are determined by the initial conditions.  We can solve for $\beta$ and the coefficients $C_{2n}$ by plugging Eq.~\ref{eq:floquet_soln} back into Eq.~\ref{eq:hyperbolic_mathieu}.  We then find
\begin{eqnarray}
	\label{eq:mathieu_soln_1}
	   && A \sum_{n=-\infty}^{\infty} C_{2n} \left( 2n + \beta \right)^{2} \cos((2n+\beta) \tau) \nonumber \\
	   && + \ii B \sum_{n=-\infty}^{\infty} C_{2n} \left( 2n + \beta \right)^{2} \sin((2n+\beta) \tau) \nonumber \\
	& = & 2 q_{x} A \sum_{n=-\infty}^{\infty} C_{2n} \cos(2 \tau) \cos((2n+\beta) \tau) \nonumber \\
	   && + \ii 2 q_{x} B \sum_{n=-\infty}^{\infty} C_{2n} \cos(2 \tau) \sin((2n+\beta) \tau) .
\end{eqnarray}
Using the trigonometric product-to-sum relation for cosine\footnote{The product-to-sum relation is $\cos(\alpha) \cos(\beta) = \frac{1}{2}\left( \cos(\alpha + \beta) + \cos(\alpha - \beta) \right)$.}, we get
\begin{eqnarray}
	\label{eq:mathieu_soln_2}
	   && A \sum_{n=-\infty}^{\infty} C_{2n} \left( 2n + \beta \right)^{2} \cos((2n+\beta) \tau) \nonumber \\
	   && + \ii B \sum_{n=-\infty}^{\infty} C_{2n} \left( 2n + \beta \right)^{2} \sin((2n+\beta) \tau) \nonumber \\
	& = & q_{x} A \sum_{n=-\infty}^{\infty} C_{2n} \left( \cos((2n + \beta)\tau + 2\tau) + \cos((2n + \beta)\tau - 2\tau) \right) \nonumber \\
	   && + \ii 2 q_{x} B \sum_{n=-\infty}^{\infty} C_{2n} \cos(2 \tau) \sin((2n+\beta) \tau) \nonumber \\
	& = & q_{x} A \sum_{n=-\infty}^{\infty} C_{2n - 2} \cos((2n + \beta)\tau) \nonumber \\
	   && + q_{x} A \sum_{n=-\infty}^{\infty} C_{2n + 2} \cos((2n + \beta)\tau) \nonumber \\
	   && + \ii 2 q_{x} B \sum_{n=-\infty}^{\infty} C_{2n} \cos(2 \tau) \sin((2n+\beta) \tau) ,
\end{eqnarray}
where in the last step we just altered the indexing (possible since the sum goes between $\pm \infty$).  This allows us to easily match the cosine terms, yielding a recursion relation:
\begin{equation}
	\label{eq:recursion}
	-K_{2n} C_{2n} + C_{2n-2} + C_{2n+2} = 0, \mbox{ with } K_{2n} = \frac{(2n + \beta)^2}{q_{x}} .
\end{equation}
This recursion relation allows us to calculate some useful relations.  By setting $n = 0$ in Eq.~\ref{eq:recursion}, we obtain an expression for $\beta$:
\begin{equation}
	\label{eq:get_beta}
	K_0 = \frac{\beta^2}{q_{x}} = \frac{C_{-2} + C_2}{C_0} .
\end{equation}
Rearranging Eq.~\ref{eq:recursion}, we also find
\begin{eqnarray}
	\label{eq:cont_frac_1}
	\frac{C_{2n}}{C_{2n+2}} = \frac{1}{K_{2n} - \frac{C_{2n-2}}{C_{2n}}}
\end{eqnarray}
and we can then plug Eq.~\ref{eq:cont_frac_1} back into itself recursively, and obtain the continued fraction expression
\begin{equation}
	\label{eq:cont_frac_c2np2}
	\frac{C_{2n}}{C_{2n+2}} = \frac{1}{K_{2n} - \frac{1}{K_{2n-2} - \frac{1}{...}}} .
\end{equation}
Similarly, we also get
\begin{equation}
	\label{eq:cont_frac_c2nm2}
	\frac{C_{2n}}{C_{2n-2}} = \frac{1}{K_{2n} - \frac{1}{K_{2n+2} - \frac{1}{...}}} .
\end{equation}
Plugging Eqs.~\ref{eq:cont_frac_c2nm2} and \ref{eq:cont_frac_c2np2} into Eq.~\ref{eq:get_beta} allow the calculation of $\beta$ to any order in $q_{x}$.  Explicitly, we have
\begin{eqnarray}
	\label{eq:beta_all_orders}
	\beta^2 & = & q_{x} \left( \frac{C_{-2}}{C_0} + \frac{C_{2}}{C_{0}} \right) \nonumber \\
					& = & q_{x} \left( \frac{1}{K_{-2} - \frac{1}{K_{-4} - \frac{1}{...}}} + \frac{1}{K_{2} - \frac{1}{K_{4} - \frac{1}{...}}} \right) .
\end{eqnarray}

The solution to the Mathieu equation has been shown to be stable for $q_{x}$ less than about 0.9.\cite{mclachlan:mathieu}  To lowest order in $q_{x}$, we find
\begin{eqnarray}
	\label{eq:beta_lowest_order1}
	\beta^2 & \approx & q_{x} \left( \frac{1}{K_{-2}} + \frac{1}{K_{2}} \right) \nonumber \\
					& \approx & q_{x} \left( \frac{q_{x}}{4} + \frac{q_{x}}{4} \right) \nonumber \\
					& = & \frac{q_{x}^2}{2} ,
\end{eqnarray}
and thus
\begin{equation}
	\label{eq:beta_lowest_order2}
	\beta \approx \frac{q_{x}}{\sqrt{2}} .
\end{equation}

Given the above expressions for $\beta$ and the coefficients $C_{2n}$, we can solve for the trajectory of the ion, $x(\tau)$.  Assuming $q_{x} \ll 1$, we will take $C_{\pm4} \approx 0$.  By assuming the initial condition $B = 0$, from Eq.~\ref{eq:floquet_soln} we then obtain
\begin{equation}
	\label{eq:ion_trajectory_1}
	x(\tau) \approx A C_{0} \cos(\beta \tau) + A C_{-2} \cos((\beta - 2) \tau) + A C_{2} \cos((\beta + 2) \tau) .
\end{equation}
Since $q_{x} \ll 1$, Eqs.~\ref{eq:cont_frac_c2np2} and \ref{eq:cont_frac_c2nm2} can be approximated as
\begin{equation}
	\label{eq:cont_frac_approx}
	C_{0} \approx \frac{C_{\pm 2}}{K_{0} + \frac{1}{K_{\mp 2}}} = \frac{C_{\pm 2}}{\frac{\beta^2}{q_{x}} - \frac{1}{\frac{(\beta \mp 2)^{2}}{q_{x}}}} \approx \frac{C_{\pm 2}}{\frac{q_{x}}{2} - \frac{q_{x}}{4}} = \frac{4 C_{\pm 2}}{q_{x}} ,
\end{equation}
where in the second-to-last step, we used the approximation of Eq.~\ref{eq:beta_lowest_order2} and $(\beta \mp~2)^{2} \approx~4$.  Plugging this value for $C_{\pm 2}$ into Eq.~\ref{eq:ion_trajectory_1} yields
\begin{eqnarray}
	\label{eq:ion_trajectory_2}
	x(\tau) & \approx & A C_{0} \left[ \cos(\beta \tau) + \frac{q_{x}}{4} \left( \cos((\beta - 2) \tau) + \cos((\beta + 2) \tau) \right) \right] \nonumber \\
					& = & A C_{0} \left[ \cos(\beta \tau) + \frac{q_{x}}{2} \cos(\beta\tau) \cos(2 \tau) \right] \nonumber \\
					& = & A C_{0} \cos(\beta \tau) \left[ 1 + \frac{q_{x}}{2} \cos(2 \tau) \right] .
\end{eqnarray}
We can finally put this back in terms of $t$ by recalling our earlier substitutions:
\begin{eqnarray}
	\label{eq:ion_trajectory_3}
	x(t) & \approx & A C_{0} \cos \left( \frac{q_{x}}{2^{1/2}} \frac{\Omega_{T} t}{2} \right) \left[ 1 + \frac{q_{x}}{2} \cos(\Omega_{T} t) \right] \nonumber \\
			 & = & A C_{0} \cos \left( \frac{e V_{0}}{2^{1/2} m \Omega_{T} R^2} t \right) \left[ 1 + \frac{q_{x}}{2} \cos(\Omega_{T} t) \right] \nonumber \\
			 & = & A C_{0} \cos(\omega_{x} t) \left[ 1 + \frac{q_{x}}{2} \cos(\Omega_{T} t) \right] .
\end{eqnarray}
In the last step we have defined the ``secular frequency'' of the ion as:
\begin{equation}
	\label{eq:secular_freq}
	\omega_{x} = \frac{e V_{0}}{2^{1/2} m \Omega_{T} R^{2}} .
\end{equation}
The secular frequency characterizes the relatively slow harmonic motion of a charged particle confined by an oscillating quadrupole field.  In analogy, the harmonic motion executed by the charged particle can be derived from an effective harmonic pseudopotential produced by the quadrupole field:\cite{dehmelt:iontrap}
\begin{equation}
	\label{eq:pseudopotential}
	\psi_{p} = \frac{e E_{0}^{2}}{4 m \Omega_{T}^{2}} .
\end{equation}
The secular motion of a collection of ions has been utilized to perform deterministic quantum information processing tasks in several experiments.\cite{blatt:ions_entangled}

Superimposed on the secular motion in Eq.~\ref{eq:ion_trajectory_3} is a modulation at the driving frequency of the trap, $\Omega_{T}$.  The oscillation at $\Omega_{T}$ is known as the ``micromotion,'' and has an amplitude proportional to the distance from the center of the quadrupole field.  Given that $q_{x} \ll 1$, the amplitude of the micromotion is typically much smaller than the secular motion.  However, if a static offset field causes the secular motion to be centered a considerable distance from the center of quadrupole field, then it is possible for the amplitude of the micromotion to be much larger than the secular amplitude.  In practice, static offset fields are carefully compensated to ensure the particle executes its secular motion about the center of the quadrupole field.\cite{berkeland:micromotion}
\begin{figure}
	\centering
	\includegraphics[width=0.7\columnwidth,keepaspectratio]{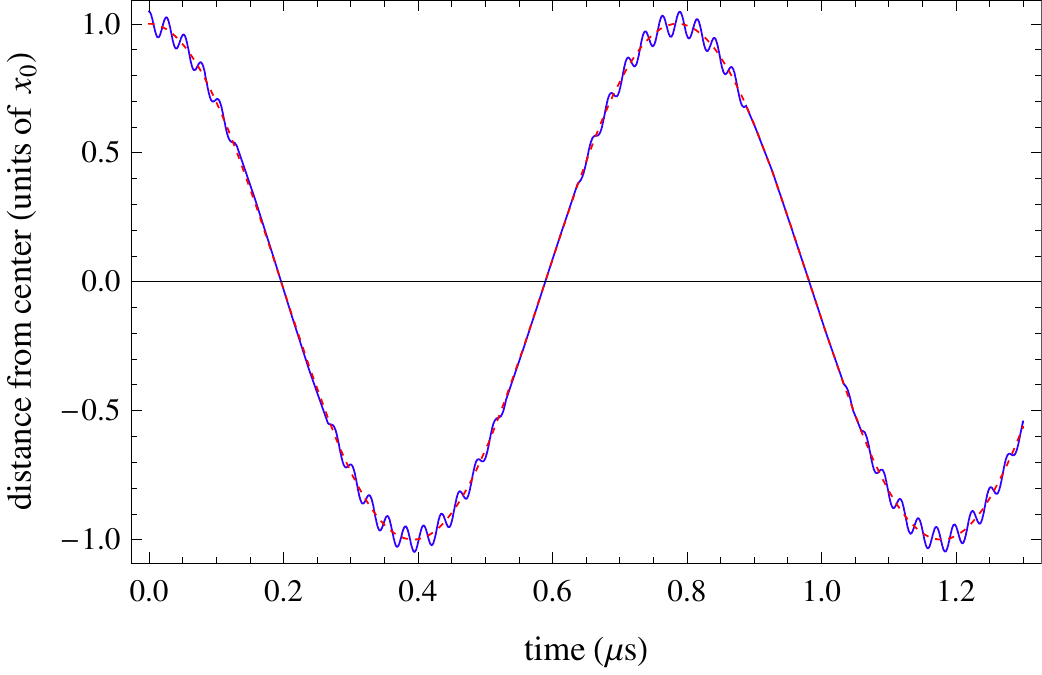}
	\caption{The motion of a trapped charged particle, as given by Eq.~\ref{eq:ion_trajectory_3}, with $R = 0.46$ mm, $\Omega_{T}/(2 \pi) = 38$ MHz, $m = 171$ amu, and $V_{0} = 1$ kV.  The slower, larger amplitude oscillation at $\omega_{x}/(2 \pi) = 1.3$ MHz is the secular motion of the particle, resulting from the time-averaged force of the inhomogenous electric field (\textcolor{red}{red} dashed line).  Superimposed on the secular motion is the micromotion that occurs at the drive frequency of trap ($\Omega_{T}/(2 \pi) = 38$ MHz) with amplitude proportional to the excursion of the particle from the center of the quadrupole field (\textcolor{blue}{blue} solid line).}
	\label{fig:ion_trajectory}
\end{figure}

A representative picture of the motion of a particle confined by the two-dimensional quadrupole field of Eq.~\ref{eq:hyperbolic_E-field} is shown in Figure~\ref{fig:ion_trajectory}.  In this case, we have taken $R = 0.46$ mm, $\Omega_{T}/(2 \pi) = 38$ MHz, $m = 171$ amu, and $V_{0} = 1$ kV.  By the equations above, we see this results in $\omega_{x}/(2 \pi) = 1.3$ MHz and $q_{x}$ = 0.1.

The derivation above can be extended to confinement of the charged particle in three dimensions by either the addition of another rf electrode along the $z$-axis or by capping the third dimension with static potentials.  In the latter option, the hyperbolic electrodes can be extended along the $z$-axis to produce an rf nodal line in this dimension.  By making the static confinement of the particles along $z$ weaker than the rf confinement in the $x,y$ directions, the charged particles can be arranged in a linear crystal.  In such a ``linear trap'' all of the particles execute their secular motion about the center of the quadrupole field.  The addition of a static potential along $z$ produces a potential near the center of the trap of the form:\cite{wineland:nist-jnl}
\begin{align}
	\label{eq:static_poten}
	\phi_{static} & \approx \frac{U_{0}}{z_0^2} \left( z^{2} - \frac{x^{2} + y^{2}}{2} \right) \nonumber \\
								& = \frac{m \omega_z^2}{2 e} \left( z^{2} - \frac{x^{2} + y^{2}}{2} \right) .
\end{align}
Here we have defined the distance to the static electrode as $z_0$, and the oscillation frequency along $z$ as $\omega_z = \sqrt{2 e U_0/(m z_0^2)}$.  In addition to confining the particle along $z$, this static potential also alters the potential in $x$ and $y$, so that the secular frequency of the ion becomes
\begin{align}
	\label{eq:secular_freq_static}
	\tilde{\omega}_{x} = \sqrt{\omega_x^2 - \frac{1}{2} \omega_z^2} .
\end{align}
Thus, the static potential along $z$ weakens the tranverse ($x,y$) confinement of a charge particle.  However, the rf potential is typically much larger than the applied static potential, so the trap remains stable.

The two-dimensional hyperbolic electrode structure presented in the previous section produces the ideal quadrupole potential for trapping charged particles.  Of course, in practice we are often required to alter the structure of the trap electrodes from the ideal hyperbolic structure to conform to other design parameters, such as optical access, multiple trapping zones, or simplicity of construction.  Nevertheless, as long as we retain the overall symmetry presented above, then near the trap center the potential can be approximated by the hyperbolic potential given in Eq.~\ref{eq:hyperbolic_poten}.\cite{wineland:nist-jnl}  The effect of non-hyperbolic electrodes can be characterized by the addition of a geometric scaling factor $\eta_{sc}$ (generally of order unity), so that the modified potential near the center of the trap is approximately represented by:\cite{madsen:planar_trap}
\begin{equation}
	\label{eq:non-hyperbolic_poten}
	\phi_{nonhyp} = \frac{\eta_{sc} V_{0}}{2} \cos(\Omega_{T} t) \left( 1 + \frac{x^{2} - y^{2}}{R^{2}} \right) .
\end{equation}
Since $\eta_{sc}$ is just a constant factor, it carries through the rest of the equations derived in the previous section.

\begin{figure}
	\centering
	\includegraphics[width=0.5\columnwidth,keepaspectratio]{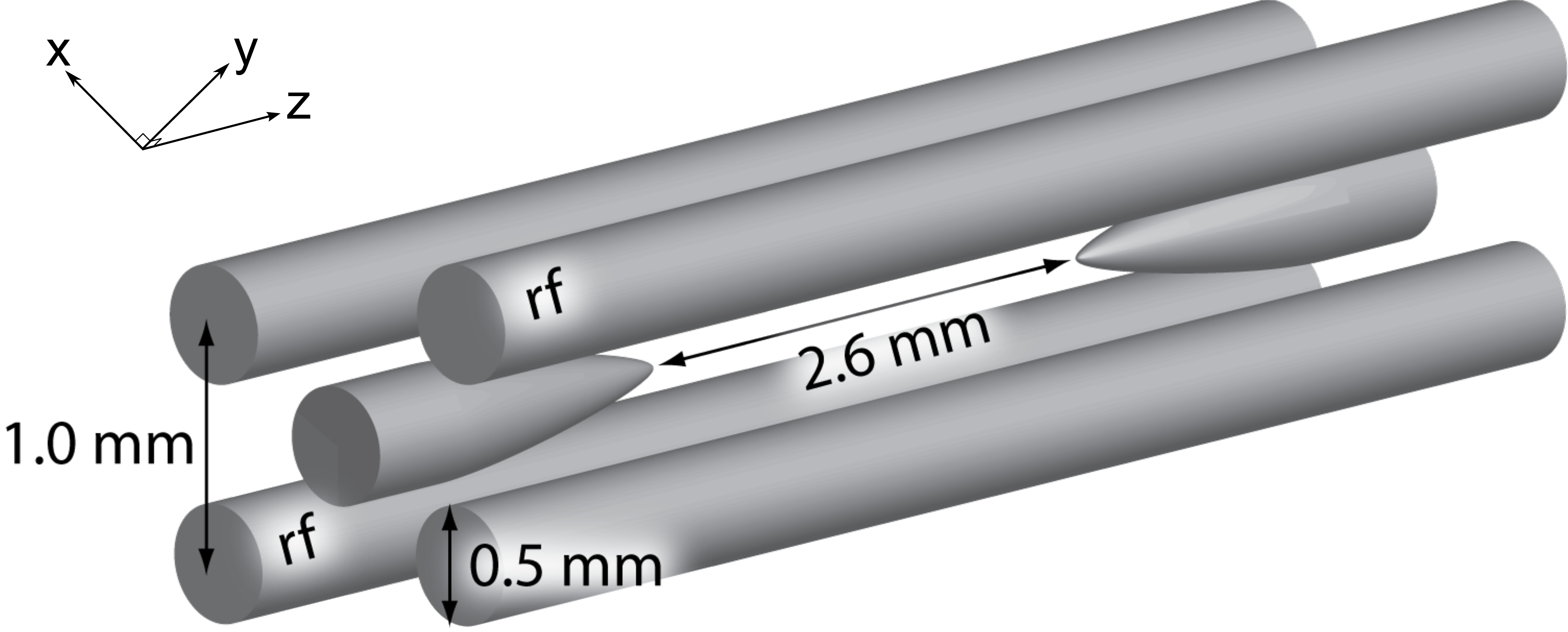}
	\caption{Four rod trap.  The trap design shown here is the type used for all experiments presented in this work.  Two of the four rods have radiofrequency (rf) potentials applied to create an oscillating quadrupole field for confinement in the $x,y$-plane.  The other two rods are held at ground.  Static potentials are applied to the needle electrodes to provide confinement along the $z$-axis of the trap.}
	\label{fig:four_rod_trap}
\end{figure}

\begin{figure}
	\centering
	\includegraphics[width=0.9\columnwidth,keepaspectratio]{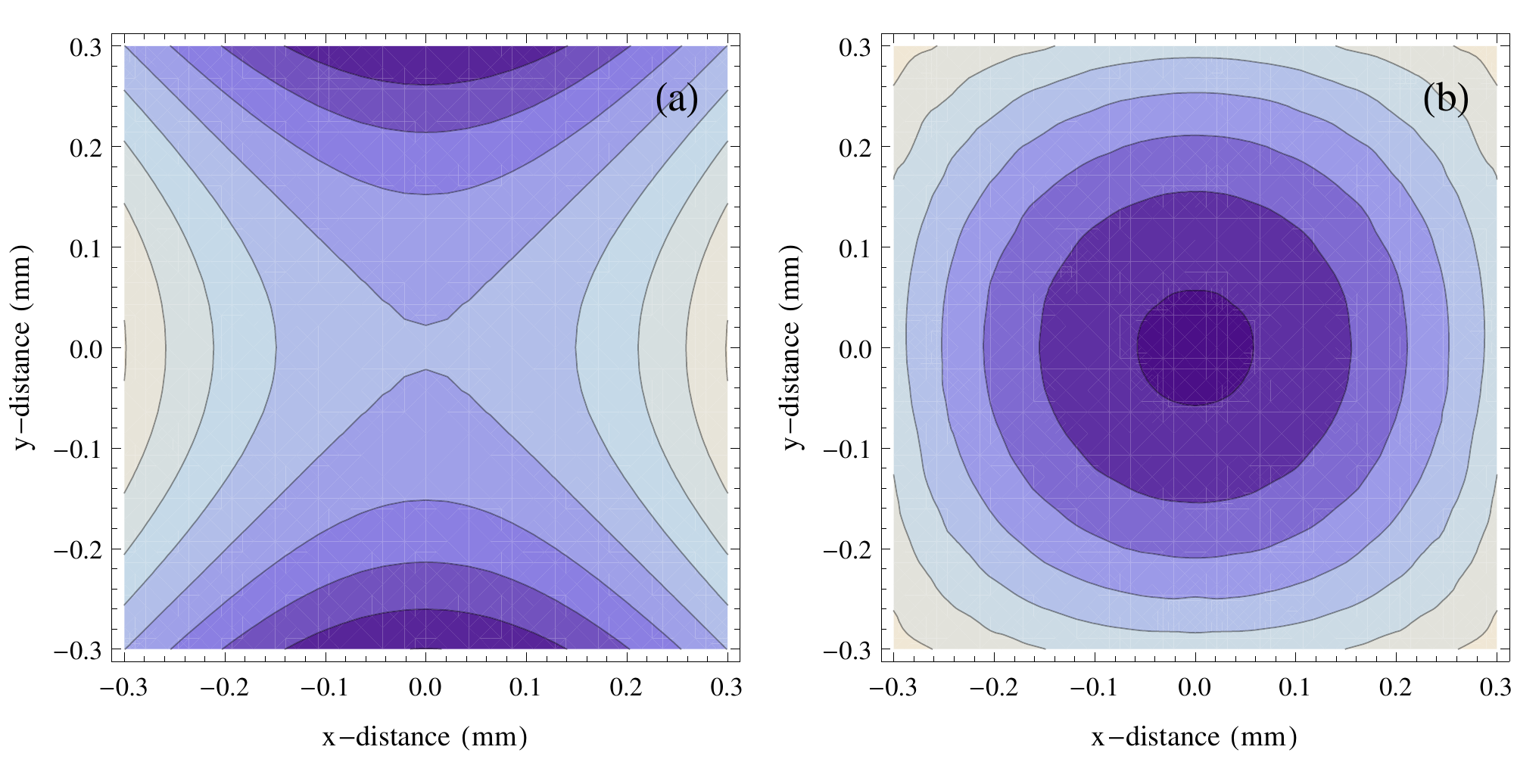}
	\caption{Numerical simulation of four-rod trap potentials. (a) The quadrupole potential produced by application of voltage to the two rf rods, while the other electrodes are held at ground (shown for the point of maximum amplitude in the oscillation).  (b) The pseudopotential derived from the quadrupole potential of (a), using Eq.~\ref{eq:pseudopotential}.  The numerical simulation was performed using the boundary element method (BEM) electrostatic modeling software CPO-3D from Charged Particle Optics, Ltd.}
	\label{fig:simulated_potentials}
\end{figure}

Numerical simulations of the candidate ion trap can be used to ensure the expected properties are consistent with the requirements of the experiment and the capabilities of the available equipment.  Given that the wavelength of oscillation of the potential applied to the electrodes (order of meters) is typically much larger than the size of the trap ($\leq 1$ mm), electrostatic simulations are sufficient.  As an example, consider the four-rod trap illustrated in Fig.~\ref{fig:four_rod_trap}; this trap design is used in all of the experiments presented in the following sections.  The simulated quadrupole potential for the four-rod trap is shown in Fig.~\ref{fig:simulated_potentials}(a), with the resulting pseudopotential illustrated in Fig.~\ref{fig:simulated_potentials}(b).  Comparing Fig.~\ref{fig:simulated_potentials}(a) with the ideal quadrupole potential of Fig.~\ref{fig:hyperbolic_elec}, we see that near the center of the trap the four rods produce a potential nearly identical to that generated by hyperbolic electrodes.

The motion of a charged particle in a trap can be viewed as a three-dimensional, uncoupled harmonic oscillator along the principal axes of the trapping potential.  In other words, motion of the particle along one of the principal axes of a trap is independent of the other two principal axes.  Therefore, the principal axes of the trap present a natural coordinate system for the electrode structure.  In addition, knowledge of the orientation of the principal axes is vital for efficient laser cooling; if the incident light is perpendicular to one of the principal axes, the particle will not be cooled along that direction.  While in simple trap structures such as our four-rod trap, the orientation of the principal axes is clear from the symmetry of the electrodes, in general we determine the principal axes of the trap by use of the Hessian matrix of the potential, which in two-dimensions is:\cite{fitzpatrick:adv_calc}
\begin{equation}
	\label{eq:hessian}
	H(\phi_{sim}(x_{0},y_{0})) = 
	\left(
	\begin{array}{cc}
	\frac{\partial^2 \phi_{sim}}{\partial x^2} (x_{0},y_{0}) & \frac{\partial^2 \phi_{sim}}{\partial x \partial y} (x_{0},y_{0}) \\
	\frac{\partial^2 \phi_{sim}}{\partial y \partial x} (x_{0},y_{0}) & \frac{\partial^2 \phi_{sim}}{\partial y^2} (x_{0},y_{0})
	\end{array}
	\right) .
\end{equation}
Here $\phi_{sim}$ is the simulated trap potential (pseudopotential plus static), and $x_{0}$ and $y_{0}$ are the coordinates where the Hessian matrix is evaluated, which in our case would be the center of the trap.\footnote{In a linear trap, the third principal axis is always clear from structure of the trap.  In Fig.~\ref{fig:four_rod_trap}, it is parallel to the rods of the trap (along the line defined by the needle electrodes).}  The eigenvalues of this matrix are related to the angle by which the principal axes are rotated with respect to the coordinate axes (used in the partial derivatives).  The Hessian matrix seeks out the directions of greatest and least curvature, which are precisely the principal axes.\footnote{Degeneracy of the principle axes can be broken by the addition of a static potential along either $x$ or $y$, as noted in Ref.~\refcite{wineland:nist-jnl}.}

The four-rod trap can now be evaluated for specific parameters, and compared to the ideal hyperbolic case to determine the geometric scale factor, $\eta_{sc}$.  Taking $V_{rf} = 1$ kV, $V_{dc} = 80$ V, $m = 171$ amu, $R = 0.5 \sqrt{2} - 0.25 = 0.46$ mm, and $\Omega_{T} = 38$ MHz, a numercial simulation determines a secular frequency $\omega_{x,sim}/(2 \pi) = 1.17$ MHz for the four-rod trap.  Plugging these same parameters into Eq.~\ref{eq:secular_freq} for the ideal hyperbolic trap, we find $\omega_{x,hyp}/(2 \pi) = 1.26$ MHz.  Thus, the geometric scale factor is $\eta_{sc} = \omega_{x,sim}/\omega_{x,hyp} = 0.93$.  The total depth of the trap can also be determined by the simulation from the depth of the pseudopotential-well, and for these parameters we find the trap depth to be approximately 10 eV.  This trap depth corresponds to a temperture of more than $10^{5}$ K, reiterating the fact that the rf trap provides excellent confinement of charged particles.

\section{Ytterbium Ions}
	\label{sec:ybion}

Trapped atomic ions have long been recognized as a promising implementation of quantum bits (qubits) for quantum information processing,\cite{cirac:cold-ions,wineland:nist-jnl} due in part to long trapping lifetimes, long coherence times of particular internal electronic states, and the exquisite control attained over both the internal and external degress of freedom.  The hydrogen-like ions that have been directly cooled and manipulated for applications in quantum information include Ba${}^{+}$,\cite{devoe:ba_heating,dubin:ion_two-photon,dietrich:ba_qc} Be${}^{+}$,\cite{monroe:qgate} Ca${}^{+}$,\cite{benhelm:ca43,home:ca_entangle,schulz:cooling_multitrap,schuck:atom-sdcp,toyoda:ca_coher} Cd${}^{+}$,\cite{lee:qubit_eom} Mg${}^{+}$,\cite{barrett:symp_be_mg,friedenauer:q-sim_magnet} Sr${}^{+}$,\cite{labaziewicz:heating} and \ybion.\cite{balzer:ybqip,olmschenk:state-detect}  In Table~\ref{tab:ion_compare} various properties of some atomic ions are compared.
%
\begin{table}[ph]
	\tbl{Ion qubit comparison.  The NIST database\protect\cite{sansonetti:nist_database} was used to determine many of the numbers quoted here.}{
	\centering
		\begin{tabular}{|c||ccccccccc|}
			\hline
										 & Be${}^{+}$ & Mg${}^{+}$ & Ca${}^{+}$ & Zn${}^{+}$ & Sr${}^{+}$ & Cd${}^{+}$ & Ba${}^{+}$ 	 & \ybion 	& Hg${}^{+}$ \\
			\hline\hline
			 isotope (amu) & 9 					& 25				 & 40, 43			& 67				 & 87, 88			& 111, 113	 & 135, 137, 138 & 171, 173 & 199, 201 \\
			 nuclear spin	 & 3/2				& 5/2				 & --, 7/2		& 5/2				 & 9/2, --		& 1/2, 1/2	 & 3/2, 3/2, --	 & 1/2, 5/2 & 1/2, 3/2 \\
			 ${}^{2}S_{1/2}$ hfs (GHz) & 1.25 & 1.8  & --, 3.2 		& 7.2 				 & 5, -- 			& 14.5, 15.3 & 7.2, 8, --		 & 12.6, 10.5 & 40.5, 30 \\
			 $P$ fs (THz)  & 0.2				& 2.75			 & 6.7				& 26.2			 & 24					& 75				 & 50.7					 & 100			& 274			 \\
			 ${}^{2}S_{1/2} \leftrightarrow {}^{2}P_{1/2}$ & 313.2 & 280.4 & 397 & 206.3 & 421.7 & 226.5 & 493.5				 & 369.5		& 194.2		 \\
			 ${}^{2}S_{1/2} \leftrightarrow {}^{2}P_{3/2}$ & 313.1 & 279.6 & 393.5 & 202.6 & 407.9 & 214.5 & 455.5			 & 329			& 165			 \\
			 ${}^{2}D_{3/2} \leftrightarrow {}^{2}P_{1/2}$ & -- & -- & 866.5 & -- & 1091.8	& --				 & 649.9				 & 2438			& 10747		 \\
			 ${}^{2}D_{3/2} \leftrightarrow {}^{2}P_{3/2}$ & -- & -- & 850	 & -- & 1003.9	& --				 & 585.5				 & 1350			& 994.7		 \\
			 ${}^{2}D_{3/2} \leftrightarrow {}^{3}[3/2]_{1/2}$ & -- & -- & -- & -- & -- 		& --				 & --						 & 935.2		&	--			 \\
			 ${}^{2}D_{5/2} \leftrightarrow {}^{2}P_{3/2}$ & -- & -- & 854.4 & -- & 1033		& --				 & 614.3				 & 1650			& 398.5		 \\
			\hline
		\end{tabular}
	\label{tab:ion_compare}}
\end{table}
%

The ytterbium ion (\ybion) has several advantages.  The strong ${}^2S_{1/2} \leftrightarrow {}^2P_{1/2}$ electronic transition near 369.5 nm is suitable for use with optical fibers, making schemes that require the coupling of atomic (hyperfine) qubits to photonic (optical) qubits feasible.\cite{duan:robust_qip}  Moreover, the large fine structure splitting of \ybion makes it amenable to fast manipulation with broadband laser pulses.\cite{poyatos:strong-excite,garcia-ripoll:fast-gates,duan:fast-gates,madsen:ultrafast-rabi}  Finally, the spin-1/2 nucleus of ${}^{171}$\ybion allows for simple, fast, and efficient preparation and detection of the ground state hyperfine levels.\cite{olmschenk:state-detect}

\begin{figure}
	\centering
	\includegraphics[width=0.7\columnwidth,keepaspectratio]{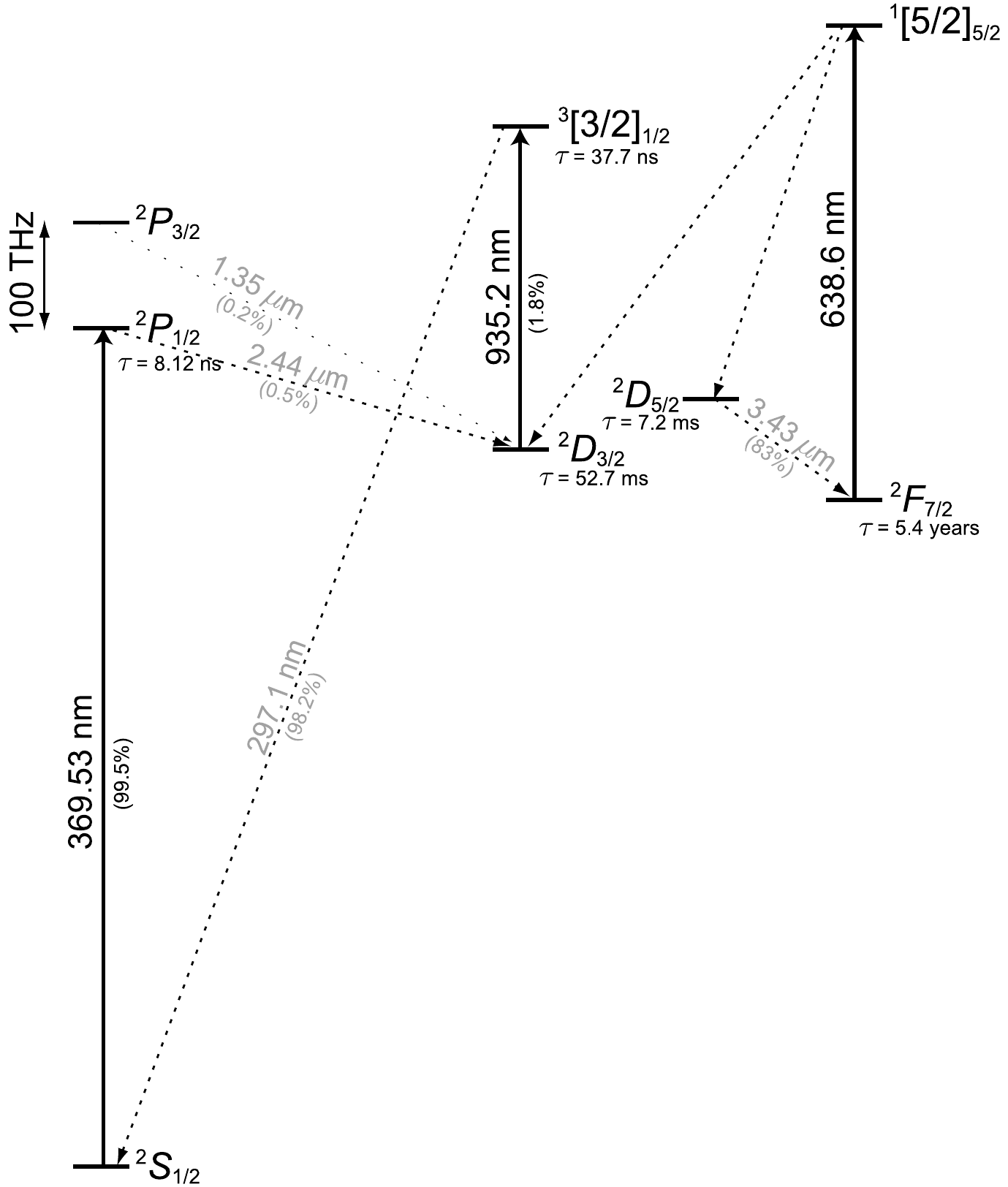}
	\caption{Relevant levels of the \ybion atom (to scale).  Transitions shown with solid lines are driven by laser sources in the experiment.  Numbers given in parentheses are branching ratios.  The lifetimes of some of the excited states are also given.  Measurement of the lifetime and branching ratio for the ${}^{2}P_{1/2}$ level is given in Ref.~\protect\refcite{olmschenk:yb_lifetime} and Ref.~\protect\refcite{olmschenk:state-detect}, respectively.  Wavelengths for decays shown in gray are from Ref.~\protect\refcite{sansonetti:nist_database}.  Lifetimes of the ${}^{2}D_{3/2}$, ${}^{2}D_{5/2}$, ${}^{3}[3/2]_{1/2}$, and ${}^{2}F_{7/2}$ levels are from Refs.~\protect\refcite{yu:lifetime}, \protect\refcite{taylor:d52}, \protect\refcite{berends:beam-laser-YbII}, and \protect\refcite{roberts:171f72}, respectively.  Branching ratios out of ${}^{3}[3/2]_{1/2}$ and ${}^{2}P_{3/2}$ are from Ref.~\protect\refcite{biemont:lifetimeybcalc}, while the ${}^{2}D_{5/2}$ branching ratio is from Ref.~\protect\refcite{taylor:d52}.}
	\label{fig:yb_ion_levels}
\end{figure}

In the experiments presented in the following sections, an ytterbium ion (\ybion) is produced by photoionization of neutral ytterbium using a two-photon, dichroic, resonantly--assisted process.  The resulting \ybion atom is confined in the four-rod rf trap described in the previous section.  The ion is Doppler-cooled by light at 369.5 nm, which is slightly red-detuned of the ${}^2S_{1/2} \leftrightarrow {}^2P_{1/2}$ transition depicted in Fig.~\ref{fig:yb_ion_levels}.  Efficient cooling of the atom requires addtional light, because the ${}^2P_{1/2}$ state also decays to the metastable ${}^2D_{3/2}$ level with a measured probability of about 0.005.\cite{olmschenk:state-detect}  Light at 935.2 nm is used to drive the atom from the ${}^2D_{3/2}$ to the ${}^3[3/2]_{1/2}$ level, from which it quickly returns to the ${}^2S_{1/2}$ ground state.\cite{bell:four-level}  An additional complication arises from the presence of the low-lying ${}^2F_{7/2}$ state.  Despite the fact that there are no allowed decays from the four levels used in cooling to ${}^2F_{7/2}$, the ion falls into this state a few times per hour, probably due to collisions with residual background gas.\cite{lehmitz:pop-trap,bauch:pop-trap,schauer:yb_coll_pop}  Laser light near 638.6 nm depopulates the ${}^2F_{7/2}$ level, returning the ion to the four-level cooling scheme.  Trapping times of several months have been observed.

An external magnetic field of about 5 gauss is applied to provide a quantization axis for definition of the polarization of the photons emitted by the atom, break the degeneracy of the atomic states, and suppress coherent dark state trapping.\cite{berkeland:coherent-pop}  The qubit states are chosen to be the first-order magnetic field-insensitive hyperfine ``clock'' states of the ${}^2S_{1/2}$ level, $\ket{F=0,m_F=0}$ and $\ket{F=1,m_F=0}$, which are separated by 12.6 GHz and defined to be $\ket{0}$ and $\ket{1}$, respectively; here, $F$ is the total angular momentum of the atom, and $m_F$ is its projection along the quantization axis.  Applying light at 369.5 nm resonant with the ${}^2S_{1/2} \ket{F=1} \leftrightarrow {}^2P_{1/2} \ket{F=1}$ transition initializes the qubit to the state $\ket{0}$ by optical pumping.  Subsequently, any combination of $\ket{0}$ and $\ket{1}$ can be prepared by application of microwave radiation at 12.6 GHz with controlled duration and phase.  The state of the atomic qubit is determined by using standard fluorescence techniques.  By illuminating the ion with light at 369.5 nm resonant with the ${}^2S_{1/2} \ket{F=1} \leftrightarrow {}^2P_{1/2} \ket{F=0}$ transition, an ion measured to be in the state $\ket{1}$ scatters many photons, while for an ion in the state $\ket{0}$ the light is off-resonance and almost no photons are scattered.  We can then discriminate between the two qubit states with better than 98\% fidelity.\cite{olmschenk:state-detect}

\section{Two Photon Interference}
	\label{sec:photon_interference}

The interference of identical single photons was first observed by Hong, Ou, and Mandel,\cite{hong:HOM} and Shih and Alley.\cite{shih:HOM}  In these experiments, pairs of photons created by parametric down-conversion were directed to interfere at a beamsplitter.  It was observed that the interference of these identical photons at the beamsplitter resulted in two simultaneously impinging photons always exiting the beamsplitter by the same port.  Thus, coincident detections behind the beamsplitter were highly suppressed.  This quantum interference effect is an essential component of the remote atom entanglement protocols reviewed in the subsequent section.

\subsection{Theoretical interference of photons at a beamsplitter}
	\label{sec:photon-bs-theory}
\begin{figure}
	\centering
	\includegraphics[width=0.5\columnwidth,keepaspectratio]{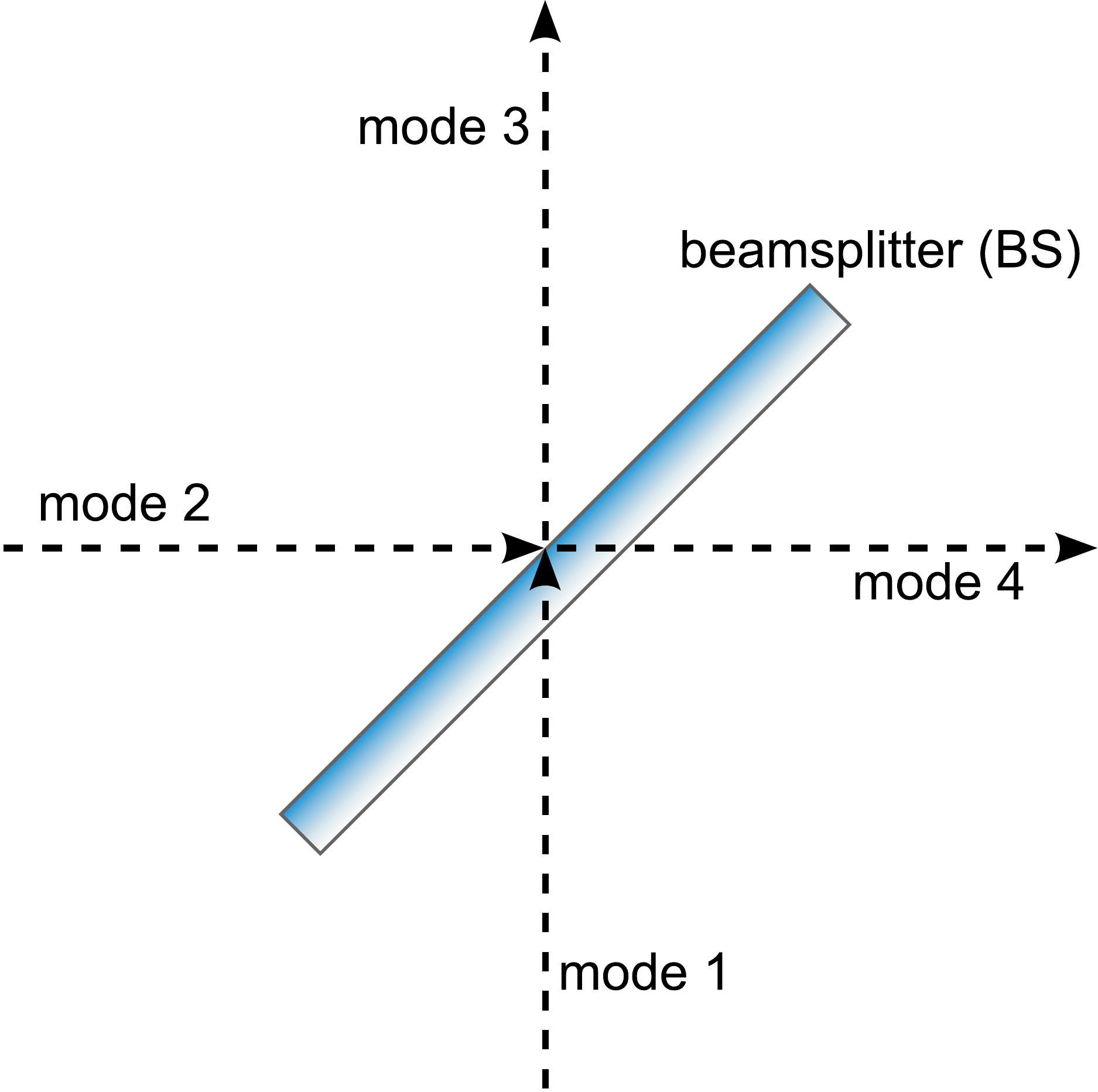}
	\caption{Schematic of a 50:50, nonpolarizing beamsplitter (BS).  Modes 1 and 2 are the two input paths to the beamsplitter, while modes 3 and 4 are the exit paths from the beamsplitter.}
	\label{fig:beamsplitter}
\end{figure}

We consider photons impinging on a 50:50, nonpolarizing beamsplitter, such as the one shown schematically in Fig.~\ref{fig:beamsplitter}.
We can describe photons and the effect of the beamsplitter using the usual photon creation operator $\aup{}$ and the vacuum state $\ket{0}$.  A single photon in mode $n$ (as in Fig.~\ref{fig:beamsplitter}) can then be written as $\ket{1_{n}} = \aup{n} \ket{0}$.  The operation of the beamsplitter is then encompassed by\cite{legero:photon-interfere}
\begin{eqnarray}
	\label{eq:beamsplitter}
	\aup{3} & = & \frac{1}{\sqrt{2}} (\aup{1} - \aup{2}) \nonumber \\
	\aup{4} & = & \frac{1}{\sqrt{2}} (\aup{1} + \aup{2}) .
\end{eqnarray}
Of note is the sign difference between the decomposition of modes 3 and 4 in terms of modes 1 and 2.  The minus sign is the result of the $\pi$-phase shift experienced by light reflected at an interface where the index of refraction changes from low to high\cite{jackson:em} and ensures energy conservation.  Equation~\ref{eq:beamsplitter} may instead be expressed as
\begin{eqnarray}
	\label{eq:ph-beamsplitter}
	\aup{1} & = & \frac{1}{\sqrt{2}} (\aup{3} + \aup{4}) \nonumber \\
	\aup{2} & = & \frac{1}{\sqrt{2}} (-\aup{3} + \aup{4}) ,
\end{eqnarray}
which is effectively a decomposition of incoming photons in terms of the exiting photons.

The above formalism makes it easy to see the effect of the beamsplitter on impinging photons.  We first consider the trivial case of a single photon in mode 1.
\begin{eqnarray}
	\label{eq:single_photon_bs}
	\ket{1_{1}} & = & \aup{1} \ket{0} \nonumber \\
							& = & \frac{1}{\sqrt{2}} (\aup{3} + \aup{4}) \ket{0} \nonumber \\
							& = & \frac{1}{\sqrt{2}} (\ket{1_{3} 0_{4}} + \ket{0_{3} 1_{4}}) .
\end{eqnarray}
As expected, there is an equal (50\%) probability of detecting the photon in mode 3 as in mode 4.\footnote{On a sidenote, notice that Eq.~\ref{eq:single_photon_bs} appears to indicate that the effect of the beamsplitter is to produce an entangled state, yet only one photon is involved.  The remedy to this apparent paradox is to realize that the entanglement here is between two modes (mode 3 and mode 4),\cite{enk:single-particle_entanglment} although there is some controversy over this interpretation.\cite{drezet:comment_single-particle_entanglement,enk:reply_single-particle_entanglment}}
With only a single photon impinging on the beamsplitter, there is zero probability of getting a detection in both mode 3 and mode 4 simultaneously.

We next consider two photons impinging on the beamsplitter: one in mode 1 and one in mode 2.  Following the same procedure as above, we then find
\begin{eqnarray}
	\label{eq:two-photon_bs}
	\ket{1_{1} 1_{2}} & = & \aup{1} \aup{2} \ket{0} \nonumber \\
										& = & \frac{1}{2} (\aup{3} + \aup{4})(-\aup{3} + \aup{4}) \ket{0} \nonumber \\
										& = & \frac{1}{2} ((\aup{4})^{2} - (\aup{3})^{2}) \ket{0} \nonumber \\
										& = & \frac{1}{\sqrt{2}} (\ket{0_{3} 2_{4}} - \ket{2_{3} 0_{4}})
\end{eqnarray}
and again there are no simultaneous detection of photons in modes 3 and 4.  Instead, both photons either go into mode 3, or both photons go into mode 4, with equal probability.  This simple analysis qualitatively explains the two-photon interference effect observed by Hong, Ou, and Mandel,\cite{hong:HOM} and Shih and Alley.\cite{shih:HOM}

Now we generalize this formalism in a way that enables us to describe the ideal expected signal of the experiment.  We closely follow the treatment presented in Ref.~\refcite{legero:photon-interfere}, but expand it to explicitly derive the signal expected from our experimental implementation.  As described in the following subsection, the experiment to demonstrate this two-photon interference effect will consist of repetitive fast excitation of two trapped atomic ions, and detection of the spontaneously emitted photons behind a beamsplitter.  If we assume that the repetitions of the experiment are well-separated (as, indeed, is the design), then we can write the electric field operator of the mode $j$ for the $n^{\mbox{th}}$ repetition as
\begin{equation}
	\label{eq:efield-op_n}
	E_{j}^{(n) +}(t) = \frac{A}{\sqrt{\tau}} \xi(r,t) e^{-\frac{1}{2}(t - n t_{p})/\tau} \Theta(t - n t_{p}) \adown{j}^{(n)} ,
\end{equation}
where $t_{p}$ is the time between repetitions, $\tau$ is the natural lifetime of the excited state, and $\Theta$ is the Heaviside step function.  The exponential decay factor accounts for the probability to detect a photon a given time interval following excitation.  The $1/\sqrt{\tau}$ factor ensures that the total probability of detecting a photon is unitless (integral over time of Eq.~\ref{eq:1st_corr_1ph} below).  The factor $A$ is an amplitude that depends on the photon generation and collection efficiency.  The function $\xi(r,t)$ describes the spatial mode of the photons, and can also be used to account for temporal offsets.  
For the remainder of this section, we ignore both $A$ and $\xi(r,t)$.\footnote{We will find these factors to be useful later, when we calculate the error in the fidelity of the quantum gate and teleportation protocol due to mismatch of the photon spatial modes at the beamsplitter.}
The full electric field operator for the mode $j$ is then simply
\begin{equation}
	\label{eq:efield-op}
	E_{j}^{+}(t) = \sum_{n} E_{j}^{(n) +}(t) .
\end{equation}
The electric field operator describes the probability of detecting a photon at any given time.

The $N+1$ photons produced in the mode $j$ by $N+1$ repetitions of the experiment can be represented by the wavefunction
\begin{equation}
	\label{eq:photon_train}
	\ket{\psi_{N}}_{j} = \left( \prod_{n=0}^{N} \hat{a}_{j}^{(n) \dagger} \right) \ket{0} ,
\end{equation}
where we have assumed that repetitions are sufficiently separated in time so that overlap of photons produced by different repetitions of the experiment is negligible; this allows us to write photons produced by different repetitions as separate ``repetition modes'' denoted by $n$ with the same spatial mode $j$.

\begin{figure}
	\centering
	\includegraphics[width=0.7\columnwidth,keepaspectratio]{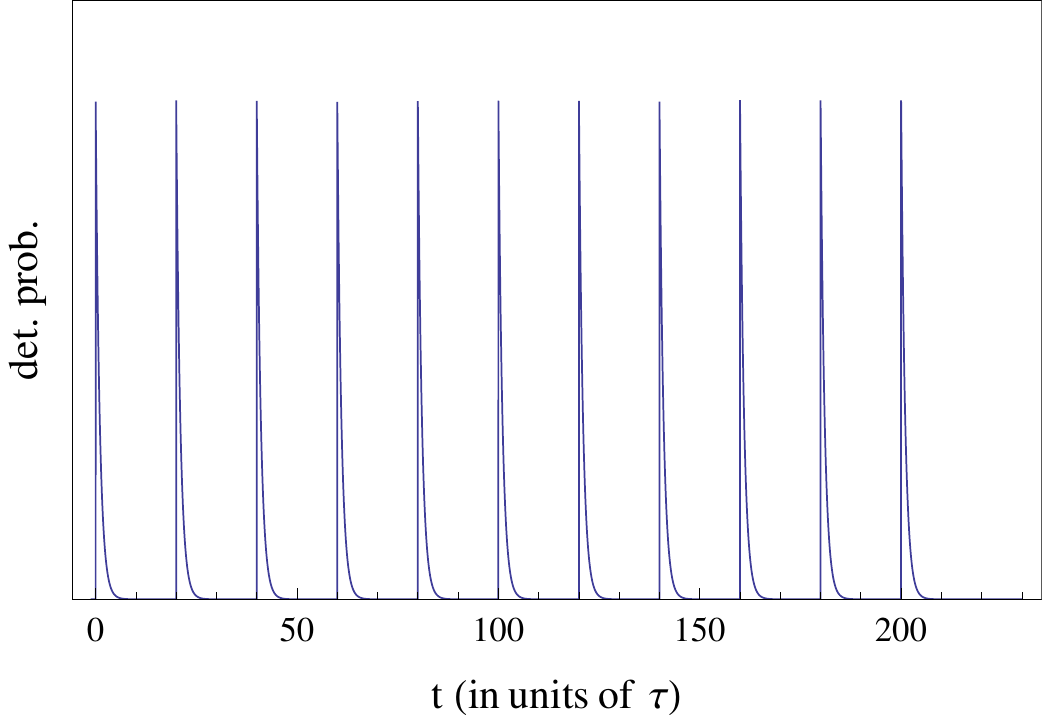}
	\caption{Theoretical single photon first-order correlation function, given by Eq.~\ref{eq:1st_corr_1ph}.  This shows the relative probability of detecting a photon in mode $j$ at time $t$ following a series of 11 excitations ($N = 10$).  The charateristic exponential decay after each repetition/excitation in the experiment is determined by the natural lifetime of the excited state of the atom producing the spontaneously emitted photons.  The repetition rate shown is 20 times the natural lifetime ($\tau$).}
	\label{fig:th_1st_corr_ph}
\end{figure}

\begin{figure}
	\centering
	\includegraphics[width=0.7\columnwidth,keepaspectratio]{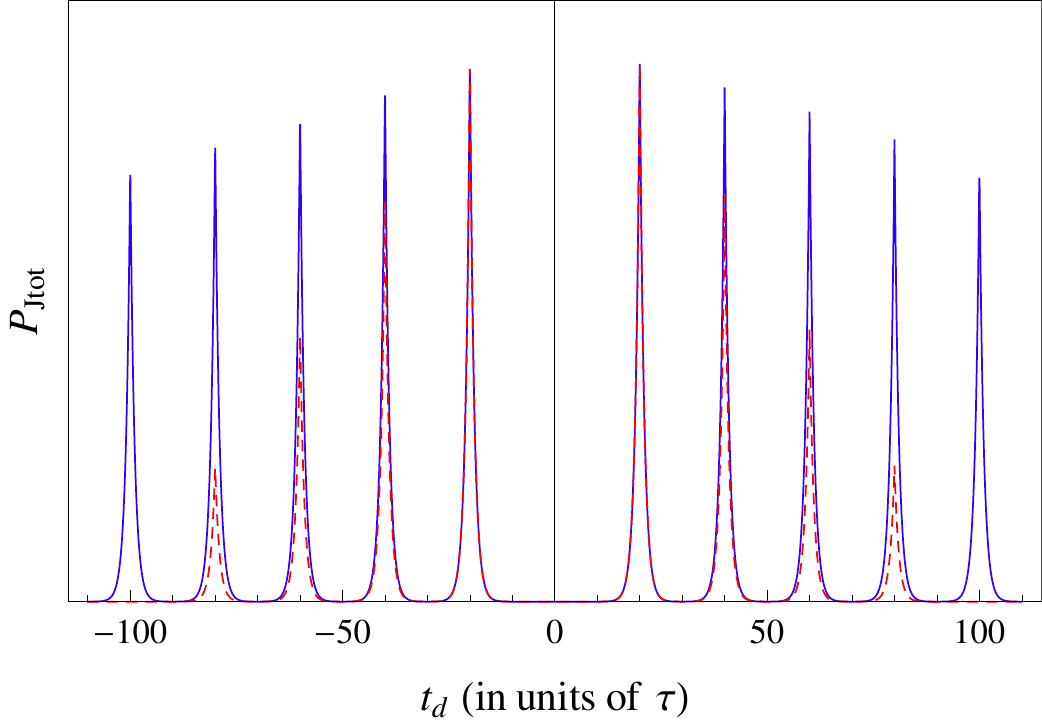}
	\caption{Theoretical single photon second-order correlation function, given by Eq.~\ref{eq:2nd_corr_1ph_last}.  This shows the relative probability of detecting a photon in both mode 3 and mode 4 as a function of the time delay ($t_{d}$) between the two photons for $N = 4$ (\textcolor{red}{red}, dashed) and $N = 20$ (\textcolor{blue}{blue}, solid).  The repetition rate shown is 20 times the natural lifetime of the atomic excited state ($\tau$).  The decrease in the peak amplitude of the probability of joint detection for increasing delay time results from having a finite train of impinging photons.  The lack of a peak at delay time $t_{d} = 0$ is characteristic of a single-photon source.}
	\label{fig:th_single-ph_g2}
\end{figure}

If a train of photons is produced in mode $j$, then the probability of detecting a photon in mode $j$ at time $t$ is simply given by the first-order correlation function
\begin{eqnarray}
	\label{eq:1st_corr_1ph}
	{}_{j}\bra{\psi_{N}} E_{j}^{-}(t) E_{j}^{+}(t) \ket{\psi_{N}}_{j} 
	& = & \bra{0} \left( \prod_{n=0}^{N} \hat{a}_{j}^{(n)} \right) \left( \sum_{k=0}^{N} \frac{1}{\sqrt{\tau}} e^{-\frac{1}{2}(t - k t_{p})/\tau} \Theta(t - k t_{p}) \hat{a}_{j}^{(k) \dagger} \right) \nonumber \\
	&& \times \left( \sum_{m=0}^{N} \frac{1}{\sqrt{\tau}} e^{-\frac{1}{2}(t - m t_{p})/\tau} \Theta(t - m t_{p}) \adown{j}^{(m)} \right) \left( \prod_{n=0}^{N} \hat{a}_{j}^{(n) \dagger} \right) \ket{0} \nonumber \\
	& = & \frac{1}{\tau} \sum_{k,m=0}^{N} e^{-\frac{1}{2}(t - k t_{p})/\tau} e^{-\frac{1}{2}(t - m t_{p})/\tau} \Theta(t - k t_{p}) \Theta(t - m t_{p}) \nonumber \\
	&& \times \bra{0} \left( \prod_{n=0}^{N} \hat{a}_{j}^{(n)} \right) \hat{a}_{j}^{(k) \dagger} \adown{j}^{(m)} \left( \prod_{n=0}^{N} \hat{a}_{j}^{(n) \dagger} \right) \ket{0} \nonumber \\
	& = & \frac{1}{\tau} \sum_{k,m=0}^{N} e^{-\frac{1}{2}(t - k t_{p})/\tau} e^{-\frac{1}{2}(t - m t_{p})/\tau} \Theta(t - k t_{p}) \Theta(t - m t_{p}) \delta_{km} \nonumber \\
	& = & \frac{1}{\tau} \sum_{k=0}^{N} e^{-(t - k t_{p})/\tau} \Theta(t - k t_{p}) ,
\end{eqnarray}
where, as mentioned above, we have ignored both $A$ and $\xi(r,t)$ from Eq.~\ref{eq:efield-op_n}.  Figure~\ref{fig:th_1st_corr_ph} illustrates this result for the case $N = 10$.  As shown in the figure, the probability of detecting a photon from the $n^{\mbox{th}}$ repetition of the experiment decays exponentially with decay constant given by the natural lifetime of the excited state ($\tau$).
The probability for a joint detection in both mode 3 and mode 4, given a train of photons being produced in mode 1, is given by the second-order correlation function
\begin{eqnarray}
	\label{eq:2nd_corr_1ph_1}
	P_{J}(t,t_{d}) & = & {}_{1}\bra{\psi_{N}} E_{3}^{-}(t) E_{4}^{-}(t + t_{d}) E_{4}^{+}(t + t_{d}) E_{3}^{+}(t) \ket{\psi_{N}}_{1} \nonumber \\
										 & = & \bra{0} \left( \prod_{m=0}^{N} \hat{a}_{1}^{(m)} \right)
										       \left( \sum_{k=0}^{N} \frac{1}{\sqrt{\tau}} e^{-\frac{1}{2}(t - k t_{p})/\tau} \Theta(t - k t_{p}) \hat{a}_{3}^{(k) \dagger} \right) \nonumber \\
										    && \times \left( \sum_{n=0}^{N} \frac{1}{\sqrt{\tau}} e^{-\frac{1}{2}(t + t_{d} - n t_{p})/\tau} \Theta(t + t_{d} - n t_{p}) \hat{a}_{4}^{(n) \dagger} \right) \nonumber \\
										    && \times \left( \sum_{n=0}^{N} \frac{1}{\sqrt{\tau}} e^{-\frac{1}{2}(t + t_{d} - n t_{p})/\tau} \Theta(t + t_{d} - n t_{p}) \adown{4}^{(n)} \right) \nonumber \\
										    && \times \left( \sum_{k=0}^{N} \frac{1}{\sqrt{\tau}} e^{-\frac{1}{2}(t - k t_{p})/\tau} \Theta(t - k t_{p}) \adown{3}^{(k)} \right) \left( \prod_{m=0}^{N} \hat{a}_{1}^{(m) \dagger} \right) \ket{0} ,
\end{eqnarray}
where $t_d$ is the time delay between the two detections.  The creation and annhilation operators for modes 3 and 4 can be decomposed in terms of modes 1 and 2, using Eq.~\ref{eq:beamsplitter}.  Since in the above equation there are no photons in mode 2, it is clear that any term with an annihilation operator of mode 2 will immediately vanish.  This allows us to simpifly Eq.~\ref{eq:2nd_corr_1ph_1} to obtain
\begin{eqnarray}
	\label{eq:2nd_corr_1ph_2}
	P_{J}(t,t_{d}) & = & \frac{1}{4 {\tau}^{2}} \bra{0} \left( \prod_{m=0}^{N} \hat{a}_{1}^{(m)} \right)
										       \left( \sum_{k=0}^{N} e^{-\frac{1}{2}(t - k t_{p})/\tau} \Theta(t - k t_{p}) \hat{a}_{1}^{(k) \dagger} \right) \nonumber \\
										    && \times \left( \sum_{n=0}^{N} e^{-\frac{1}{2}(t + t_{d} - n t_{p})/\tau} \Theta(t + t_{d} - n t_{p}) \hat{a}_{1}^{(n) \dagger} \right) \nonumber \\
										    && \times \left( \sum_{n=0}^{N} e^{-\frac{1}{2}(t + t_{d} - n t_{p})/\tau} \Theta(t + t_{d} - n t_{p}) \adown{1}^{(n)} \right) \nonumber \\
										    && \times \left( \sum_{k=0}^{N} e^{-\frac{1}{2}(t - k t_{p})/\tau} \Theta(t - k t_{p}) \adown{1}^{(k)} \right) \left( \prod_{m=0}^{N} \hat{a}_{1}^{(m) \dagger} \right) \ket{0} \nonumber \\
										 & = & \frac{1}{4 {\tau}^{2}} \sum_{k=0}^{N} \sum_{n=0}^{N} e^{-(t - k t_{p})/\tau} e^{-(t + t_{d} - n t_{p})/\tau} \Theta(t - k t_{p}) \Theta(t + t_{d} - n t_{p}) \nonumber \\
										    && \times \bra{0} \left( \prod_{m=0}^{N} \hat{a}_{1}^{(m)} \right)
										 		   \hat{a}_{1}^{(k) \dagger} 
										 		   \hat{a}_{1}^{(n) \dagger}
										 		   \adown{1}^{(n)}
										 		   \adown{1}^{(k)}
										 		   \left( \prod_{m=0}^{N} \hat{a}_{1}^{(m) \dagger} \right) \ket{0} \nonumber \\
										 & = & \frac{1}{4 {\tau}^{2}} \sum_{k=0}^{N} \sum_{n=0}^{N} e^{-(t - k t_{p})/\tau} e^{-(t + t_{d} - n t_{p})/\tau} \Theta(t - k t_{p}) \Theta(t + t_{d} - n t_{p}) \nonumber \\
										    && \times \left( 1 - \delta_{kn} \right) ,
\end{eqnarray}
where the last step is simply a consequence of having an annihilation operator for both $k$ and $n$, but only one creation operator for $m$; thus, if $k = n$ the term goes to zero.  Of course, we are interested in the total probability of a joint detection as a function of the time delay $t_{d}$ between the two photon detections.  In order to obtain this function, we integrate Eq.~\ref{eq:2nd_corr_1ph_2} over $t$:
\begin{eqnarray}
	\label{eq:2nd_corr_1ph_3}
	P_{Jtot}(t_{d}) & = & \int_{-\infty}^{\infty} P_{joint}(t,t_{d}) \mbox{ d}t \nonumber \\
									& = & \frac{1}{4 {\tau}^{2}} \sum_{k=0}^{N} \sum_{n=0}^{N} \left( 1 - \delta_{kn} \right) \nonumber \\
									   && \times \int_{-\infty}^{\infty} e^{-(t - k t_{p})/\tau} e^{-(t + t_{d} - n t_{p})/\tau} \Theta(t - k t_{p}) \Theta(t + t_{d} - n t_{p}) \mbox{ d}t \nonumber \\
									& = & \frac{1}{4 {\tau}^{2}} \sum_{k=0}^{N} \sum_{n=0}^{N} \left( 1 - \delta_{kn} \right) e^{-(t_{d} - t_{p} (k + n))/\tau} \nonumber \\
									   && \times \int_{-\infty}^{\infty} e^{-2t/\tau} \Theta(t - k t_{p}) \Theta(t + t_{d} - n t_{p}) \mbox{ d}t \nonumber \\
									& = & \frac{1}{4 {\tau}^{2}} \sum_{k=0}^{N} \sum_{n=0}^{N} \left( 1 - \delta_{kn} \right) e^{-(t_{d} - t_{p} (k + n))/\tau} \nonumber \\
									   && \times \int_{k t_{p}}^{\infty} e^{-2t/\tau} \Theta(t + t_{d} - n t_{p}) \mbox{ d}t \nonumber \\
									& = & \frac{1}{4 {\tau}^{2}} \sum_{k=0}^{N} \sum_{n=0}^{N} \left( 1 - \delta_{kn} \right) e^{(t_{d} + t_{p} (k - n))/\tau} \nonumber \\
									   && \times \int_{t_{d} + t_{p} (k - n)}^{\infty} e^{-2u/\tau} \Theta(u) \mbox{ d}u .
\end{eqnarray}
In the last line we've set $u = t + t_{d} - n t_{p}$.  The integration is completed by looking at two cases:
\begin{eqnarray}
	\label{eq:2nd_corr_1ph_4}
	P_{Jtot}(t_{d} \leq - t_{p} (k - n)) & = & \frac{1}{4 {\tau}^{2}} \sum_{k=0}^{N} \sum_{n=0}^{N} \left( 1 - \delta_{kn} \right) e^{(t_{d} + t_{p} (k - n))/\tau} \nonumber \\
									   										 && \times \int_{0}^{\infty} e^{-2u/\tau} \Theta(u) \mbox{ d}u \nonumber \\
									   								  & = & \frac{1}{4 {\tau}^{2}} \sum_{k=0}^{N} \sum_{n=0}^{N} \left( 1 - \delta_{kn} \right) e^{(t_{d} + t_{p} (k - n))/\tau} \nonumber \\
									   										 && \times \left[ - \frac{\tau}{2} e^{-2u/\tau} \right]_{0}^{\infty} \nonumber \\
									   								  & = & \frac{1}{8 \tau} \sum_{k=0}^{N} \sum_{n=0}^{N} \left( 1 - \delta_{kn} \right) e^{(t_{d} + t_{p} (k - n))/\tau}
\end{eqnarray}
and
\begin{eqnarray}
	\label{eq:2nd_corr_1ph_5}
	P_{Jtot}(t_{d} \geq - t_{p} (k - n)) & = & \frac{1}{4 {\tau}^{2}} \sum_{k=0}^{N} \sum_{n=0}^{N} \left( 1 - \delta_{kn} \right) e^{(t_{d} + t_{p} (k - n))/\tau} \nonumber \\
									   										 && \times \int_{t_{d} + t_{p} (k - n)}^{\infty} e^{-2u/\tau} \Theta(u) \mbox{ d}u \nonumber \\
									   								  & = & \frac{1}{4 {\tau}^{2}} \sum_{k=0}^{N} \sum_{n=0}^{N} \left( 1 - \delta_{kn} \right) e^{(t_{d} + t_{p} (k - n))/\tau} \nonumber \\
									   										 && \times \left[ - \frac{\tau}{2} e^{-2u/\tau} \right]_{t_{d} + t_{p} (k - n)}^{\infty} \nonumber \\
									   								  & = & \frac{1}{8 \tau} \sum_{k=0}^{N} \sum_{n=0}^{N} \left( 1 - \delta_{kn} \right) e^{-(t_{d} + t_{p} (k - n))/\tau} .
\end{eqnarray}
By inspection, Eq.~\ref{eq:2nd_corr_1ph_4} and \ref{eq:2nd_corr_1ph_5} can be written succinctly as
\begin{equation}
	\label{eq:2nd_corr_1ph_last}
	P_{Jtot}(t_{d}) = \frac{1}{8 \tau} \sum_{k=0}^{N} \sum_{n=0}^{N} \left( 1 - \delta_{kn} \right) e^{-\vert t_{d} + t_{p} (k - n) \vert / \tau} .
\end{equation}
This result is plotted in Fig.~\ref{fig:th_single-ph_g2}.
As expected from Eq.~\ref{eq:single_photon_bs}, the probability of simultaneously detecting a photon in both mode 3 and mode 4 is zero, indicated by the lack of a peak at delay time $t_{d} = 0$.  Experimentally, the number of detection events at $t_{d} = 0$ relative to the height of the adjacent peaks can be used to evaluate the fidelity of single-photon generation.\cite{yuan:single-photon,maunz:interference}

We now look at the signal we would expect from photons impinging on both input ports of the beamsplitter.  Equation~\ref{eq:photon_train} is used to express the photonic wavefunction for a train of pairs of photons in modes 1 and 2 as
\begin{eqnarray}
	\label{eq:two_photon_train}
	\ket{\psi_{N}}_{1,2} & = & \left( \prod_{n=0}^{N} \hat{a}_{1}^{(n) \dagger} \right) \left( \prod_{m=0}^{N} \hat{a}_{2}^{(m) \dagger} \right) \ket{0} \nonumber \\
											 & = & \left( \prod_{n=0}^{N} \hat{a}_{1}^{(n) \dagger} \hat{a}_{2}^{(n) \dagger} \right) \ket{0} ,
\end{eqnarray}
where the second line follows from the fact that $[\hat{a}_{1}^{(n) \dagger}, \hat{a}_{2}^{(m) \dagger}] = 0$ $\forall$ $m,n$.  Here we assume that the ``repetition modes'' are common to the two pulse trains; in other words, we assume to have pairs of identical photons impinging on the beamsplitter.\footnote{Slight temporal mismatch between photons in modes 1 and 2 could be accounted for in the definition of the electric field operator, Eq.~\ref{eq:efield-op_n}.  Since at the present we are interested in deriving the ideal signal that would be measured by this kind of photon interference experiment, we assume the ideal case of photons from the same repetition having perfect temporal overlap.}

The second-order correlation function for pairs of photons can then be calculated in a way similar to Eq.~\ref{eq:2nd_corr_1ph_1}.  
\begin{eqnarray}
	\label{eq:2nd_corr_2ph_1}
	P_{2ph,J}(t,t_{d}) & = & {}_{1,2}\bra{\psi_{N}} E_{3}^{-}(t) E_{4}^{-}(t + t_{d}) E_{4}^{+}(t + t_{d}) E_{3}^{+}(t) \ket{\psi_{N}}_{1,2} \nonumber \\
										 & = & \bra{0} \left( \prod_{m=0}^{N} \hat{a}_{1}^{(m)} \hat{a}_{2}^{(m)} \right)
										       \left( \sum_{k=0}^{N} \frac{1}{\sqrt{\tau}} e^{-\frac{1}{2}(t - k t_{p})/\tau} \Theta(t - k t_{p}) \hat{a}_{3}^{(k) \dagger} \right) \nonumber \\
										    && \times \left( \sum_{n=0}^{N} \frac{1}{\sqrt{\tau}} e^{-\frac{1}{2}(t + t_{d} - n t_{p})/\tau} \Theta(t + t_{d} - n t_{p}) \hat{a}_{4}^{(n) \dagger} \right) \nonumber \\
										    && \times \left( \sum_{n=0}^{N} \frac{1}{\sqrt{\tau}} e^{-\frac{1}{2}(t + t_{d} - n t_{p})/\tau} \Theta(t + t_{d} - n t_{p}) \adown{4}^{(n)} \right) \nonumber \\
										    && \times \left( \sum_{k=0}^{N} \frac{1}{\sqrt{\tau}} e^{-\frac{1}{2}(t - k t_{p})/\tau} \Theta(t - k t_{p}) \adown{3}^{(k)} \right) \nonumber \\
										    && \times \left( \prod_{m=0}^{N} \hat{a}_{1}^{(m) \dagger} \hat{a}_{2}^{(m) \dagger} \right) \ket{0} .
\end{eqnarray}
Of course, the time dependence of Eq.~\ref{eq:2nd_corr_2ph_1} appears identical to Eq.~\ref{eq:2nd_corr_1ph_2}.  Thus, the integration over time will be identical to the single-photon case, and so for the present calculation we only need to focus on the effect of the additional creation and annihilation operators.  Writing the operators for modes 3 and 4 in terms of modes 1 and 2 yields
\begin{eqnarray}
	\label{eq:2nd_corr_2ph_2}
	P_{2ph,J}(t,t_{d}) & \propto & \bra{0} \left( \prod_{m=0}^{N} \hat{a}_{1}^{(m)} \hat{a}_{2}^{(m)} \right)
										 		   \hat{a}_{3}^{(k) \dagger} 
										 		   \hat{a}_{4}^{(n) \dagger}
										 		   \adown{4}^{(n)}
										 		   \adown{3}^{(k)}
										 		   \left( \prod_{m=0}^{N} \hat{a}_{1}^{(m) \dagger} \hat{a}_{2}^{(m) \dagger} \right) \ket{0} \nonumber \\
										 & = & \frac{1}{4} 
										 			 \bra{0} \left( \prod_{m=0}^{N} \hat{a}_{1}^{(m)} \hat{a}_{2}^{(m)} \right)
										 		   \left( \hat{a}_{1}^{(k) \dagger} - \hat{a}_{2}^{(k) \dagger} \right)
										 		   \left( \hat{a}_{1}^{(n) \dagger} + \hat{a}_{2}^{(n) \dagger} \right) \nonumber \\
										 		&& \times 
										 		   \left( \adown{1}^{(n)} + \adown{2}^{(n)} \right)
										 		   \left( \adown{1}^{(k)} - \adown{2}^{(k)} \right)
										 		   \left( \prod_{m=0}^{N} \hat{a}_{1}^{(m) \dagger} \hat{a}_{2}^{(m) \dagger} \right) \ket{0} \nonumber \\
										 & = & \frac{1}{4}
										 			 \bra{0} \left( \prod_{m=0}^{N} \hat{a}_{1}^{(m)} \hat{a}_{2}^{(m)} \right)
										 			 \left( \hat{a}_{1}^{(k) \dagger} \hat{a}_{1}^{(n) \dagger} \adown{1}^{(n)} \adown{1}^{(k)}
										 			 - \hat{a}_{2}^{(k) \dagger} \hat{a}_{1}^{(n) \dagger} \adown{2}^{(n)} \adown{1}^{(k)} \right. \nonumber \\
										 	  && + \hat{a}_{2}^{(k) \dagger} \hat{a}_{1}^{(n) \dagger} \adown{1}^{(n)} \adown{2}^{(k)}
										 	     + \hat{a}_{1}^{(k) \dagger} \hat{a}_{2}^{(n) \dagger} \adown{2}^{(n)} \adown{1}^{(k)}
										 	     - \hat{a}_{1}^{(k) \dagger} \hat{a}_{2}^{(n) \dagger} \adown{1}^{(n)} \adown{2}^{(k)} \nonumber \\
										 	  && \left. + \hat{a}_{2}^{(k) \dagger} \hat{a}_{2}^{(n) \dagger} \adown{2}^{(n)} \adown{2}^{(k)} \right)
										 	     \left( \prod_{m=0}^{N} \hat{a}_{1}^{(m) \dagger} \hat{a}_{2}^{(m) \dagger} \right) \ket{0} \nonumber \\
										 & = & \left( 1 - \delta_{kn} \right) - \delta_{kn} + 1 + 1 - \delta_{kn} + (1 - \delta_{kn}) \nonumber \\
										 & = & 4 \left( 1 - \delta_{kn} \right) ,
\end{eqnarray}
where the sum over $k$ and $n$ is implicit.  The final solution, after integrating over $t$ (with integrals identical to Eq.~\ref{eq:2nd_corr_1ph_3}), is
\begin{equation}
	\label{eq:2nd_corr_2ph_last}
	P_{2ph,Jtot}(t_{d}) = \frac{1}{2 \tau} \sum_{k=0}^{N} \sum_{n=0}^{N} \left( 1 - \delta_{kn} \right) e^{-\vert t_{d} + t_{p} (k - n) \vert / \tau} ,
\end{equation}
which is just Eq.~\ref{eq:2nd_corr_1ph_last} multiplied by a factor of 4.  This indicates that there is zero probability of obtaining simultaneous dectections in both modes 3 and 4 if the incoming photons in modes 1 and 2 are identical, as was determined qualitatively in Eq.~\ref{eq:two-photon_bs}.  The additional factor of 4 results because in this case one photon in mode 3 and one photon in mode 4 separated by non-zero delay time $t_{d}$ can be the result of: 2 sequential photons in mode 1; 2 sequential photons in mode 2; 1 photon in mode 1 followed by 1 photon in mode 2; 1 photon in mode 2 followed by 1 photon in mode 1.

Note that the fidelity of a two-photon interference experiment can be evaluated by taking the ratio of the number of detection events at $t_{d} = 0$ to $1/2$ the number detected at a multiple of the experiment repetition rate $t_{d} = n t_{p}$.  The factor of $1/2$ is derived from the second-order correlation function for an incoming train of \textit{distinguishable} photons.  Assuming the incoming photons in mode 1 can be distinguished from those in mode 2 (for instance, by polarization), then the two-photon wavefunction from Eq.~\ref{eq:two_photon_train} is
\begin{equation}
	\label{eq:two_photon_train_distinct}
	\ket{\psi_{N}}_{1a,2b} = \left( \prod_{n=0}^{N} \hat{a}_{1}^{(n) \dagger} \hat{b}_{2}^{(n) \dagger} \right) \ket{0} ,
\end{equation}
with $[\aup{i},\bdown{j}] = 0$ $\forall$ $i,j$.  The four possible (distinguishable) cases are: detecting an $\aup{}$ photon in mode 3 and a $\bup{}$ photon in mode 4; detecting a $\bup{}$ photon in mode 3 and an $\aup{}$ photon in mode 4; detecting an $\aup{}$ photon in both mode 3 and mode 4; and detecting a $\bup{}$ photon in both mode 3 and mode 4.  Thus, the second-order correlation function is
\begin{eqnarray}
	\label{eq:2nd_corr_ab_1}
	P_{a,b,J}(t,t_{d}) & = & {}_{1a,2b}\bra{\psi_{N}} E_{3a}^{-}(t) E_{4b}^{-}(t + t_{d}) E_{4b}^{+}(t + t_{d}) E_{3a}^{+}(t) \ket{\psi_{N}}_{1a,2b} \nonumber \\
												&& + {}_{1a,2b}\bra{\psi_{N}} E_{3b}^{-}(t) E_{4a}^{-}(t + t_{d}) E_{4a}^{+}(t + t_{d}) E_{3b}^{+}(t) \ket{\psi_{N}}_{1a,2b} \nonumber \\
												&& + {}_{1a,2b}\bra{\psi_{N}} E_{3a}^{-}(t) E_{4a}^{-}(t + t_{d}) E_{4a}^{+}(t + t_{d}) E_{3a}^{+}(t) \ket{\psi_{N}}_{1a,2b} \nonumber \\
												&& + {}_{1a,2b}\bra{\psi_{N}} E_{3b}^{-}(t) E_{4b}^{-}(t + t_{d}) E_{4b}^{+}(t + t_{d}) E_{3b}^{+}(t) \ket{\psi_{N}}_{1a,2b} \nonumber \\
										 & = & 2 {}_{1a,2b}\bra{\psi_{N}} E_{3a}^{-}(t) E_{4b}^{-}(t + t_{d}) E_{4b}^{+}(t + t_{d}) E_{3a}^{+}(t) \ket{\psi_{N}}_{1a,2b} \nonumber \\
										    && + 2 {}_{1a,2b}\bra{\psi_{N}} E_{3a}^{-}(t) E_{4a}^{-}(t + t_{d}) E_{4a}^{+}(t + t_{d}) E_{3a}^{+}(t) \ket{\psi_{N}}_{1a,2b} ,
\end{eqnarray}
where the last line follows from symmetry.  In addition, since $[\aup{i},\bdown{j}] = 0$ $\forall$ $i,j$, the second term in Eq.~\ref{eq:2nd_corr_ab_1} is equal to (two times) the second-order correlation function for a train of single-photons (Eq.~\ref{eq:2nd_corr_1ph_1}).  On the other hand, the first term is
\begin{eqnarray}
	\label{eq:2nd_corr_ab_2}
	&& 2 {}_{1a,2b}\bra{\psi_{N}} E_{3a}^{-}(t) E_{4b}^{-}(t + t_{d}) E_{4b}^{+}(t + t_{d}) E_{3a}^{+}(t) \ket{\psi_{N}}_{1a,2b} \nonumber \\
										 & = & 2 \bra{0} \left( \prod_{m=0}^{N} \hat{a}_{1}^{(m)} \hat{b}_{2}^{(m)} \right)
										       \left( \sum_{k=0}^{N} \frac{1}{\sqrt{\tau}} e^{-\frac{1}{2}(t - k t_{p})/\tau} \Theta(t - k t_{p}) \hat{a}_{3}^{(k) \dagger} \right) \nonumber \\
										    && \times \left( \sum_{n=0}^{N} \frac{1}{\sqrt{\tau}} e^{-\frac{1}{2}(t + t_{d} - n t_{p})/\tau} \Theta(t + t_{d} - n t_{p}) \hat{b}_{4}^{(n) \dagger} \right) \nonumber \\
										    && \times \left( \sum_{n=0}^{N} \frac{1}{\sqrt{\tau}} e^{-\frac{1}{2}(t + t_{d} - n t_{p})/\tau} \Theta(t + t_{d} - n t_{p}) \bdown{4}^{(n)} \right) \nonumber \\
										    && \times \left( \sum_{k=0}^{N} \frac{1}{\sqrt{\tau}} e^{-\frac{1}{2}(t - k t_{p})/\tau} \Theta(t - k t_{p}) \adown{3}^{(k)} \right) \left( \prod_{m=0}^{N} \hat{a}_{1}^{(m) \dagger} \hat{b}_{2}^{(m) \dagger} \right) \ket{0} \nonumber \\
										 & = & \frac{2}{\tau^{2}} \sum_{k=0}^{N} \sum_{n=0}^{N} e^{-(t - k t_{p})/\tau} e^{-(t + t_{d} - n t_{p})/\tau} \Theta(t - k t_{p}) \Theta(t + t_{d} - n t_{p}) \nonumber \\
										    && \times
										    	 \bra{0} \left( \prod_{m=0}^{N} \hat{a}_{1}^{(m)} \hat{b}_{2}^{(m)} \right)
										       \hat{a}_{3}^{(k) \dagger}
										       \hat{b}_{4}^{(n) \dagger}
										       \bdown{4}^{(n)}
										       \adown{3}^{(k)} 
										       \left( \prod_{m=0}^{N} \hat{a}_{1}^{(m) \dagger} \hat{b}_{2}^{(m) \dagger} \right) \ket{0} .
\end{eqnarray}
Writing the operators for modes 3 and 4 in terms of 1 and 2 yields
\begin{eqnarray}
	\label{eq:2nd_corr_ab_3}
	&& 2 {}_{1a,2b}\bra{\psi_{N}} E_{3a}^{-}(t) E_{4b}^{-}(t + t_{d}) E_{4b}^{+}(t + t_{d}) E_{3a}^{+}(t) \ket{\psi_{N}}_{1a,2b} \nonumber \\
										 & = & \frac{1}{2 \tau^{2}} \sum_{k=0}^{N} \sum_{n=0}^{N} e^{-(t - k t_{p})/\tau} e^{-(t + t_{d} - n t_{p})/\tau} \Theta(t - k t_{p}) \Theta(t + t_{d} - n t_{p}) \nonumber \\
										    && \times
										    	 \bra{0} \left( \prod_{m=0}^{N} \hat{a}_{1}^{(m)} \hat{b}_{2}^{(m)} \right)
										       \left( \hat{a}_{1}^{(k) \dagger} - \hat{a}_{2}^{(k) \dagger} \right)
										       \left( \hat{b}_{1}^{(n) \dagger} + \hat{b}_{2}^{(n) \dagger} \right) \nonumber \\
										    && \times \left( \bdown{1}^{(n)} + \bdown{2}^{(n)} \right)
										       \left( \adown{1}^{(k)} - \adown{2}^{(k)} \right)
										       \left( \prod_{m=0}^{N} \hat{a}_{1}^{(m) \dagger} \hat{b}_{2}^{(m) \dagger} \right) \ket{0} \nonumber \\
										 & = & \frac{1}{2 \tau^{2}} \sum_{k=0}^{N} \sum_{n=0}^{N} e^{-(t - k t_{p})/\tau} e^{-(t + t_{d} - n t_{p})/\tau} \Theta(t - k t_{p}) \Theta(t + t_{d} - n t_{p}) \nonumber \\
										    && \times
										    	 \bra{0} \left( \prod_{m=0}^{N} \hat{a}_{1}^{(m)} \hat{b}_{2}^{(m)} \right)
										    	 \hat{a}_{1}^{(k) \dagger}
										    	 \hat{b}_{2}^{(n) \dagger}
										    	 \bdown{2}^{(n)}
										    	 \adown{1}^{(k)}
										    	 \left( \prod_{m=0}^{N} \hat{a}_{1}^{(m) \dagger} \hat{b}_{2}^{(m) \dagger} \right) \ket{0} \nonumber \\
										 & = & \frac{1}{2 \tau^{2}} \sum_{k=0}^{N} \sum_{n=0}^{N} e^{-(t - k t_{p})/\tau} e^{-(t + t_{d} - n t_{p})/\tau} \Theta(t - k t_{p}) \Theta(t + t_{d} - n t_{p}) .
\end{eqnarray}
\begin{figure}
	\centering
	\includegraphics[width=0.7\columnwidth,keepaspectratio]{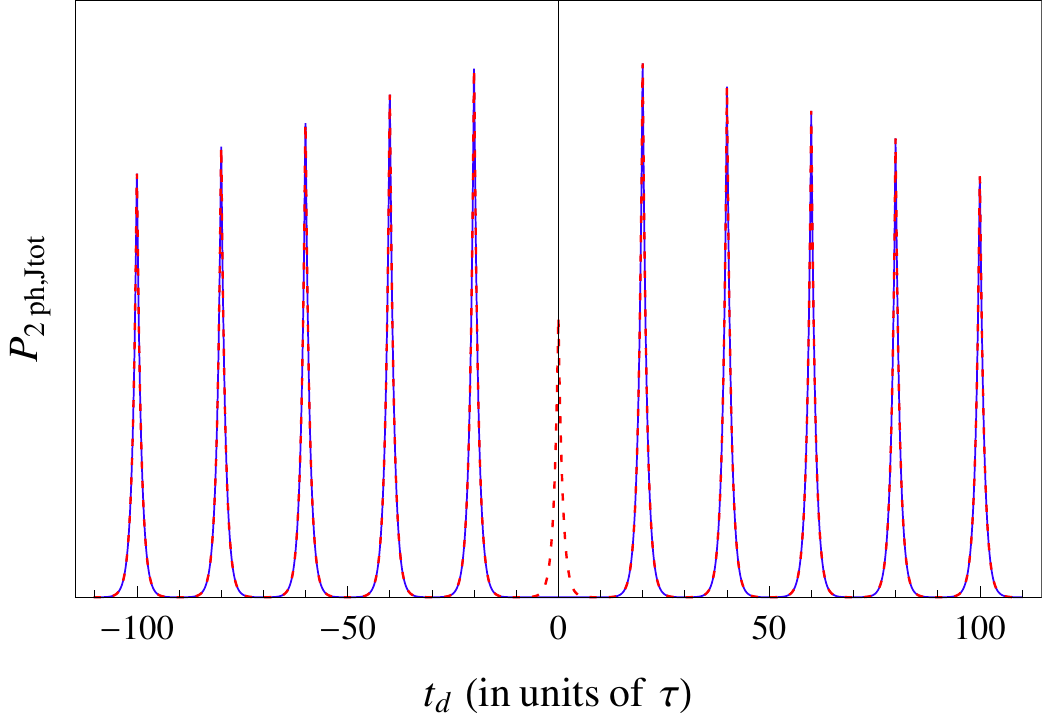}
	\caption[Theoretical two-photon second-order correlation function]{Theoretical two-photon second-order correlation function, for identical (\textcolor{blue}{blue}, solid line; Eq.~\ref{eq:2nd_corr_2ph_last}) and distinguishable (\textcolor{red}{red}, dotted line; Eq.~\ref{eq:2nd_corr_ab_last}) photons.  Plotted is the relative probability of detecting a photon in both mode 3 and mode 4 as a function of the time delay ($t_{d}$) between the two photons for $N = 20$.  The repetition rate shown is 20 times the natural lifetime of the atomic excited state ($\tau$).  The decrease in the peak amplitude of the probability of joint detection for increasing delay time results from having a finite train of impinging photons.  Interference between identical photons impinging on the beamsplitter results in no events at zero time delay $t_{d} = 0$ (\textcolor{blue}{blue}, solid line).  In contrast, distinguishable photons do not interfere, and thus coincident detections are allowed, as evidenced by the peak at $t_{d} = 0$ (\textcolor{red}{red}, dotted line).}
	\label{fig:th_two-ph_g2}
\end{figure}
Plugging this into Eq.~\ref{eq:2nd_corr_ab_1} and integrating over $t$ (again, integrals identical to Eq.~\ref{eq:2nd_corr_1ph_2}), we obtain the final form for the second-order correlation function for distinguishable photons:
\begin{eqnarray}
	\label{eq:2nd_corr_ab_last}
	P_{2ph,Jtot}(t_{d}) = \frac{1}{4 \tau} \sum_{k=0}^{N} \sum_{n=0}^{N} \left( 2 - \delta_{kn} \right) e^{-\vert t_{d} + t_{p} (k - n) \vert / \tau} .
\end{eqnarray}
This result for distinguishable photons impinging on the beamsplitter is plotted against the case of identical photons (Eq.~\ref{eq:2nd_corr_2ph_last}) in Fig.~\ref{fig:th_two-ph_g2}.

\subsection{Experimental interference of photons at a beamsplitter}
	\label{sec:photon-bs-expt}
\begin{figure}
	\centering
	\includegraphics[width=1.0\columnwidth,keepaspectratio]{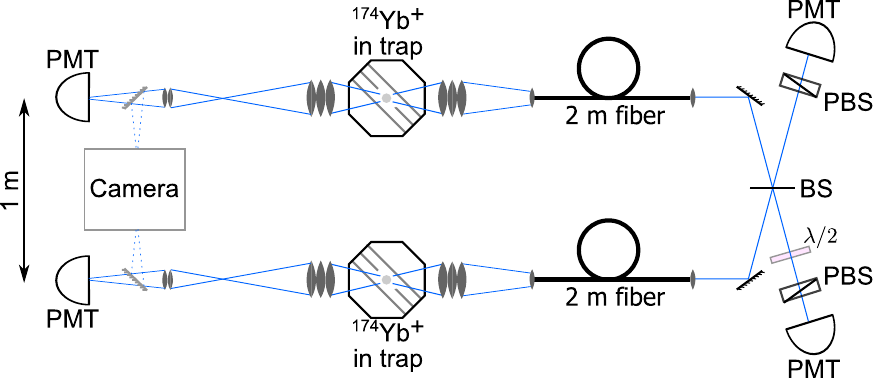}
	\caption{Experimental setup for the two-photon interference experiment.  Fluorescence of an \ybion atom during Doppler cooling can be monitored by either a photon-counting photomultiplier (PMT) or a camera (the grayed mirror is removable by a motorized flip-mount).  The second-order correlation function (joint detection probability) is measured by coupling spontaneously emitted photons from the ion into a single-mode optical fiber, the output of which is directed to a 50:50 nonpolarizing beamsplitter (BS).  Polarizers (PBS) and a $\lambda/2$-waveplate ($\lambda/2$) are removable for the measurement of unpolarized, parallel polarized, or perpendicularly polarized photons.}
	\label{fig:interference_expt_setup}
\end{figure}

The interference of single photons at a beamsplitter was accomplished by using two trapped ${}^{174}$\ybion atoms as single photon emitters.\cite{maunz:interference}  A single ytterbium ion is confined in each of two nearly identical four-rod rf traps that are contained in independent vacuum chambers.  As described in Sec.~\ref{sec:ybion}, the ions are Doppler-cooled by cw laser light at 369.5 nm, and the presence of a single atom in each trap is confirmed by imaging this fluorescence on a camera.  In addition to the objective lens used to observe the ion fluorescence on the camera, we position a second objective with effective numerical aperture $\sim$0.3 on the other side of the vacuum chamber, as illustrated in Fig.~\ref{fig:interference_expt_setup}.  Photons collected by this objective are coupled into a single-mode optical fiber and directed to a 50:50 nonpolarizing beamsplitter.\footnote{The optical fiber is StockerYale NUV-320-K1.  The fiber is single-mode at 370 nm, with a specified attenuation of approximately 0.1 dB/meter.}  The output of the fiber is incident on the beamsplitter at a small angle (about $10^{\circ}$) to obtain polarization independence.  The photons are detected by two PMTs, positioned at each exit port of the beamsplitter.  Polarizers on motorized flip-mounts can be added to the beam path between the beamsplitter and each PMT to enable detection of only identically (parallel) polarized photons.  Conversely, distinguishable (perpendicularly polarized) photons are detected by the addition of a $\lambda/2$-waveplate to one of these beam paths.  The arrival times of the photons are recorded by a time-to-digital converter (TDC) with 4 ps resolution.\footnote{The TDC is a PicoHarp 300, made by PicoQuant.}

Ultrafast excitation of an \ybion atom from ${}^{2}S_{1/2}$ to ${}^{2}P_{1/2}$ results in the generation of a single spontaneously emitted photon.  We use pulses of about 1 ps duration with central wavelength at 369.5 nm to drive the ${}^{2}S_{1/2} \leftrightarrow {}^{2}P_{1/2}$ transition in the atom.  Since the duration of the pulse is much shorter than the excited state lifetime (8.12 ns),\cite{olmschenk:yb_lifetime} the probability that the atom scatters two photons during a single pulse is approximately given by the probability of one spontaneous decay over the duration of the pulse: $1 - e^{-(1 \mbox{ ps})/(8.12 \mbox{ ns})} \approx 10^{-4}$.

\begin{figure}
	\centering
	\includegraphics[width=0.7\columnwidth,keepaspectratio]{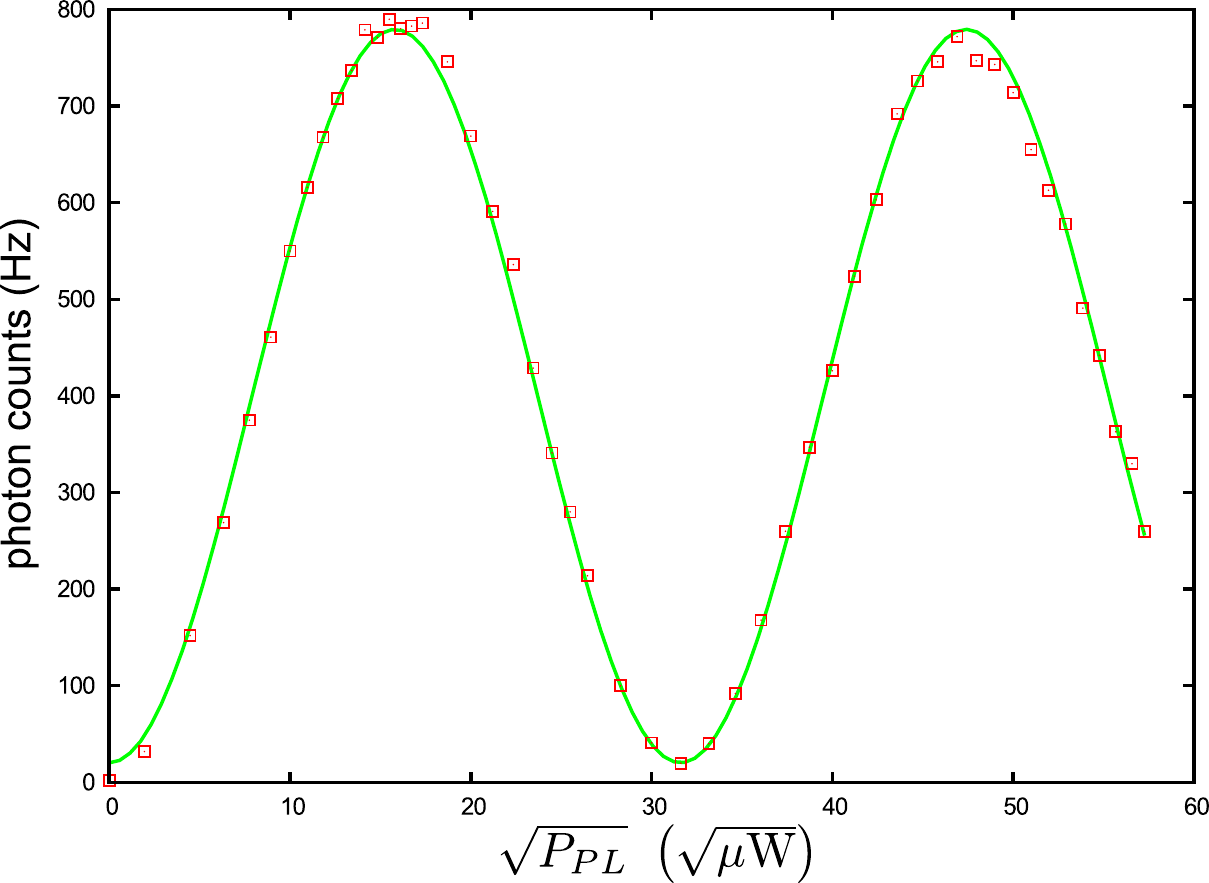}
	\caption{Resonant picosecond Rabi flopping.  The scattering rate of a single ${}^{174}$\ybion atom is monitored as a train of 1 ps pulses of variable power excite the atom.  The Rabi frequency for driving the atom from ${}^{2}S_{1/2}$ to ${}^{2}P_{1/2}$ is proportional to the square-root of the incident pulse power.  At maximum pulse energy, we achieve about a $3.5\pi$ rotation.  The \textcolor{green}{green} curve is a fit to the function $A \sin^{2}(B \sqrt{P_{PL}}/2)$, where $P_{PL}$ is the average power of the pulsed laser incident on the ion, and $A$ and $B$ are the fit parameters.}
	\label{fig:ps_rabi_flopping_sqrt_power}
\end{figure}

The 1 ps ultrafast pulses are produced by an actively mode-locked Ti:sapphire laser (Spectra-Physics Tsunami) operating at a central wavelength of 739 nm and a repetition rate of about 81 MHz.  This pulsed laser is pumped by approximately 6 W of 532 nm light (Spectra-Physics Millennia), producing an average output power of 1 W at 739 nm.  We measured the width of the pulses to be 1 ps by using a simple autocorrelator, consisting of a Michelson interferometer and a lithium triborate (LBO) crystal with a photodetector.  The pulses are passed through an electro-optic pulse picker that has an average extinction ratio better than 100:1 in the infrared.  With the pulse picker synced to the intracavity acousto-optic modulator (AOM) of the pulse laser, we reduce the effective pulse repetition rate to about 8.1 MHz by allowing only one in every ten pulses to pass.  A critically phase-matched LBO crystal is used to frequency double each pulse, resulting in approximately 80 mW average power at 369.5 nm.  
This second-harmonic light is separated from the fundamental infrared by a prism.  In addition to creating the wavelength needed to resonantly excite the ion, frequency-doubling increases the effective average extinction ratio of the pulse picker to about $10^{4}$:1.\footnote{The immediately trailing pulse is not extinquished as well as the other pulses (only about $10^{3}$:1), due to the finite switching of the bias voltage applied to the pulse picker.  Averaging over all pulses though yields an extinction ratio of approximately $10^{4}$:1.}  Even though the second-harmonic generation efficiency is only about 8\%, the resulting pulses at 369.5 nm have more than enough power to excite an atom with unit probability.  This is demonstrated in Fig.~\ref{fig:ps_rabi_flopping_sqrt_power}, where the scattering rate of a single atom is measured as a function of the incident pulse power, indicating each pulse has enough power to drive about a $3.5\pi$ rotation between the ${}^{2}S_{1/2}$ and ${}^{2}P_{1/2}$ levels.  As the Rabi frequency is proportional to the square root of the incident pulse power, this means that a single pulse could be split into about 12 pulses, each with enough power to drive a $\pi$ rotation.\cite{madsen:ultrafast-rabi}  In the experiments presented here, each pulse is attenuated to achieve approximately unit excitation probability ($1\pi$ pulse).

\begin{figure}
	\centering
	\includegraphics[width=0.6\columnwidth,keepaspectratio]{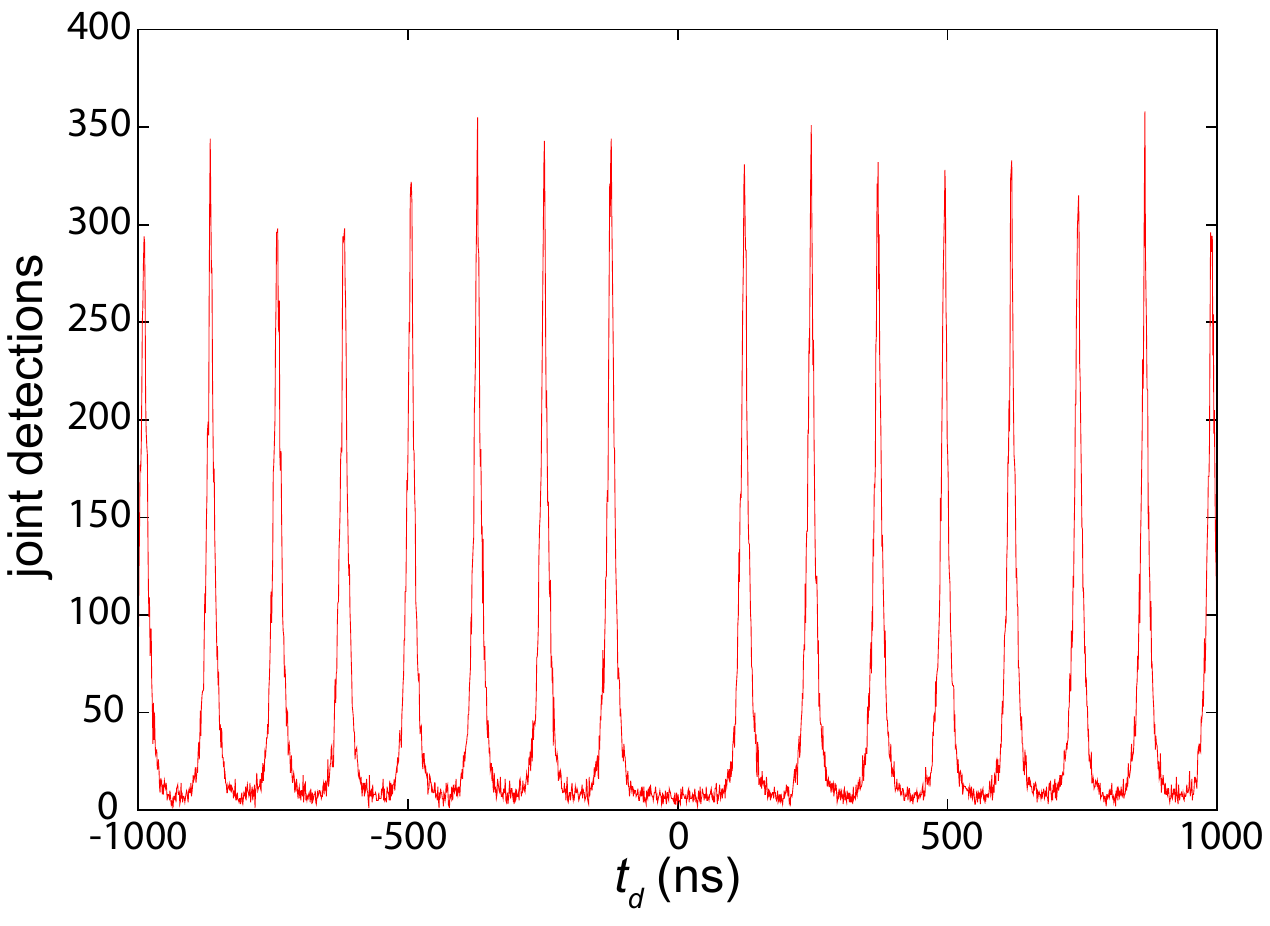}
	\caption{Experimental single-photon second-order correlation function.  The data is shown with 1 ns binning, and was integrated over 3 hours.  The lack of a peak at time delay $t_{d} = 0$ is indicative of a source of single photons, confirming that at most one photon is emitted by the atom following ultrafast excitation.\protect\cite{maunz:interference}
	}
	\label{fig:expt_single_photon_g2}
\end{figure}

Before looking at the interference of two photons, we first demonstrate that the ultrafast excitation of a single \ybion atom is an excellent source of single photons.  For this measurement, only the photons scattered by one of the two ions are incident on the beamsplitter (the other ion is blocked).  The experiment consists of a repetitive sequence of 10 $\mu$s of cooling followed by 40 $\mu$s of excitation/measurement.  During the cooling interval, the pulse picker is switched off so that the ion is illuminated solely by the cw lasers used for Doppler cooling.  For the excitation/measurement interval the 369.5 nm cw light is switched off by an AOM, and the pulse picker is gated open to allow a train of 1 ps pulses, separated in time by about 124 ns, to sequentially excite the atom from ${}^{2}S_{1/2}$ to ${}^{2}P_{1/2}$.  After each excitation, the atom will spontaneously decay while emitting a single photon.  Emitted photons at 369.5 nm are collected by the objective lens, coupled into the single-mode optical fiber, directed onto the beamsplitter, and detected by the PMTs.  The arrivals times of the detected photons allow us to determine the joint detection probability, shown in Fig.~\ref{fig:expt_single_photon_g2}.  The functional form of the experimental joint detection probability matches the theorectical single-photon source illustrated previously (Fig.~\ref{fig:th_single-ph_g2}), with the lack of a peak at delay time $t_{d} = 0$ indicating that at most a single photon is emitted after each ultrafast excitation pulse.\cite{maunz:interference}  Dark counts on the PMTs result in a small background contribution at all delay times; coupling the light from the atom into a single-mode fiber highly suppresses the contribution of background scattered light.

\begin{figure}
	\centering
	\includegraphics[width=1.0\columnwidth,keepaspectratio]{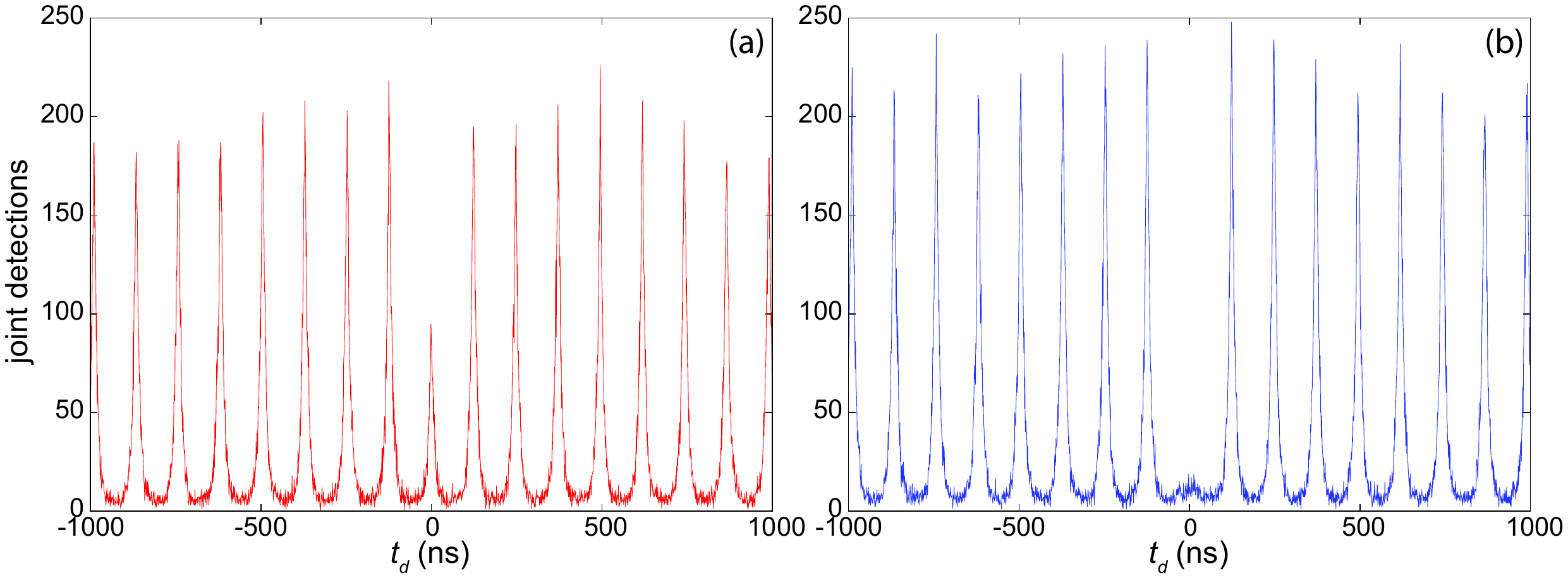}
	\caption{Experimental two-photon second-order correlation function for: (a) distinguishable (perpendicularly polarized) photons; and (b) identical (parallel polarized) photons.\protect\cite{maunz:interference}  
  Both sets of data are shown with 1 ns binning, and each data set was integrated over about 4 hours. 	As expected from the theorectical calculations of Sec.~\ref{sec:photon-bs-theory}, two identical photons incident on the beamsplitter always exit by the same port, resulting in suppression of joint detections at time delay $t_{d} = 0$ in (b).}
	\label{fig:expt_two_photon_g2}
\end{figure}

We observe the quantum two-photon interference effect by simultaneously exciting both trapped \ybion atoms, and combining the emitted photons on the beamsplitter.\footnote{Of course, simultaneously excitation is not essential; the emitted photons just need to arrive at the beamsplitter simultaneously.  As such, the path length of the excitation pulses are adjusted so that emitted photons from each ion arrive at the beamsplitter within 100 ps of each other.  Since the path lengths between each ion and the beamsplitter are approximately equal, this means the ions are also excited at about the same time.}  The joint probability of detection for identical photons is measured by placing both polarizers in the beam path, while detection of distinguishable photons is evaluated by including the $\lambda/2$-waveplate in one of the two paths.  As was seen theoretically in Fig.~\ref{fig:th_two-ph_g2}, for identical photons we expect no detections at time delay $t_{d} = 0$, while for distinguishable photons we expect a probability of joint detection half as large as that from adjacent excitations.  The data shown in Fig.~\ref{fig:expt_two_photon_g2} demonstrates this quantum two-photon interference effect.  In the measurement of parallel polarized photons, the residual counts at time delay $t_{d} = 0$ result from both dark counts on the PMTs, and imperfect spatial mode overlap of the photons on the beamsplitter.  The data shown in Fig.~\ref{fig:expt_two_photon_g2} corresponds to an interferometer contrast of approximately 95\%.\cite{maunz:interference}${}^{,}$\footnote{More recently, the contrast of our interferometer has reached $>98$\%, as measured with laser light (Sec.~\ref{subsubsec:teleport_mode-mismatch_bs}).}  The spatial filtering afforded by coupling the spontaneously emitted photons into single-mode fibers was essential to achieving this two-photon interference.
%

\section{Remote Quantum Gate}
	\label{sec:entanglement}

The standard model of quantum computation is the quantum circuit model, where a series of gate operations are performed and at the output the qubits are measured as a final step.  Deterministic, two-qubit gates that utilize the common modes of motion of trapped atomic ions have already been achieved with 99.3\% fidelity.\cite{benhelm:ms-gate}  Effort in this area is focused on scaling the system to larger numbers of qubits, as can be done, for instance, by constructing complex trap arrays where ions can be stored and operated on at various intervals and locations.\cite{kielpinski:ion_architecture}  Crucial to the success of this methodology will be understanding the motional decoherence that plagues all ion traps.\cite{wineland:nist-jnl,turchette:heating,deslauriers:needle,labaziewicz:heating}

Another approach to scalable quantum computation is the cluster-state (or one-way) quantum computing model, where the measurement process itself is an integral part of the gate operation.\cite{raussendorf:one-way_qc}  In this scheme, a large entangled state is prepared first.  The computation is then executed by successive measurements of the qubits, with classical feed-forward of prior measurement results determining the single-qubit gates to be performed on subsequent sets of qubits before their measurement.  This approach has been proven to be equivalent to the circuit model,\cite{raussendorf:one-way_qc} and allows for a new class of quantum gates to be considered for scalable quantum computation: non-unitary, measurement-based gates.\cite{duan:prob_gate}

Basic quantum operations applicable to the cluster-state model of quantum computation have been carried out using entangled photons.\cite{walther:photon_one-way_qc,lu:photon_graph_states}  However, the photonic cluster states in these experiments were generated by spontaneous parametric down-conversion, relied on post-selection, and did not utilize a quantum memory, making these implementations difficult to scale to larger systems.\cite{bodiya:linear_op_graph}

We review the implementation of a probabilistic, heralded quantum gate between two remote quantum memories.  A full set of input states is used to evaluate the operation of the gate, resulting in an average fidelity of 89(2)\%.\cite{maunz:heralded-gate}  Even though the gate is intrinsically probabilistic, the incorporation of a quantum memory allows this system to be efficiently scaled, potentially allowing for the generation of the large cluster-states necessary for one-way quantum computation.\cite{duan:prob_gate,duan:freq-qubit,duan:robust_qip}  While in this demonstration the success probability is only about $2.2 \times 10^{-8}$, it should be possible to improve the success rate for practical implementation.  The photon-mediated approach to scalable quantum computing with trapped atomic ions has the advantage that complex trap arrays may not be required, the operation is insensitive to motional decoherence, and the gate can be operated on qubits separated by a large distance.  Alternatively, spatially separated quantum registers of Coulomb-coupled ions could be linked using this heralded gate to establish a quantum network.  Moreover, since the operation is mediated by photons, this gate could be applied to any system of optically active qubits, such as neutral atoms, ions, nitrogen-vacancy centers, or quantum dots.  Indeed, hybrid systems are also envisioned, where disparate quantum systems are connected via this photon-mediated process to exploit the advantages of each individual quantum system.   

\subsection{Ion-photon entanglement}
	\label{subsec:ion-photon_entangle}
	
\begin{figure}
	\centering
	\includegraphics[width=1.0\columnwidth,keepaspectratio]{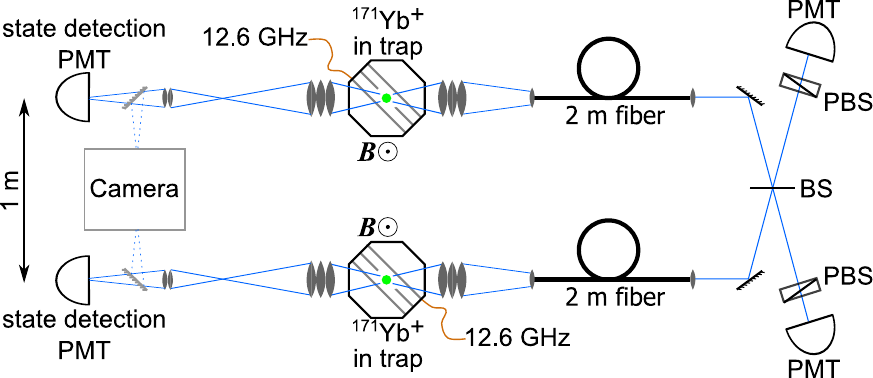}
	\caption{Experimental setup for the heralded quantum gate.  Spontaneously emitted $\pi$-polarized photons are coupled in a single-mode fiber and directed to interfere on a 50:50 non-polarizing beamsplitter (BS).  Coincident detection of two photons by photon-counting photomultipliers (PMTs) announces the success of the gate between the two ions.  Polarizers (PBSs) are used to filter the photons so that only $\pi$-polarized photons are detected.  The state of each ion is measured by state-dependent fluorescence, detected by a PMT on the opposite side of the vacuum chamber.}
	\label{fig:gate_expt_setup}
\end{figure}

The implementation of the gate begins by confining and cooling two ${}^{171}$\ybion atoms in two vacuum chambers, separated by a distance of about one meter.  The experimental setup, shown in Fig.~\ref{fig:gate_expt_setup}, is similar to that employed for the two-photon interference experiments of the previous section.  In this case, though, the 171 isotope of \ybion is used, and the ions are subjected to an external magnetic field of about 5.2 gauss that is aligned perpendicular to both the observation axis and the impinging light from the picosecond pulsed laser.  Also, the polarizers are present in the setup, whereas the optional $\lambda/2$-waveplate used previously is removed from the beam path.

After Doppler cooling, each ion is initialized by optically pumping to the state $\ket{0}$ using a 1 $\mu$s pulse of 369.5 nm light.  Each of the two ions, ion $a$ and ion $b$, are then prepared in a particular quantum state by resonant microwave radiation at 12.6 GHz.  The microwaves are applied directly to one of the trap electrodes in each vacuum chamber with controlled phase and duration (0--16 $\mu$s).  Thus, after application of the microwave radiation we write the state of the atoms as
\begin{eqnarray}
	\label{eq:ion_a_b_prep}
	\ket{\psi}_{a} & = & \alpha \ket{0}_{a} + \beta \ket{1}_{a} \nonumber \\
	\ket{\psi}_{b} & = & \gamma \ket{0}_{b} + \delta \ket{1}_{b}
\end{eqnarray}
with $\vert \alpha \vert^{2} + \vert \beta \vert^{2} = 1$ and $\vert \gamma \vert^{2} + \vert \delta \vert^{2} = 1$.  For clarity, we will assume ideal state evolution throughout this discussion.

\begin{figure}
	\centering
	\includegraphics[width=0.7\columnwidth,keepaspectratio]{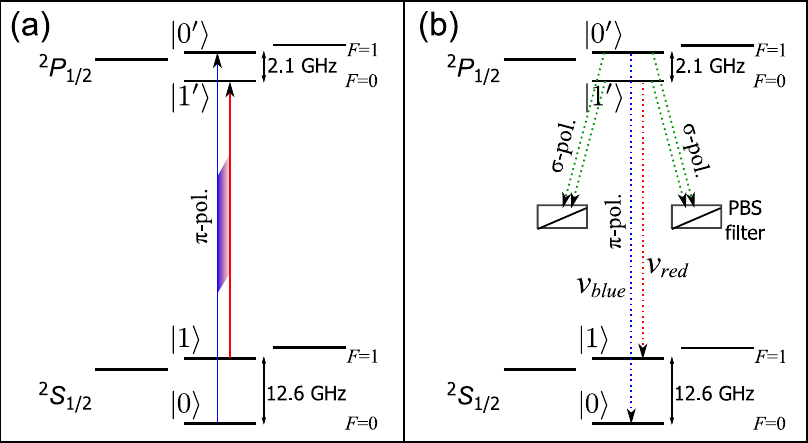}
	\caption{Pulsed excitation with $\pi$-polarized light to generate ion-photon entanglement.  (a) Due to the atomic selection rules, $\pi$-polarized light drives $\ket{0}$ to $\ket{0'}$ and $\ket{1}$ to $\ket{1'}$ only.  The pulse bandwidth of about 300 GHz allows both of these transitions to be driven simultaneously with near-unit excitation probability.  (b) After excitation, the ion spontaneously decays back to ${}^{2}S_{1/2}$ and emits a single photon at 369.5 nm.  If we consider only $\pi$-polarized photons, then the frequency of the emitted photon is entangled with the internal electronic state of the atom, with the separation of the different frequency modes equal to the sum of the ${}^{2}S_{1/2}$ and ${}^{2}P_{1/2}$ hyperfine splittings ($\nu_{blue} - \nu_{red} = 14.7$ GHz).  Polarizers (PBS) are used to filter out the $\sigma$-polarized light.}
	\label{fig:pi_pulse_excitation}
\end{figure}

After state preparation, each ion is excited with near-unit probability to the ${}^{2}P_{1/2}$ level by an ultrafast (1 ps) laser pulse having linear polarization aligned parallel to the magnetic field ($\pi$-polarized) and central wavelength at 369.5 nm.  Due to the polarization of the pulse and atomic selection rules, the broadband pulse coherently drives $\ket{0}$ to $\ket{0'} := {}^{2}P_{1/2} \ket{F = 1, m_{F} = 0}$ and $\ket{1}$ to $\ket{1'} := {}^{2}P_{1/2} \ket{F = 0, m_{F} = 0}$, as illustrated in Fig.~\ref{fig:pi_pulse_excitation}(a).  Since the bandwidth of the 1 ps pulse is approximately 300 GHz, we are able to drive both of these transitions simultaneously.\cite{madsen:ultrafast-rabi}  The 100 THz fine structure splitting of the ${}^{2}P$ levels ensures that the coupling to ${}^{2}P_{3/2}$ is negligible.

As seen in Sec.~\ref{sec:photon_interference}, each ion spontaneously emits only a single photon at 369.5 nm while returning to the ${}^{2}S_{1/2}$ ground state.  The emitted photons at 369.5 nm can each be collected along a direction perpendicular to the quantization axis by an objective lens with effective numerical aperture $\sim$0.3 and coupled into a single-mode fiber.  Observation along this direction allows for polarization filtering of the emitted photons because those produced by $\pi$ and $\sigma$ transitions appear as orthogonally polarized.\cite{blinov:ion-photon}  Considering only $\pi$ decays results in each ion being entangled with the frequency of its emitted photon such that
\begin{eqnarray}
	\label{eq:ion_photon_entangle}
	\ket{\psi}_{a} & = & \alpha \ket{0}_{a} \ket{\nu_{blue}}_{a} + \beta \ket{1}_{a} \ket{\nu_{red}}_{a} \nonumber \\
	\ket{\psi}_{b} & = & \gamma \ket{0}_{b} \ket{\nu_{blue}}_{b} + \delta \ket{1}_{b} \ket{\nu_{red}}_{b} ,
\end{eqnarray}
where $\ket{\nu_{blue}}$ and $\ket{\nu_{red}}$ are the single photon states having well-resolved frequencies $\nu_{blue}$ and $\nu_{red}$, respectively.  The bandwidth of each state is determined by the natural linewidth of the excited state to be $1/(2 \pi \tau) = 19.6$ MHz.\cite{olmschenk:yb_lifetime}  The two frequency states are separated by the sum of the ${}^{2}S_{1/2}$ and ${}^{2}P_{1/2}$ hyperfine splittings, so that $\nu_{blue} - \nu_{red} = 14.7$ GHz.  The output of the single-mode fibers is then directed to interfere at a 50:50 nonpolarizing beamsplitter.

\subsection{Ion-ion entanglement}
	\label{subsec:ion-ion_entangle}
	
Although the action of the beamsplitter on identical and distinguishable photons was reviewed in Sec.~\ref{sec:photon_interference}, the case of superposition states requires additional treatment.  In order to evaluate the action of the beamsplitter in the present scenario, the state of the two ion-photon entangled systems from the previous section is expressed in the Bell state basis of the photons.  The total state of the two ion-photon systems is simply the product of the states in Eq.~\ref{eq:ion_photon_entangle}:
\begin{eqnarray}
	\label{eq:ion-photon_in_bell}
	\ket{\Psi} & = & \ket{\psi}_{a} \otimes \ket{\psi}_{b} \nonumber \\
						 & = & \left( \alpha \ket{0}_{a} \ket{\nu_{blue}}_{a} 
										+ \beta \ket{1}_{a} \ket{\nu_{red}}_{a} \right) 
										\otimes \left( \gamma \ket{0}_{b} \ket{\nu_{blue}}_{b} 
										+ \delta \ket{1}_{b} \ket{\nu_{red}}_{b} \right) \nonumber \\
						 & = & \alpha \gamma \ket{0}_{a} \ket{0}_{b} \ket{\nu_{blue}}_{a} \ket{\nu_{blue}}_{b} 
						 				+ \alpha \delta \ket{0}_{a} \ket{1}_{b} \ket{\nu_{blue}}_{a} \ket{\nu_{red}}_{b} \nonumber \\
						 		&&	+ \beta \gamma \ket{1}_{a} \ket{0}_{b} \ket{\nu_{red}}_{a} \ket{\nu_{blue}}_{b} 
						 				+ \beta \delta \ket{1}_{a} \ket{1}_{b} \ket{\nu_{red}}_{a} \ket{\nu_{red}}_{b} \nonumber \\
						 & = & \alpha \gamma \ket{0}_{a} \ket{0}_{b} \frac{1}{\sqrt{2}} \left( \ket{\phi^{+}}_{ph} + \ket{\phi^{-}}_{ph} \right) 
						 				+ \alpha \delta \ket{0}_{a} \ket{1}_{b} \frac{1}{\sqrt{2}} \left( \ket{\psi^{+}}_{ph} + \ket{\psi^{-}}_{ph} \right) \nonumber \\
						 		&&	+ \beta \gamma \ket{1}_{a} \ket{0}_{b} \frac{1}{\sqrt{2}} \left( \ket{\psi^{+}}_{ph} - \ket{\psi^{-}}_{ph} \right)
						 				+ \beta \delta \ket{1}_{a} \ket{1}_{b} \frac{1}{\sqrt{2}} \left( \ket{\phi^{+}}_{ph} - \ket{\phi^{-}}_{ph} \right) \nonumber \\
						 & = & \frac{1}{\sqrt{2}} 
						 				\left[ 
						 				\ket{\phi^{+}}_{ph} \left( \alpha \gamma \ket{0}_{a} \ket{0}_{b} + \beta \delta \ket{1}_{a} \ket{1}_{b} \right)
						 				+ \ket{\phi^{-}}_{ph} \left( \alpha \gamma \ket{0}_{a} \ket{0}_{b} - \beta \delta \ket{1}_{a} \ket{1}_{b} \right) 
						 				\right. \nonumber \\
						 		&&	\left.
						 				+ \ket{\psi^{+}}_{ph} \left( \alpha \delta \ket{0}_{a} \ket{1}_{b} + \beta \gamma \ket{1}_{a} \ket{0}_{b} \right)
						 				+ \ket{\psi^{-}}_{ph} \left( \alpha \delta \ket{0}_{a} \ket{1}_{b} - \beta \gamma \ket{1}_{a} \ket{0}_{b} \right) 
						 				\right] .
\end{eqnarray}
The action of the beamsplitter can then be determined for the four possible Bell states.  Assuming photon $a$ enters the beamsplitter via port 1 and photon $b$ enters the beamsplitter by port 2, and associating the creation operators $\aup{}$ and $\bup{}$ with frequency modes $\ket{\nu_{red}}$ and $\ket{\nu_{blue}}$, respectively, we then find
\begin{eqnarray}
	\label{eq:bs_bell_states_phi}
	\ket{\phi^{\pm}}_{ph} & = & \frac{1}{\sqrt{2}} \left[ \ket{\nu_{blue}}_{a} \ket{\nu_{blue}}_{b} \pm \ket{\nu_{red}}_{a} \ket{\nu_{red}}_{b} \right] \nonumber \\
											 & = & \frac{1}{\sqrt{2}} \left[ \bup{1} \bup{2} \pm \aup{1} \aup{2} \right] \ket{0}_{ph} \nonumber \\
											 & = & \frac{1}{2 \sqrt{2}} \left[ \left( \bup{3} + \bup{4} \right) \left( -\bup{3} + \bup{4} \right) \pm \left( \aup{3} + \aup{4} \right) \left( -\aup{3} + \aup{4} \right) \right] \ket{0}_{ph} \nonumber \\
											 & = & \frac{1}{2 \sqrt{2}} \left[ -(\bup{3})^2 - \bup{4} \bup{3} + \bup{3} \bup{4} + (\bup{4})^2 \mp (\aup{3})^2 \mp \aup{4} \aup{3} \pm \aup{3} \aup{4} \pm (\aup{4})^2 \right] \ket{0}_{ph} \nonumber \\
											 & = & \frac{1}{2} \left[ -\ket{{2 (\nu_{blue})}_{3} 0_{4}} + \ket{0_{3} {2 (\nu_{blue})}_{4}} \mp \ket{{2 (\nu_{red})}_{3} 0_{4}} \pm \ket{0_{3} {2 (\nu_{red})}_{4}} \right]
\end{eqnarray}
for the $\ket{\phi^{\pm}}_{ph}$ state, where in the last line the notation shows the number of photons of a particular frequency in the spatial mode 3 or 4.  Similarly, for the $\ket{\psi^{+}}_{ph}$ and $\ket{\psi^{-}}_{ph}$ photon states, we find
\begin{eqnarray}
	\label{eq:bs_bell_states_psip}
	\ket{\psi^{+}}_{ph} & = & \frac{1}{\sqrt{2}} \left[ \ket{\nu_{blue}}_{a} \ket{\nu_{red}}_{b} + \ket{\nu_{red}}_{a} \ket{\nu_{blue}}_{b} \right] \nonumber \\
											& = & \frac{1}{\sqrt{2}} \left[ \bup{1} \aup{2} + \aup{1} \bup{2} \right] \ket{0}_{ph} \nonumber \\
											& = & \frac{1}{2 \sqrt{2}} \left[ (\bup{3} + \bup{4}) (-\aup{3} + \aup{4}) + (\aup{3} + \aup{4}) (-\bup{3} + \bup{4}) \right] \ket{0}_{ph} \nonumber \\
											& = & \frac{1}{2 \sqrt{2}} \left[ -\bup{3} \aup{3} - \bup{4} \aup{3} + \bup{3} \aup{4} + \bup{4} \aup{4} - \bup{3} \aup{3} - \bup{3} \aup{4} + \bup{4} \aup{3} + \bup{4} \aup{4} \right] \ket{0}_{ph} \nonumber \\
											& = & \frac{1}{\sqrt{2}} \left[ \ket{0_{3} (\nu_{blue} \nu_{red})_{4}} - \ket{(\nu_{blue} \nu_{red})_{3} 0_{4}} \right]
\end{eqnarray}
and
\begin{eqnarray}
	\label{eq:bs_bell_states_psim}
	\ket{\psi^{-}}_{ph} & = & \frac{1}{\sqrt{2}} \left[ \ket{\nu_{blue}}_{a} \ket{\nu_{red}}_{b} - \ket{\nu_{red}}_{a} \ket{\nu_{blue}}_{b} \right] \nonumber \\
											& = & \frac{1}{\sqrt{2}} \left[ \bup{1} \aup{2} - \aup{1} \bup{2} \right] \ket{0}_{ph} \nonumber \\
											& = & \frac{1}{2 \sqrt{2}} \left[ (\bup{3} + \bup{4}) (-\aup{3} + \aup{4}) - (\aup{3} + \aup{4}) (-\bup{3} + \bup{4}) \right] \ket{0}_{ph} \nonumber \\
											& = & \frac{1}{2 \sqrt{2}} \left[ -\bup{3} \aup{3} - \bup{4} \aup{3} + \bup{3} \aup{4} + \bup{4} \aup{4} + \bup{3} \aup{3} + \bup{3} \aup{4} - \bup{4} \aup{3} - \bup{4} \aup{4} \right] \ket{0}_{ph} \nonumber \\
											& = & \frac{1}{\sqrt{2}} \left[ \ket{(\nu_{blue})_{3} (\nu_{red})_{4}} - \ket{(\nu_{red})_{3} (\nu_{blue})_{4}} \right] .
\end{eqnarray}
The calculations above show the striking result that only the antisymmetric Bell state, $\ket{\psi^{-}}_{ph}$, produces a single photon in both exit ports of the beamsplitter.\cite{braunstein:bs_bell_measure}  This crucial point is the final step in the quantum gate.

We use the joint detection of two photons at the exit ports of the beamsplitter to herald the success of the quantum gate.  By the above derivation, a coincident detection signals that the photons were measured in the $\ket{\psi^{-}}_{ph}$ state.  From Eq.~\ref{eq:ion-photon_in_bell}, the coincident detection thereby projects the ions into the state
\begin{equation}
	\label{eq:coincident_detection}
	\ket{\psi}_{ions} = {}_{ph}\langle \psi^{-} \ket{\Psi} = \frac{1}{\sqrt{2 \vartheta}} \left( \alpha \delta \ket{0}_{a} \ket{1}_{b} - \beta \gamma \ket{1}_{a} \ket{0}_{b} \right) ,
\end{equation}
where the factor $\vartheta = (\vert \alpha \vert^2 \vert \delta \vert^2 + \vert \beta \vert^2 \vert \gamma \vert^2)/2$ in front is the usual renormalization term present after a measurement.\cite{nielsen:qcqi}  In the case $\vert \alpha \vert = \vert \beta \vert = \vert \gamma \vert = \vert \delta \vert = 1/\sqrt{2}$, the ions are left in a maximally entangled state.  In terms of operators, this gate is written as 
\begin{equation}
	\label{eq:heralded_gate}
	\frac{1}{2} \sigma_{3}^{a} (\sigma_{0}^{a} \sigma_{0}^{b} - \sigma_{3}^{a} \sigma_{3}^{b}) ,
\end{equation}
where $\sigma_j^i$ is the $j$th Pauli operator acting on the $i$th qubit.\cite{duan:freq-qubit}

In contrast to our earlier demonstrations of remote ion entanglement,\cite{moehring:ion-ion,matsukevich:bell_ion} the initial state amplitudes are preserved by the heralded quantum gate and determine the form of the final ion-ion entangled state.  This is a defining feature of a gate operation, and is essential to establishing entanglement between more than two qubits.  However, the quantum gate presented here is not unitary.  Indeed, for certain initial states (e.g. $\alpha = \gamma = 1$), a coincident detection should never occur.  While this behavior voids its application to the quantum circuit model, cluster states can still be generated by having all qubits initially in a superposition state.  In this case, the gate succeeds with nonvanishing probability, and scales favorably.\cite{duan:freq-qubit}

%
\begin{table}[ph]
	\tbl{Experimental characterization of the heralded quantum gate.  Listed are the input states of the two ions, the expected (ideal) output state after operation of the gate, the measurement performed, the number of heralding events, the obtained fidelity, and the measured and ideal probability for two photons to be in the antisymmetric Bell state.  The fidelity of the output states is obtained by parity measurements in the appropriate bases.  Here we have defined the parity $P_{xy}$ as the difference of the probabilities to find the two ions in the same state and in opposite states when ion $a$, $b$ is measured in the $x$, $y$ basis, respectively.  The other parity values are defined similarly ($I$ indicates independence to that qubit measurement).  From these results we calculate an average fidelity $\overline{\mathcal{F}} = 0.89(2)$.\protect\cite{maunz:heralded-gate}  The success probability of the gate is $P_{gate} = \vartheta \times (8.5) \times 10^{-8}$.}{
		\centering
		\begin{tabular}{|ccccccc|}
			\hline
									&									&							&				 &					& measured	& theory		\\
			input state	&	expected output & measurement & events & fidelity & $\vartheta$ & $\vartheta$ \\
			\hline\hline
			$(\ket{0}_{a} + \ket{1}_{a}) \otimes (\ket{0}_{b} + \ket{1}_{b})$ & $\ket{0}_{a} \ket{1}_{b} - \ket{1}_{a} \ket{0}_{b}$ & $\frac{1}{4} (1 - P_{xx} - P_{yy} - P_{zz})$ & 210 & 0.89(2) & 0.26(1) & 1/4 \\
			$(\ket{0}_{a} + \ii \ket{1}_{a}) \otimes (\ket{0}_{b} + \ket{1}_{b})$ & $\ket{0}_{a} \ket{1}_{b} - \ii \ket{1}_{a} \ket{0}_{b}$ & $\frac{1}{4} (1 - P_{xy} + P_{yx} - P_{zz})$ & 179 & 0.86(2) & 0.26(1) & 1/4 \\
			$(\ket{0}_{a} - \ket{1}_{a}) \otimes (\ket{0}_{b} + \ket{1}_{b})$ & $\ket{0}_{a} \ket{1}_{b} + \ket{1}_{a} \ket{0}_{b}$ & $\frac{1}{4} (1 + P_{xx} + P_{yy} - P_{zz})$ & 178 & 0.85(1) & 0.22(2) & 1/4 \\
			$(\ket{0}_{a} - \ii \ket{1}_{a}) \otimes (\ket{0}_{b} + \ket{1}_{b})$ & $\ket{0}_{a} \ket{1}_{b} + \ii \ket{1}_{a} \ket{0}_{b}$ & $\frac{1}{4} (1 + P_{xy} - P_{yx} - P_{zz})$ & 188 & 0.81(2) & 0.27(2) & 1/4 \\
			$(\ket{0}_{a} + \ket{1}_{a}) \otimes \ket{1}_{b}$ & $\ket{0}_{a} \ket{1}_{b}$ & $\frac{1}{4} (1 + P_{Iz} - P_{zI} - P_{zz})$ & 42 & 0.86(5) & 0.24(4) & 1/4 \\
			$\ket{0}_{a} \otimes (\ket{0}_{b} + \ket{1}_{b})$ & $\ket{0}_{a} \ket{1}_{b}$ & $\frac{1}{4} (1 + P_{Iz} - P_{zI} - P_{zz})$ & 52 & 0.90(4) & 0.20(3) & 1/4 \\
			$\ket{0}_{a} \otimes \ket{1}_{b}$ & $\ket{0}_{a} \ket{1}_{b}$ & $\frac{1}{4} (1 + P_{Iz} - P_{zI} - P_{zz})$ & 48 & 0.98(2) & 0.39(6) & 1/2 \\
			$\ket{0}_{a} \otimes \ket{0}_{b}$ & 0 & -- & 65 & -- & 0.04(1) & 0 \\
			\hline
		\end{tabular}
	\label{tab:gate_measures}}
\end{table}
%

\subsection{Gate evaluation}
	\label{subsec:gate_evaluation}
We verify the operation of the gate by implementing it on a variety of input states.  The gate is characterized by determining the fidelity of the resulting output state with respect to the ideal case described by Eq.~\ref{eq:coincident_detection}.  The fidelity is defined as the overlap of the measured state with the ideal state
\begin{equation}
	\label{eq:fidelity}
	\mathcal{F} = \bra{\psi_{ideal}} \rho \ket{\psi_{ideal}} ,
\end{equation}
where we have written our measured density matrix as $\rho$, and the ideal output state as $\ket{\psi_{ideal}}$.  The various input states used to characterize the quantum gate, together with the measurements made and the resulting fidelities for each, are given in Table~\ref{tab:gate_measures}.  As measurement of each ion is accomplished using the aforementioned state fluorescence technique, measurement in the remaining two bases requires an additional microwave pulse before detection.  We define the rotation $\{R_{y}(\pi/2), R_{x}(\pi/2), R(0)\}$ before detection to correspond to measurement in the basis $\{x, y, z\}$.  Overall, we attain an average fidelity of 89(2)\%.\cite{maunz:heralded-gate}

To evaluate the gate, we do not make measurements of all combinations of possible input states around the equator of the Bloch sphere, as input states that differ only by a global microwave phase rotation on the equator of the Bloch sphere are indistinguishable in the lab.  It is only the phase difference between the states that matters experimentally, where the microwave oscillator is used as a reference.  As an example, consider the two possible combinations of input states $1/\sqrt{2}(\ket{0}_{a} + \ket{1}_{a})$ and $1/\sqrt{2}(\ket{0}_{b} + \ii \ket{1}_{b})$ versus $1/\sqrt{2}(\ket{0}_{a} + \ii \ket{1}_{a})$ and $1/\sqrt{2}(- \ket{0}_{b} + \ket{1}_{b})$.  Experimentally, the preparation of both of these pairs of states is the same: application of a microwave $\pi/2$-pulse to ion $a$, and application of a microwave $\pi/2$-pulse to ion $b$ that is $+\pi/2$ out of phase with the pulse that was applied to ion $a$.

\begin{figure}
	\centering
	\includegraphics[width=0.7\columnwidth,keepaspectratio]{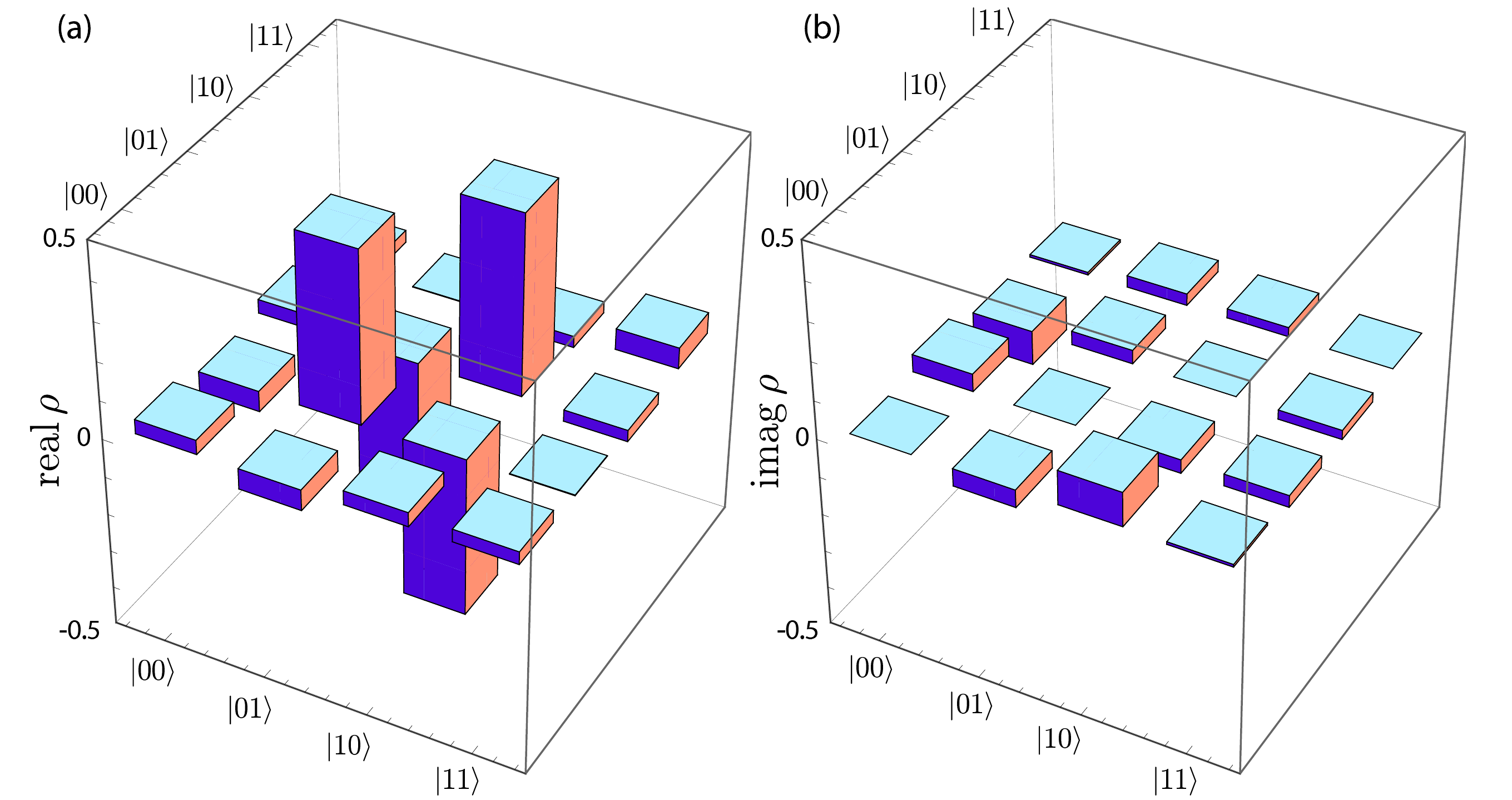}
	\caption{State tomography of $\rho$ for the case $\ket{\psi_{ideal}} = \left( \ket{0}_a \ket{1}_b - \ket{1}_a \ket{0}_b \right)/\sqrt{2}$.  The figure shows the real (a) and imaginary (b) elements of the reconstructed density matrix.  The density matrix was obtained with a maximum likelihood method from 601 events measured in 9 different bases.  From this density matrix we calculate an entangled state fidelity of $\mathcal{F} = 0.87(2)$, a concurrence of $C = 0.77(4)$ and an entanglement of formation $E_{F} = 0.69(6)$.\protect\cite{maunz:heralded-gate}}
	\label{fig:state_tomography}
\end{figure}

In addition to the fidelity measurements of Table~\ref{tab:gate_measures}, we also assess the generation of the maximally entangled antisymmetric Bell state $\ket{\psi_{ideal}} = 1/\sqrt{2} (\ket{0}_{a} \ket{1}_{b} - \ket{1}_{a} \ket{0}_{b})$ by full state tomography.  The resulting density matrix shown in Fig.~\ref{fig:state_tomography} is obtained using a maximum likelihood method.\cite{james:qubit_measure}  From this density matrix we calculate an entangled state fidelity of $\mathcal{F} = 0.87(2)$, a concurrence of $C = 0.77(4)$ and an entanglement of formation $E_{F} = 0.69(6)$.\cite{maunz:heralded-gate}

The observed fidelity is consistent with known experimental errors.  The primary contributions to the error are imperfect
state detection (3\%), spatial mode mismatch on the beamsplitter (6\%), and detection of $\sigma$-polarized light due to the finite solid angle of collection and misalignment of the magnetic field ($<$2\%). Other sources, including imperfect state preparation, pulsed excitation to the wrong atomic state, dark counts of the PMT leading to false coincidence events, and multiple excitation due to pulsed laser light leakage, are each estimated to contribute to the overall error by much less than 1\%.  Micromotion at the rf-drive frequency of the ion trap, which alters the spectrum of the emitted photons and can degrade the quantum interference, is expected to contribute to the overall error by less than 1\%. 

\subsection{Success probability}
	\label{sec:success_prob}
	
The quantum gate is a heralded, probabilistic process.  The net probability for coincident detection of two emitted photons is given by
\begin{equation}
	\label{eq:prob_gate_success}
	P_{gate} = \vartheta \left[ p_{\pi} \eta T_{fiber} T_{optics} \xi (\Delta \Omega/4 \pi) \right]^{2} \approx \vartheta \times \left( 8.5 \times 10^{-8} \right) ,
\end{equation}
where $p_{\pi} = 0.5$ is the fraction of photons with the correct polarization (half are filtered out as being produced by $\sigma$ decays); $\eta = 0.15$ is the quantum efficiency of each PMT; $T_{fiber} = 0.2$ is the coupling into, and transmission through, the single-mode optical fiber; $T_{optics} = 0.95$ is the transmission through the other optical components; $\xi = 1 - 0.005 = 0.995$, where 0.005 is the branching ratio into the ${}^{2}D_{3/2}$ level; $\Delta \Omega/4 \pi = 0.02$ is the solid angle of light collection; and $\vartheta$ is the normalization factor from Eq.~\ref{eq:coincident_detection} whose value depends on the initial amplitudes of the ion qubit states. 

In the current setup, the attempt rate was limited to about 75 kHz, due to the time required for the state preparation microwave pulse.  This resulted in about one successful gate operation every 12 minutes.  However, the expression for $P_{gate}$ reveals multiple ways to substantially increase the success rate.  The most dramatic increase would be achieved by increasing the effective solid angle of collection, which, for instance, could be accomplished by surrounding each ion with an optical cavity.  Although improvements that increase the success probability of the gate operation can enhance scalability, even with a low success probability this gate is still scalable to more complex systems.\cite{duan:freq-qubit}

\section{Quantum Teleportation}
	\label{sec:teleportation}

Quantum teleportation is the faithful transfer of quantum states between systems that relies on the prior establishment of entanglement, but uses only classical communication during the transmission.\cite{bennett:teleportation}  Teleportation, where quantum information is transferred between two disparate locations without traversing the space between the systems, is a stark realization of the counter-intuitive aspects of quantum physics.  In addition, the ability to teleport quantum information is an essential ingredient for the long-distance quantum communication afforded by quantum repeaters\cite{briegel:quantum_repeater} and may be a vital component to quantum computation.\cite{gottesman:qc_teleportation}

Previous demonstrations of quantum teleportation have been accomplished separately using optical systems over long distances,\cite{bouwmeester:photon_teleport,boschi:photon_teleport,furusawa:cont_var_teleport,sherson:light-ensemble_teleport,chen:light-ensemble_teleport} and quantum memories in close proximity.\cite{riebe:ion_teleport,barrett:ion_teleport,riebe:ion_teleport2}  In this section, we review the implementation of a heralded teleportation protocol where the advantages from both optical systems and quantum memories are combined to teleport quantum states between two trapped ytterbium ion qubits over a distance of about one meter.  The protocol is evaluated on a complete set of mutually unbiased basis states, resulting in an average teleportation fidelity of 90(2)\%.\cite{olmschenk:teleportation}  The execution of the heralded quantum gate in this basic quantum algorithm elucidates the applicability of this scheme for quantum networks.

\subsection{Teleportation protocol}
	\label{subsec:teleport_protocol}
	
The experimental setup for the teleportation protocol is identical to the one used to demonstrate the quantum gate (Fig.~\ref{fig:gate_expt_setup}).  A single ${}^{171}$\ybion atom is confined and cooled in each of two nearly-identical traps separated by a distance of about one meter.  The atoms are initialized in the state $\ket{0}$ by optical pumping via a 1 $\mu$s pulse of 369.5 nm light.  We then prepare the ions $a$ and $b$ in the states
\begin{eqnarray}
	\label{eq:teleport_prepare}
	\ket{\psi}_{a} & = & \alpha \ket{0}_{a} + \beta \ket{1}_{a} \\
	\ket{\psi}_{b} & = & \frac{1}{\sqrt{2}} \left( \ket{0}_{b} + \ket{1}_{b} \right) \nonumber
\end{eqnarray}
by applying a resonant microwave pulse of controlled phase and duration (0--16 $\mu$s) directly to one of the trap electrodes, where $\vert \alpha \vert^{2} + \vert \beta \vert^{2} = 1$.  The quantum state written into ion $a$ is the information we seek to teleport.  While in the present case the amplitudes are determined by the applied microwave radiation, in principle $\alpha$ and $\beta$ could be unknown.

Following state preparation, each atom is excited by a 1 ps pulse of $\pi$-polarized light at 369.5 nm, and this broadband excitation coherently drives the population in the hyperfine levels of the ${}^{2}S_{1/2}$ state to complementary levels in the ${}^{2}P_{1/2}$ state, as shown previously (Fig.~\ref{fig:pi_pulse_excitation}).  As each atom decays back to the ${}^{2}S_{1/2}$ levels, it emits a single photon at 369.5 nm.  Considering only $\pi$ decays results in the frequency of the emitted photon becoming entangled with the electronic state of the ion, such that:
\begin{eqnarray}
	\label{eq:teleport_ion-photon_entangle}
	\ket{\psi}_{a} & = & \alpha \ket{0}_{a} \ket{\nu_{blue}}_{a} + \beta \ket{1}_{a} \ket{\nu_{red}}_{a} \nonumber \\
	\ket{\psi}_{b} & = & \frac{1}{\sqrt{2}} \left( \ket{0}_{b} \ket{\nu_{blue}}_{b} + \ket{1}_{b} \ket{\nu_{red}}_{b} \right) .
\end{eqnarray}
Spontaneously emitted photons are collected by an objective lens, coupled into a single-mode fiber, and directed to interfere at a 50:50 non-polarizing beamsplitter.  As shown in Sec.~\ref{sec:entanglement}, due to the quantum interference of the photons at the beamsplitter, a simultaneous detection at both output ports of the beamsplitter occurs only if the photons are in the $\ket{\psi^{-}}_{ph}$ state, so that the action of this heralded quantum gate is to project the ions into the entangled state:
\begin{equation}
	\label{eq:teleport_ion-ion_entangle}
	\ket{\psi}_{ions} = {}_{ph}\bra{\psi^{-}} \left( \ket{\psi}_{a} \otimes \ket{\psi}_{b} \right) = \alpha \ket{0}_{a} \ket{1}_{b} - \beta \ket{1}_{a} \ket{0}_{b} .
\end{equation}
Even though the gate is a probabilistic process, we do not require postselection because the coincident detection of the two photons is a heralding event that announces the success of the quantum gate.  In the teleportation protocol $\vartheta = 1/4$ in Eq.~\ref{eq:prob_gate_success}, so that the overall success probability for any $\alpha$ and $\beta$ is about $2.2 \times 10^{-8}$, limited by the efficiency of collecting and detecting both spontaneously emitted photons.  Therefore, the previous steps (state preparation and pulsed excitation) are repeated at a rate of 40 to 75 kHz, including intermittent cooling, until the gate operation is successful (once every 12 minutes, on average).

After the success of the quantum gate has been confirmed by the heralding event, ion $a$ is subjected to another pulse of microwaves to execute the rotation $R_{y}(\pi/2)$.  The microwave rotation transforms the state given in Eq.~\ref{eq:teleport_ion-ion_entangle} to:
\begin{eqnarray}
	\label{eq:teleport_rotate_a}
	\ket{\psi}_{ions} & = & \frac{\alpha}{\sqrt{2}} \left( \ket{0}_{a} + \ket{1}_{a} \right) \ket{1}_{b} - \frac{\beta}{\sqrt{2}} \left( -\ket{0}_{a} + \ket{1}_{a} \right) \ket{0}_{b} \nonumber \\
										& = & \frac{1}{\sqrt{2}} \left( \alpha \ket{1}_{b} + \beta \ket{0}_{b} \right) \ket{0}_{a} + \frac{1}{\sqrt{2}} \left( \alpha \ket{1}_{b} - \beta \ket{0}_{b} \right) \ket{1}_{a} .
\end{eqnarray}

We then measure ion $a$ using the state dependent fluorescence technique discussed earlier.  As is apparent from Eq.~\ref{eq:teleport_rotate_a}, measuring ion $a$ projects ion $b$ into one of the two states:
\begin{eqnarray}
	\label{eq:teleport_ion_b_projection}
	\mbox{If measured } \ket{0}_{a} \Rightarrow \ket{\psi}_{b} & = & \alpha \ket{1}_{b} + \beta \ket{0}_{b} \nonumber \\
	\mbox{If measured } \ket{1}_{a} \Rightarrow \ket{\psi}_{b} & = & \alpha \ket{1}_{b} - \beta \ket{0}_{b} .
\end{eqnarray}

The result of the measurement on ion $a$ is relayed through a classical communication channel and used to determine the necessary phase of a conditional microwave $\pi$ pulse applied to ion $b$ to recover the state initially written to ion $a$.  If $\ket{0}_{a}$ is measured, the rotation $R_{x}(\pi)$ is applied to ion $b$ so that:
\[
	R_{x}(\pi) \left( \alpha \ket{1}_{b} + \beta \ket{0}_{b} \right) = \alpha \ket{0}_{b} + \beta \ket{1}_{b} .
\]
On the other hand, if $\ket{1}_{a}$ is measured, the rotation $R_{y}(\pi)$ is applied to ion $b$, yielding
\[
	R_{y}(\pi) \left( \alpha \ket{1}_{b} - \beta \ket{0}_{b} \right) = \alpha \ket{0}_{b} + \beta \ket{1}_{b} ,
\]
where we have ignored any global phase factors.  Thus, the state amplitudes initially written to ion $a$ are now stored in ion $b$, which completes the teleportation of the quantum state between the two distant matter qubits.

\subsection{Experimental evaluation}
	\label{subsec:teleport_results}
\begin{figure}
	\centering
	\includegraphics[width=0.8\columnwidth,keepaspectratio]{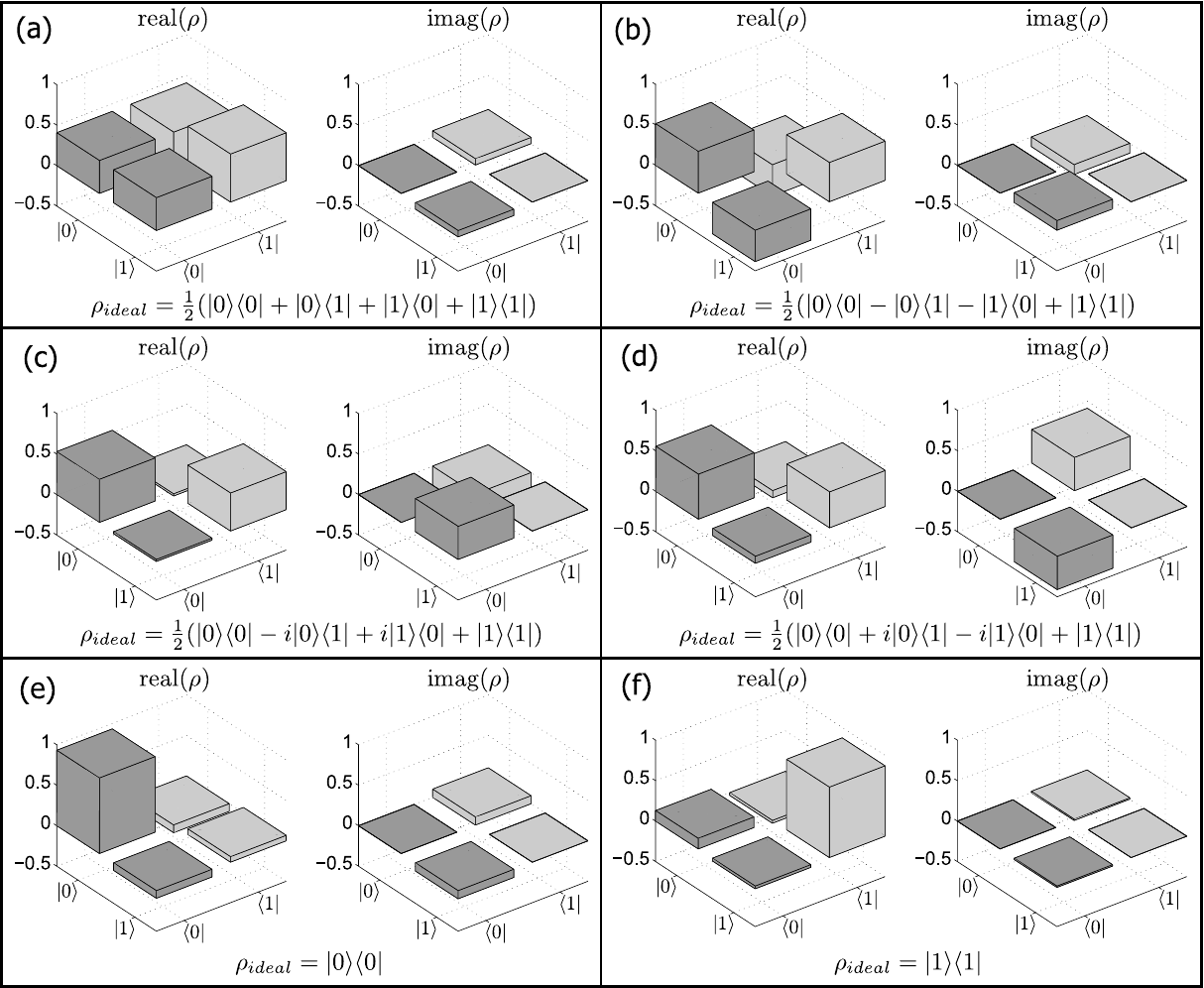}
	\caption{Tomography of the teleported quantum states.  The reconstructed density matrices for the six unbiased basis states teleported from ion $a$ to ion $b$: (a) $\ket{\psi_{ideal}} = 1/\sqrt{2} (\ket{0} + \ket{1})$ teleported with fidelity $\mathcal{F} = 0.91(3)$, (b) $\ket{\psi_{ideal}} = 1/\sqrt{2} (\ket{0} - \ket{1})$ teleported with fidelity $\mathcal{F} = 0.88(4)$, (c) $\ket{\psi_{ideal}} = 1/\sqrt{2} (\ket{0} + \ii \ket{1})$ teleported with fidelity $\mathcal{F} = 0.92(4)$, (d) $\ket{\psi_{ideal}} = 1/\sqrt{2} (\ket{0} - \ii \ket{1})$ teleported with fidelity $\mathcal{F} = 0.91(4)$, (e) $\ket{\psi_{ideal}} = \ket{0}$ teleported with fidelity $\mathcal{F} = 0.93(4)$, and (f) $\ket{\psi_{ideal}} = \ket{1}$ teleported with fidelity $\mathcal{F} = 0.88(4)$.  These measurements yield an average teleportation fidelity $\overline{\mathcal{F}} = 0.90(2)$.  The data shown is from 1285 events recorded over 253 hours.}
	\label{fig:teleport_state_tomography}
\end{figure}

We execute the teleportation protocol on a set of six mutually unbiased basis states $\ket{\psi}_{ideal} \in \{ 1/\sqrt{2} (\ket{0} + \ket{1}), 1/\sqrt{2} (\ket{0} - \ket{1}), 1/\sqrt{2} (\ket{0} + \ii \ket{1}), 1/\sqrt{2} (\ket{0} - \ii \ket{1}), \ket{0}, \ket{1} \}$, and evaluate the process by performing state tomography on each teleported state.  A single-qubit density matrix can be reconstructed by measuring the state in three mutually unbiased measurement bases.  Of course, measurement of the ion occurs via the aforementioned state fluorescence technique, and therefore only distinguishes between $\ket{0}$ and $\ket{1}$ ($z$-basis); two states such as $1/\sqrt{2} (\ket{0} + \ket{1})$ and $1/\sqrt{2} (\ket{0} - \ket{1})$ ($x$-basis) are not distinguishable by fluorescence alone.  Measurement in the remaining two bases requires an additional microwave pulse before detection.  As before, rotation $\{R_{y}(\pi/2), R_{x}(\pi/2), R(0)\}$ before detection to correspond to measurement in the basis $\{x, y, z\}$.  These measurements allow reconstruction of the single-qubit density matrix, $\rho$, for each teleported state using a simple analytical expression,\cite{altepeter:tomography} with the results shown in Fig.~\ref{fig:teleport_state_tomography}.  The fidelity of the teleportation protocol, defined as the overlap of the ideal and measured density matrices $\mathcal{F} = tr(\rho_{ideal} \rho) = {}_{ideal}\bra{\psi} \rho \ket{\psi}_{ideal}$, for this set of states is measured to be $\mathcal{F} = \{ 0.91(3), 0.88(4),
0.92(4), 0.91(4), 0.93(4), 0.88(4) \}$.  This yields an average teleportation fidelity of $\overline{\mathcal{F}} = 0.90(2)$.\cite{olmschenk:teleportation}  The reconstructed density matrices also facilitate characterization of the protocol by quantum process tomography.\cite{nielsen:qcqi}  Using a maximum likelihood method,\cite{obrien:process_tomography} we determine the process fidelity $\mathcal{F}_{process} = tr(\chi_{ideal} \chi) = 0.84(2)$, which is consistent with the aforementioned average fidelity of the teleportation protocol.\cite{horodecki:teleport_channel}
%

\subsection{Error analysis}
	\label{subsec:teleport_error_analysis}
	
The experimental errors in the teleportation protocol differ slightly from the implementation of the quantum gate.  The primary sources of error that reduce the average fidelity are photon mode mismatch at the 50:50 beamsplitter (4\%), imperfect state detection (3.5\%), and polarization--mixing resulting from the nonzero numerical aperture of the objective lens and from misalignment with respect to the magnetic field (2\%).  Other sources, including incomplete state preparation, pulsed excitation to the wrong atomic state, dark counts of the PMT leading to false coincidence events, photon polarization rotation while traversing the optical fiber, and multiple excitation resulting from pulsed laser light leakage, are each expected to contribute to the error by much less than 1\%. Residual micromotion at the rf-drive frequency of the ion trap, which alters the spectrum of the emitted photons and degrades the quantum interference, reduces the average fidelity by less than 1\%.  Detailed calculations of the two most prominent error contributions are presented below.

\subsubsection{Photon spatial mode-mismatch at the beamsplitter}
	\label{subsubsec:teleport_mode-mismatch_bs}
We determine the effect of imperfect spatial mode-overlap of two photons incident on the beamsplitter by utilizing the formalism presented in Sec.~\ref{sec:photon_interference}.  In this case, only two photons impinging on the beamsplitter are considered, rather than the train of photons described previously. Given that photons from different excitation pulses are well-separated in time, this simplification can be made without loss of generality.

To determine the decrease in fidelity resulting from spatial mode-mismatch of the photons at the beamsplitter, we calculate the effect of the electric field operators on the four possible input states $\ket{\nu_{red}}_{a} \ket{\nu_{red}}_{b}$, $\ket{\nu_{red}}_{a} \ket{\nu_{blue}}_{b}$, $\ket{\nu_{blue}}_{a} \ket{\nu_{red}}_{b}$, and $\ket{\nu_{blue}}_{a} \ket{\nu_{blue}}_{b}$.  As an attempt to unclutter the expressions, henceforth the states will not be labeled with $a$ or $b$ unless there are possible ambiguities.  Instead, we will write, e.g. $\ket{\nu_{red}}_{a} \ket{\nu_{blue}}_{b} = \ket{RB}$, where the ordering of the frequencies determine to which subsystem they apply (e.g. in this case ``R'' indicates a ``red'' photon as part of system $a$).

The four possible ways to get a photon in each exit port of the beamsplitter are
\begin{equation}
	\label{eq:teleport_electric_field_op}
	E_{3b}^{+} E_{4b}^{+}(t_d) \mbox{, } E_{3b}^{+} E_{4a}^{+}(t_d) \mbox{, } E_{3a}^{+} E_{4b}^{+}(t_d) \mbox{, } E_{3a}^{+} E_{4a}^{+}(t_d) ,
\end{equation}
where the electric field operators are defined as in Eq.~\ref{eq:efield-op_n}, with $\bup{}$ and $\aup{}$ as the creation operators for a $\ket{\nu_{blue}} = \ket{B}$ and $\ket{\nu_{red}} = \ket{R}$ photon, respectively (as in Sec.~\ref{sec:entanglement}).  In the subsequent calculation, we suppress the exponential decay factor, the Heaviside step function, and the normalization factors of Eq.~\ref{eq:efield-op_n}.  The present concern is the resulting density matrix if a coincident detection of two photons occurs, and in this derivation these other factors may be regarded as part of a normalization factor.

There are then 16 terms to calcuate; fortunately, many of them have an identical form.
\begin{eqnarray}
	\label{eq:teleport_e_op_rrrr}
	E_{3a}^{+} E_{4a}^{+}(t_d) \ket{RR} & = & \frac{1}{2} \left[ \xi_1 \adown{1} + \xi_2 \adown{2} \right] \left[ -\xi_1(t_d) \adown{1} + \xi_2(t_d) \adown{2} \right] \aup{1} \aup{2} \ket{0} \nonumber \\
				& = & \frac{1}{2} \left[ -\xi_1 \xi_1(t_d) (\adown{1})^2 - \xi_2 \xi_1(t_d) \adown{2} \adown{1} \right. \nonumber \\
					 && \left. + \xi_1 \xi_2(t_d) \adown{1} \adown{2} + \xi_2 \xi_2(t_d) (\adown{2})^2 \right] \aup{1} \aup{2} \ket{0} \nonumber \\
				& = & \frac{1}{2} \left[ \xi_1 \xi_2(t_d) - \xi_2 \xi_1(t_d) \right] \ket{0}
\end{eqnarray}
This will also be the result for $E_{3b}^{+} E_{4b}^{+}(t_d) \ket{BB}$, as can be seen by switching $\adown{}$ with $\bdown{}$, and $\ket{R}$ with $\ket{B}$.

Next, we find
\begin{equation}
	\label{eq:teleport_e_op_bbrr}
	E_{3b}^{+} E_{4b}^{+}(t_d) \ket{RR} = \frac{1}{2} \left[ \xi_1 \bdown{1} + \xi_2 \bdown{2} \right] \left[ -\xi_1(t_d) \bdown{1} + \xi_2(t_d) \bdown{2} \right] \aup{1} \aup{2} \ket{0} = 0 ,
\end{equation}
because $\bdown{} \aup{} \ket{0} = 0$.  Of course, the term will vanish whenever there are more annihilation operators of a specific mode than creation operators.  As such, the following terms also vanish:
\begin{eqnarray}
	\label{eq:teleport_e_op_uneq}
	0 & = & E_{3a}^{+} E_{4a}^{+}(t_d) \ket{BB} = E_{3a}^{+} E_{4b}^{+}(t_d) \ket{BB} = E_{3a}^{+} E_{4b}^{+}(t_d) \ket{RR} \nonumber \\
		& = & E_{3b}^{+} E_{4a}^{+}(t_d) \ket{BB} = E_{3b}^{+} E_{4a}^{+}(t_d) \ket{RR} = E_{3a}^{+} E_{4a}^{+}(t_d) \ket{RB} \nonumber \\
		& = & E_{3a}^{+} E_{4a}^{+}(t_d) \ket{BR} = E_{3b}^{+} E_{4b}^{+}(t_d) \ket{BR} = E_{3b}^{+} E_{4b}^{+}(t_d) \ket{RB} .
\end{eqnarray}

Of the other combinations we need to calculate, we find
\begin{eqnarray}
	\label{eq:teleport_e_op_rbrb}
	E_{3a}^{+} E_{4b}^{+}(t_d) \ket{RB} & = & \frac{1}{2} \left[ \xi_1 \adown{1} + \xi_2 \adown{2} \right]  \left[ -\xi_1(t_d) \bdown{1} + \xi_2(t_d) \bdown{2} \right] \aup{1} \bup{2} \ket{0} \nonumber \\
					& = & \frac{1}{2} \left[ -\xi_1 \xi_1(t_d) \adown{1} \bdown{1} - \xi_2 \xi_1(t_d) \adown{2} \bdown{1} \right. \nonumber \\
						 && \left. + \xi_1 \xi_2(t_d) \adown{1} \bdown{2} + \xi_2 \xi_2(t_d) \adown{2} \bdown{2} \right] \aup{1} \bup{2} \ket{0} \nonumber \\
					& = & \frac{1}{2} \xi_1 \xi_2(t_d) .
\end{eqnarray}
The same answer will result from $E_{3b}^{+} E_{4a}^{+}(t_d) \ket{BR}$.  Finally,
\begin{eqnarray}
	\label{eq:teleport_e_op_rbbr}
	E_{3a}^{+} E_{4b}^{+}(t_d) \ket{BR} & = & \frac{1}{2} \left[ \xi_1 \adown{1} + \xi_2 \adown{2} \right]  \left[ -\xi_1(t_d) \bdown{1} + \xi_2(t_d) \bdown{2} \right] \bup{1} \aup{2} \ket{0} \nonumber \\
					& = & \frac{1}{2} \left[ -\xi_1 \xi_1(t_d) \adown{1} \bdown{1} - \xi_2 \xi_1(t_d) \adown{2} \bdown{1} \right. \nonumber \\
						 && \left. + \xi_1 \xi_2(t_d) \adown{1} \bdown{2} + \xi_2 \xi_2(t_d) \adown{2} \bdown{2} \right] \bup{1} \aup{2} \ket{0} \nonumber \\
					& = & -\frac{1}{2} \xi_2 \xi_1(t_d) .
\end{eqnarray}
The result of Eq.~\ref{eq:teleport_e_op_rbbr} will also be the answer obtained for the term $E_{3b}^{+} E_{4a}^{+}(t_d) \ket{RB}$.

The terms calculated above can be used to analyze the two ion-photon systems.  As seen previously, the system after excitation and spontaneous emission of a $\pi$-polarized photon is:
\begin{equation}
	\ket{\psi}_{ion,ph} = \frac{1}{\sqrt{2}} \left[ \alpha \ket{00} \ket{BB} + \beta \ket{10} \ket{RB} + \alpha \ket{01} \ket{BR} + \beta \ket{11} \ket{RR} \right] .
\end{equation}
To determine the density matrix of the two ions after a coincident detection of two photons, we use terminology similar to that in Ref.~\refcite{nielsen:qcqi} to define the joint electric field measurement operator as
\begin{equation}
	M_{ph,jk} = E_{3j}^{+} E_{4k}^{+}(t_d) ,
\end{equation}
where $j,k \in \{a,b\}$.  The joint detection probability from Sec.~\ref{sec:photon_interference} is then
\begin{eqnarray}
	P_{J} & = & {}_{ph}\bra{\psi} E_{4k}^{-} (t_d) E_{3j}^{-} E_{3j}^{+} E_{4k}^{+} (t_d) \ket{\psi}_{ph} \nonumber \\
				& = & {}_{ph}\bra{\psi} M_{ph,jk}^{\dagger} M_{ph,jk} \ket{\psi}_{ph} \nonumber \\
				& = & tr(M_{ph,jk}^{\dagger} M_{ph,jk} \ket{\psi}_{ph} \bra{\psi}) ,
\end{eqnarray}
where in the last two lines the definition of the electric field measurement operator and the usual properties of the trace were used.\cite{nielsen:qcqi}

In this formalism, the density matrix of the two ions will be given by a partial trace over the photon states of the electric field operators acting on the two-ion-photon system:
\begin{eqnarray}
	\label{eq:teleport_ion_density_matrix}
	\rho_{ions} & = & tr_{ph} \left[ \sum_{j,k} M_{ph,jk}^{\dagger} M_{ph,jk} \ket{\psi}_{ion,ph} \bra{\psi} \right] \nonumber \\
							& = & \frac{1}{2} \left[ \vert \alpha \vert^2 \ket{00} \bra{00} \bra{BB} M_{ph,jk}^{\dagger} M_{ph,jk} \ket{BB} \right. \nonumber \\
							&& + \alpha \beta^{\ast} \ket{00} \bra{10} \bra{BB} M_{ph,jk}^{\dagger} M_{ph,jk} \ket{RB} + \vert \alpha \vert^2 \ket{00} \bra{01} \bra{BB} M_{ph,jk}^{\dagger} M_{ph,jk} \ket{BR} \nonumber \\
							&& + \alpha \beta^{\ast} \ket{00} \bra{11} \bra{BB} M_{ph,jk}^{\dagger} M_{ph,jk} \ket{RR} + \beta \alpha^{\ast} \ket{10} \bra{00} \bra{RB} M_{ph,jk}^{\dagger} M_{ph,jk} \ket{BB} \nonumber \\
							&& + \vert \beta \vert^2 \ket{10} \bra{10} \bra{RB} M_{ph,jk}^{\dagger} M_{ph,jk} \ket{RB} + \beta \alpha^{\ast} \ket{10} \bra{01} \bra{RB} M_{ph,jk}^{\dagger} M_{ph,jk} \ket{BR} \nonumber \\
							&& + \vert \beta \vert^2 \ket{10} \bra{11} \bra{RB} M_{ph,jk}^{\dagger} M_{ph,jk} \ket{RR} + \vert \alpha \vert^2 \ket{01} \bra{00} \bra{BR} M_{ph,jk}^{\dagger} M_{ph,jk} \ket{BB} \nonumber \\
							&& + \alpha \beta^{\ast} \ket{01} \bra{10} \bra{BR} M_{ph,jk}^{\dagger} M_{ph,jk} \ket{RB} + \vert \alpha \vert^2 \ket{01} \bra{01} \bra{BR} M_{ph,jk}^{\dagger} M_{ph,jk} \ket{BR} \nonumber \\
							&& + \alpha \beta^{\ast} \ket{01} \bra{11} \bra{BR} M_{ph,jk}^{\dagger} M_{ph,jk} \ket{RR} + \beta \alpha^{\ast} \ket{11} \bra{00} \bra{RR} M_{ph,jk}^{\dagger} M_{ph,jk} \ket{BB} \nonumber \\
							&& + \vert \beta \vert^2 \ket{11} \bra{10} \bra{RR} M_{ph,jk}^{\dagger} M_{ph,jk} \ket{RB} + \beta \alpha^{\ast} \ket{11} \bra{01} \bra{RR} M_{ph,jk}^{\dagger} M_{ph,jk} \ket{BR} \nonumber \\
							&& + \left. \vert \beta \vert^2 \ket{11} \bra{11} \bra{RR} M_{ph,jk}^{\dagger} M_{ph,jk} \ket{RR} \right] ,
\end{eqnarray}
where in the last several lines, the sum $\sum_{j,k}$ is implicit.  Using the results for the action of the different measurement operators on the states yields:
\begin{eqnarray}
	\label{eq:teleport_ion_density_matrix2}
	\rho_{ions} & = & \frac{1}{8} \left[ \vert \alpha \vert^2 \ket{00} \bra{00} \vert \xi_1 \xi_2(t_d) - \xi_2 \xi_1(t_d) \vert^2 + 0 + 0 + 0 + 0 \right. \nonumber \\
								 && + \vert \beta \vert^2 \ket{10} \bra{10} \left( \vert \xi_1 \xi_2(t_d) \vert^2 + \vert \xi_2 \xi_1(t_d) \vert^2 \right) \nonumber \\
								 && + \beta \alpha^{\ast} \ket{10} \bra{01} \left( -\xi_1^{\ast} \xi_2^{\ast}(t_d) \xi_2 \xi_1(t_d) - \xi_1 \xi_2(t_d) \xi_2^{\ast} \xi_1^{\ast}(t_d) \right) + 0 \nonumber \\
								 && + 0 + \alpha \beta^{\ast} \ket{01} \bra{10} \left( -\xi_1 \xi_2(t_d) \xi_2^{\ast} \xi_1^{\ast}(t_d) - \xi_1^{\ast} \xi_2^{\ast}(t_d) \xi_2 \xi_1(t_d) \right) \nonumber \\
								 && \vert \alpha \vert^2 \ket{01} \bra{01} \left( \vert \xi_2 \xi_1(t_d) \vert^2 + \vert \xi_1 \xi_2(t_d) \vert^2 \right) + 0 + 0 \nonumber \\
								 && + 0 + 0 + \left. \vert \beta \vert^2 \ket{11} \bra{11} \vert \xi_1 \xi_2(t_d) - \xi_2 \xi_1(t_d) \vert^2 \right] .
\end{eqnarray}

Following a coincident detection in the teleportation protocol, ion $a$ is rotated with a $R_y(\pi/2)$ microwave pulse, and then measured.  The ion-ion density matrix after rotation of ion $a$ is:
\begin{eqnarray}
	\label{eq:teleport_ion_density_matrix3}
	\left(R_y(\pi/2)\right)_{a} & \rho_{ions} & \left(R_y^{\dagger}(\pi/2)\right)_{a} \nonumber \\
			 & = & \frac{1}{16} \left[ \vert \alpha \vert^2 \vert \xi_1 \xi_2(t_d) - \xi_2 \xi_1(t_d) \vert^2 (\ket{0} + \ket{1}) \ket{0} (\bra{0} + \bra{1}) \bra{0} \right. \nonumber \\
					&& + \vert \beta \vert^2 \left( \vert \xi_1 \xi_2(t_d) \vert^2 + \vert \xi_2 \xi_1(t_d) \vert^2 \right) (-\ket{0} + \ket{1}) \ket{0} (-\bra{0} + \bra{1}) \bra{0} \nonumber \\
					&& + \beta \alpha^{\ast} \left( -\xi_1^{\ast} \xi_2^{\ast}(t_d) \xi_2 \xi_1(t_d) - \xi_1 \xi_2(t_d) \xi_2^{\ast} \xi_1^{\ast}(t_d) \right) \nonumber \\
					&& \times (-\ket{0} + \ket{1}) \ket{0} (\bra{0} + \bra{1}) \bra{1} \nonumber \\
					&& + \alpha \beta^{\ast} \left( -\xi_1 \xi_2(t_d) \xi_2^{\ast} \xi_1^{\ast}(t_d) - \xi_1^{\ast} \xi_2^{\ast}(t_d) \xi_2 \xi_1(t_d) \right) \nonumber \\
					&& \times (\ket{0} + \ket{1}) \ket{1} (-\bra{0} + \bra{1}) \bra{0} \nonumber \\
					&& + \vert \alpha \vert^2 \left( \vert \xi_2 \xi_1(t_d) \vert^2 + \vert \xi_1 \xi_2(t_d) \vert^2 \right) (\ket{0} + \ket{1}) \ket{1} (\bra{0} + \bra{1}) \bra{1} \nonumber \\
					&& \left. \vert \beta \vert^2 \vert \xi_1 \xi_2(t_d) - \xi_2 \xi_1(t_d) \vert^2 (-\ket{0} + \ket{1}) \ket{1} (-\bra{0} + \bra{1}) \bra{1} \right] .
\end{eqnarray}

Analogous to the electric field measurement operator, we define an operator for the measurement of the quantum state of the ion.  In this calculation, we assume ideal measurement of the ion.  The influence of imperfect state detection on the fidelity is calculated in the next section.\footnote{As both are independently small, it is a good approximation to consider each separately.}  Define the (ideal) measurement operators for the ion state as\cite{nielsen:qcqi}
\begin{equation}
	\label{eq:teleport_ion_measurement_op_ideal}
	M_{0j} = \ket{0}_{j} \bra{0} \mbox{, } M_{1j} = \ket{1}_{j} \bra{1} ,
\end{equation}
where here $j \in \{a,b\}$ denotes operation on the $j$th ion.  Since with the final rotation on ion $b$, conditioned on the measurement of ion $a$, the states will be equivalent, only the case of $\ket{0}_{a}$ is presented below.  Measurement of ion $a$ is defined similarly to the case for the photons, where the density matrix of ion $b$ is given by the partial trace of $a$ on the measurment operator and the quantum state of the ions.
\begin{eqnarray}
	\label{eq:teleport_ion_b_density_matrix}
	\rho_b & = & tr_a \left( M_{0a}^{\dagger} M_{0a} \left(R_y(\pi/2)\right)_{a} \rho_{ions} \left(R_y^{\dagger}(\pi/2)\right)_{a} \right) \nonumber \\
				 & = & \frac{1}{16} \left[ \left( \vert \alpha \vert^2 \vert \xi_1 \xi_2(t_d) - \xi_2 \xi_1(t_d) \vert^2 + \vert \beta \vert^2 \left( \vert \xi_1 \xi_2(t_d) \vert^2 + \vert \xi_2 \xi_1(t_d) \vert^2 \right) \right) \ket{0} \bra{0} \right. \nonumber \\
				 		&& - \beta \alpha^{\ast} \left( -\xi_1^{\ast} \xi_2^{\ast}(t_d) \xi_2 \xi_1(t_d) - \xi_1 \xi_2(t_d) \xi_2^{\ast} \xi_1^{\ast}(t_d) \right) \ket{0} \bra{1} \nonumber \\
				 		&& - \alpha \beta^{\ast} \left( -\xi_1 \xi_2(t_d) \xi_2^{\ast} \xi_1^{\ast}(t_d) - \xi_1^{\ast} \xi_2^{\ast}(t_d) \xi_2 \xi_1(t_d) \right) \ket{1} \bra{0} \nonumber \\
				 		&& + \left. \left( \vert \alpha \vert^2 \left( \vert \xi_2 \xi_1(t_d) \vert^2 + \vert \xi_1 \xi_2(t_d) \vert^2 \right) + \vert \beta \vert^2 \vert \xi_1 \xi_2(t_d) - \xi_2 \xi_1(t_d) \vert^2 \right) \ket{1} \bra{1} \right] .
\end{eqnarray}
Recall that in the teleportation protocol if $\ket{0}_{a}$ is measured then ion $b$ has a $R_x(\pi)$ microwave pulse applied to it.  Since this just flips the state of the ion, the final density matrix for ion $b$ is given by:
\begin{eqnarray}
	\label{eq:teleport_ion_b_density_matrix2}
	\rho_b & = & \frac{1}{16} \left[ \left( \vert \alpha \vert^2 \left( \vert \xi_2 \xi_1(t_d) \vert^2 + \vert \xi_1 \xi_2(t_d) \vert^2 \right) + \vert \beta \vert^2 \vert \xi_1 \xi_2(t_d) - \xi_2 \xi_1(t_d) \vert^2 \right) \ket{0} \bra{0} \right. \nonumber \\
				 		&& - \alpha \beta^{\ast} \left( -\xi_1 \xi_2(t_d) \xi_2^{\ast} \xi_1^{\ast}(t_d) - \xi_1^{\ast} \xi_2^{\ast}(t_d) \xi_2 \xi_1(t_d) \right) \ket{0} \bra{1} \nonumber \\
				 		&& - \beta \alpha^{\ast} \left( -\xi_1^{\ast} \xi_2^{\ast}(t_d) \xi_2 \xi_1(t_d) - \xi_1 \xi_2(t_d) \xi_2^{\ast} \xi_1^{\ast}(t_d) \right) \ket{1} \bra{0} \nonumber \\
				 		&& + \left. \left( \vert \alpha \vert^2 \vert \xi_1 \xi_2(t_d) - \xi_2 \xi_1(t_d) \vert^2 + \vert \beta \vert^2 \left( \vert \xi_1 \xi_2(t_d) \vert^2 + \vert \xi_2 \xi_1(t_d) \vert^2 \right) \right) \ket{1} \bra{1} \right] .
\end{eqnarray}

Evaluating this expression requires determination of the spatial mode factors $\xi$ in terms of a measured quantity, such as the visibility of the interferometer.  The visibility is defined as
\begin{equation}
	\label{eq:teleport_interferometer_visibility}
	V = \frac{I_{max} - I_{min}}{I_{max} + I_{min}} ,
\end{equation}
where $I_{max}$ and $I_{min}$ are the maximum and minimum intensities of the incident light.  If two fields are incident, with amplitudes $E_{1}$ and $E_{2}$, then $I_{max} = \vert E_{1} + E_{2} \vert^2$ and $I_{min} = \vert E_{1} - E_{2} \vert^2$.  On the other hand, the intensity of a single field is simply $I_1 = \vert E_{1} \vert^2$.

Suppose a function $e(r)$ describes the amplitude of the incident light as a function of position.  However, the detector measures the intensity $I$ incident over the area of the detector.  Assuming the detector area completely covers the incident mode, then the registered intensity will be the integral over space:
\begin{equation}
	I_e = \int \vert e(r) \vert^2 \,dr .
\end{equation}
If two fields are incident, $e(r)$ and $f(r)$, then we can define the maximum and minimum intensities as
\begin{eqnarray}
	I_{max} & = & \int \vert e(r) + f(r) \vert^2 \,dr \nonumber \\
	I_{min} & = & \int \vert e(r) - f(r) \vert^2 \,dr ,
\end{eqnarray}
and thus,
\begin{eqnarray}
	I_{max} & = & \int \left( e(r) + f(r) \right)\left( e^{\ast}(r) + f^{\ast}(r) \right) \,dr \nonumber \\
					& = & \int \left( \vert e(r) \vert^2 + \vert f(r) \vert^2 + f(r) e^{\ast}(r) + e(r) f^{\ast}(r) \right) \,dr \nonumber \\
					& = & I_e + I_f + \int \left( f(r) e^{\ast}(r) + e(r) f^{\ast}(r) \right) \,dr .
\end{eqnarray}
Similarly,
\begin{equation}
	I_{min} = I_e + I_f - \int \left( f(r) e^{\ast}(r) + e(r) f^{\ast}(r) \right) \,dr .
\end{equation}
Therefore, the visibility of the interferometer is:
\begin{equation}
	\label{eq:teleport_interferometer_visibility2}
	V = \frac{\int \left( f(r) e^{\ast}(r) + e(r) f^{\ast}(r) \right) \,dr}{I_e + I_f} .
\end{equation}
Of course, the relative phase between $e(r)$ and $f(r)$ was given by the definitions of the maximum and minimum intensities, and therefore $f(r)e^{\ast}(r)$ is real.  Assuming that the total incident intensities of the two fields are equal, $I_e = I_f = I$, the visibility of the interferometer is simplified to:
\begin{equation}
	\label{eq:teleport_interferometer_visibility3}
	V = \frac{\int f(r) e^{\ast}(r) \,dr}{I} .
\end{equation}
This expression allows us to derive a number of relations,
\begin{equation}
	\label{eq:teleport_visibility_relations1}
	\int\int \vert e(x) f(y) \vert^2 \,dx \,dy = \int \vert e(x) \vert^2 \,dx \int \vert f(y) \vert^2 \,dy = I^2
\end{equation}
\begin{equation}
	\label{eq:teleport_visibility_relations2}
	\int\int e^{\ast}(x) f(x) e(y) f^{\ast}(y) \,dx \,dy = \int e^{\ast}(x) f(x) \,dx \int e(y) f^{\ast}(y) \,dy = I^2 V^2
\end{equation}
\begin{eqnarray}
	\label{eq:teleport_visibility_relations3}
	\int\int \vert e(x) f(y) - f(x) e(y) \vert^2 \,dx \,dy & = & \int \int \left(e(x) f(y) - f(x) e(y) \right) \nonumber \\
						 && \times \left( e^{\ast}(x) f^{\ast}(y) - f^{\ast}(x) e^{\ast}(y) \right) \,dx \,dy \nonumber \\
					& = & \int \int \left( \vert e(x) \vert^2 \vert f(y) \vert^2 + \vert f(x) \vert^2 \vert e(y) \vert^2 \right. \nonumber \\
						 && - e^{\ast}(x) f(x) e(y) f^{\ast}(y) \nonumber \\
						 && \left. - e(x) f^{\ast}(x) e^{\ast}(y) f(y) \right) \,dx \,dy \nonumber \\
					& = & 2I^2 - 2 I^2 V^2 \nonumber \\
					& = & 2 I^2 \left( 1 - V^2 \right) ,
\end{eqnarray}
that will be useful below.

The spatial modes $\xi$ appearing in the density matrix of ion $b$ can now be expressed in terms of the measureable parameters $I$ and $V$.  As above, we will assume the incident intensities are equal.
\begin{equation}
	\label{eq:teleport_ion_b_density_matrix3}
	\rho_{b} = \frac{I^2}{8}
	\left( 
			\begin{array}{cc}
			\vert \alpha \vert^2 + \vert \beta \vert^2 \left( 1 - V^2 \right) & \alpha \beta^{\ast} V^2 \\
			\beta \alpha^{\ast} V^2 & \vert \beta \vert^2 + \vert \alpha \vert^2 \left( 1 - V^2 \right)
			\end{array}
		\right) .
\end{equation}
Of course, the factor of $I^2/8$ is part of the normalization factor neglected throughout this derivation; it is thereby discarded, and the condition $tr(\rho_b) = 1$ is used to determine the proper normalization.  After doing so, we finally end up with the density matrix for ion $b$ as:
\begin{equation}
	\label{eq:teleport_ion_b_density_matrix4}
	\rho_{b} = \frac{1}{2 - V^2}
	\left( 
			\begin{array}{cc}
			\vert \alpha \vert^2 + \vert \beta \vert^2 \left( 1 - V^2 \right) & \alpha \beta^{\ast} V^2 \\
			\beta \alpha^{\ast} V^2 & \vert \beta \vert^2 + \vert \alpha \vert^2 \left( 1 - V^2 \right)
			\end{array}
		\right) .
\end{equation}
The above enables calculation of the expected reduction in fidelity for any teleported state as a function of the visibility of the interferometer.  For any given input state, the ideal density matrix is simply:
\begin{equation}
	\label{eq:teleport_ion_b_density_matrix_ideal}
	\rho_{ideal} =
	\left( 
			\begin{array}{cc}
			\vert \alpha \vert^2 & \alpha \beta^{\ast} \\
			\beta \alpha^{\ast} & \vert \beta \vert^2
			\end{array}
		\right) .
\end{equation}
Therefore, the fidelity $\mathcal{F} = tr\left( \rho_{b} \rho_{ideal} \right)$ is:
\begin{eqnarray}
	\label{eq:teleport_ion_b_fidelity}
	\mathcal{F} & = & tr \left[ 
			\frac{1}{2 - V^2}
		\left( 
			\begin{array}{cc}
			\vert \alpha \vert^2 + \vert \beta \vert^2 \left( 1 - V^2 \right) & \alpha \beta^{\ast} V^2 \\
			\beta \alpha^{\ast} V^2 & \vert \beta \vert^2 + \vert \alpha \vert^2 \left( 1 - V^2 \right)
			\end{array}
		\right)
		\left( 
			\begin{array}{cc}
			\vert \alpha \vert^2 & \alpha \beta^{\ast} \\
			\beta \alpha^{\ast} & \vert \beta \vert^2
			\end{array}
		\right)
	 \right] \nonumber \\
	 						& = & \frac{1}{2 - V^2} \left( \vert \alpha \vert^4 + \vert \beta \vert^4 + 2 \vert \alpha \vert^2 \vert \beta \vert^2 (1 - V^2) + 2 \vert \alpha \vert^2 \vert \beta \vert^2 V^2 \right) \nonumber \\
	 						& = & \frac{1}{2 - V^2} \left( \vert \alpha \vert^2 + \vert \beta \vert^2 \right) \nonumber \\
	 						& = & \frac{1}{2 - V^2} .
\end{eqnarray}
We see that if the visibility of the interferometer is perfect, $V = 1$, then the fidelity is 1; whereas if the spatial mode-overlap is nonexistent, $V = 0$, then the fidelity drops to $1/2$ (as expected for a totally mixed state).

In the experiment, we measure the visibility of the interferometer by coupling laser light into the single-mode fibers used for transferring the spontaneously emitted photons from the atom to the beamsplitter.  We find a visibility $V > 0.98$.  By the above derivation, we can thereby estimate that the spatial mode-mismatch at the beamsplitter reduces the fidelity of the teleportation protocol by at most about 4\%.

\subsubsection{Imperfect state detection}
	\label{subsubsec:teleport_imperfect_state_detection}
	
We calculate the expected degradation in fidelity due to imperfect state detection of the atomic qubit using the formalism of the measurement operators, as above.  Let $\epsilon_{j}$ be the error in the measurement of the qubit state of ion $j = a,b$.  The measurement operators may then be defined as:
\begin{eqnarray}
	\label{eq:teleport_imperfect_measurement_ops}
	M_{0j} & = & \sqrt{1-\epsilon_j} \ket{0} \bra{0} + \sqrt{\epsilon_j} \ket{1} \bra{1} \nonumber \\
	M_{1j} & = & \sqrt{\epsilon_j} \ket{0} \bra{0} + \sqrt{1-\epsilon_j} \ket{1} \bra{1} .
\end{eqnarray}
Note that these satisfy the completeness relation $I_j = \sum_{m} M_{mj}^{\dagger} M_{mj}$.

In this calculation we assume perfect interference of the photons at the beamsplitter, so that the state of the two ions after a coincident detection of the photons and rotation of ion $a$ by $R_y(\pi/2)$, yields the state (Eq.~\ref{eq:teleport_rotate_a})
\begin{eqnarray}
	\ket{\psi}_{ions} & = & \frac{1}{\sqrt{2}} \ket{0}_{a} \left( \alpha \ket{1}_{b} + \beta \ket{0}_{b} \right) + \frac{1}{\sqrt{2}} \ket{1}_{a} \left( \alpha \ket{1}_{b} - \beta \ket{0}_{b} \right) \nonumber \\
										& = & \frac{1}{\sqrt{2}} \ket{0} \left( \alpha \ket{1} + \beta \ket{0} \right) + \frac{1}{\sqrt{2}} \ket{1} \left( \alpha \ket{1} - \beta \ket{0} \right) ,
\end{eqnarray}
where in the second line we have again invoked the shorter notation of the prior section and suppressed the subscripts.

The explicit derivation for measuring the state $\ket{0}_{a}$ is presented below; the derivation for $\ket{1}_{a}$ is analogous, and after the conditional microwave rotation on ion $b$, yields exactly the same result.  As in the previous section, the measurement is completed by taking the partial trace over the measurement operator and the quantum state, so that following measurment of ion $a$ the density matrix for $b$ is given by:
\begin{eqnarray}
	\label{eq:teleport_imperfect_measure_0a}
	\rho_b & = & tr_a\left( M_{0a}^{\dagger} M_{0a} \ket{\psi}_{ions} \bra{\psi} \right) \nonumber \\
				 & = & \left[ \sqrt{1 - \epsilon_a} ({}_{a}\bra{0}) \left( \alpha \ket{1}_b + \ket{0}_b \right) + \sqrt{\epsilon_a} ({}_{a}\bra{1}) \left( \alpha \ket{1}_b - \beta \ket{0}_b \right) \right] \nonumber \\
				 		&& \otimes \left[ \sqrt{1 - \epsilon_a} \ket{0}_a \left( \alpha^{\ast} {}_b\bra{1} + \beta^{\ast} {}_b\bra{0} \right) + \sqrt{\epsilon_a} \ket{1}_a \left( \alpha^{\ast} {}_b\bra{1} - \beta^{\ast} {}_b\bra{0} \right) \right] \nonumber \\
				 & = & (1 - \epsilon_a) \left( \vert \alpha \vert^2 \ket{1} \bra{1} + \beta \alpha^{\ast} \ket{0} \bra{1} + \alpha \beta^{\ast} \ket{1} \bra{0} + \vert \beta \vert^2 \ket{0} \bra{0} \right) \nonumber \\
				 		&& + \epsilon_a \left( \vert \alpha \vert^2 \ket{1} \bra{1} - \beta \alpha^{\ast} \ket{0} \bra{1} - \alpha \beta^{\ast} \ket{1} \bra{0} + \vert \beta \vert^2 \ket{0} \bra{0} \right) \nonumber \\
				 & = & \vert \beta \vert^2 \ket{0} \bra{0} + (1 - 2 \epsilon_a) \beta \alpha^{\ast} \ket{0} \bra{1} + (1 - 2 \epsilon_a) \alpha \beta^{\ast} \ket{1} \bra{0} + \vert \alpha \vert^2 \ket{1} \bra{1} .
\end{eqnarray}
After measurement of ion $a$, a microwave pulse conditioned upon the measurement is applied to ion $b$.  In the case of measuring $\ket{0}_a$, the rotation $R_x(\pi)$ is applied to ion $b$.  The density matrix is then:
\begin{equation}
	\label{eq:teleport_imperfect_measure_rotate_b}
	\rho_b = 
	\left( 
			\begin{array}{cc}
			\vert \alpha \vert^2 & (1 - 2 \epsilon_a) \alpha \beta^{\ast} \\
			(1 - 2 \epsilon_a) \beta \alpha^{\ast} & \vert \beta \vert^2
			\end{array}
		\right) .
\end{equation}
Notice that here the fidelity of the teleported state depends critically upon the state amplitudes.  As can be seen in Eq.~\ref{eq:teleport_imperfect_measure_rotate_b}, if either $\alpha$ or $\beta$ is zero, then the imperfect detection of ion $a$ plays no role in the final density matrix of ion $b$.  Intuitively, this is correct because the influence of imperfect measurement of ion $a$ is propagated by the conditional rotation on ion $b$.  In the case that $\alpha$ or $\beta$ is zero, then the two possible rotations $R_x(\pi)$ and $R_y(\pi)$ perform the same action, and the measurement error of ion $a$ is inconsequential.

The fidelity of the operation with imperfect measurements can be evaluated by reconstructing the density matrix following measurement of ion $b$ in the same fashion as is done with the actual experimental data.  We define the probability of measuring a particular state $\ket{\psi}$ as $P_{\ket{\psi}}$.  Since all the measurements occur via state dependent fluorescence, measurement in bases other than the $z$-basis are completed by performing a microwave rotation prior to detection.  The six possible probabilities are then given by:
\begin{eqnarray}
	P_{\ket{0}} & = & tr\left( M_{0b}^{\dagger} M_{0b} \rho_b \right) \nonumber \\
	P_{\ket{1}} & = & tr\left( M_{1b}^{\dagger} M_{1b} \rho_b \right) \nonumber \\
	P_{\ket{0} - \ket{1}} & = & tr\left( M_{0b}^{\dagger} M_{0b} \left[ R_y(\pi/2) \rho_b R_y^{\dagger}(\pi/2) \right] \right) \nonumber \\
	P_{\ket{0} + \ket{1}} & = & tr\left( M_{1b}^{\dagger} M_{1b} \left[ R_y(\pi/2) \rho_b R_y^{\dagger}(\pi/2) \right] \right) \nonumber \\
	P_{\ket{0} - \ii \ket{1}} & = & tr\left( M_{1b}^{\dagger} M_{1b} \left[ R_x(\pi/2) \rho_b R_x^{\dagger}(\pi/2) \right] \right) \nonumber \\
	P_{\ket{0} + \ii \ket{1}} & = & tr\left( M_{0b}^{\dagger} M_{0b} \left[ R_x(\pi/2) \rho_b R_x^{\dagger}(\pi/2) \right] \right) .
\end{eqnarray}
These probabilities allow us to calculate the Stokes parameters:
\begin{eqnarray}
	S_0 & = & P_{\ket{1}} + P_{\ket{0}} = 1 \nonumber \\
	S_1 & = & P_{\ket{0} + \ket{1}} - P_{\ket{0} - \ket{1}} \nonumber \\
	S_2 & = & P_{\ket{0} + \ii \ket{1}} - P_{\ket{0} - \ii \ket{1}} \nonumber \\
	S_3 & = & P_{\ket{0}} - P_{\ket{1}} .
\end{eqnarray}
These parameters are the coefficients of the Pauli matrices in a simple analytical formula for the reconstruction of the density matrix:\cite{altepeter:tomography}
\begin{equation}
	\rho_{b,recon} = \frac{1}{2} \sum_{j=0}^{3} S_{j} \hat{\sigma}_{j} .
\end{equation}
The reconstructed density matrix contains the effect of imperfect measurement on ion $b$.  We can then calculate the fidelity of the teleported state, taking into account imperfect detection, by $\mathcal{F} = tr(\rho_{b,recon} \rho_{ideal})$.  Taking the state detection fidelities $\epsilon_a = 0.985$ and $\epsilon_b = 0.975$, we calculate the fidelity of the states \{ $\ket{0}$, $\ket{1}$, $(\ket{0} + \ket{1})/\sqrt{2}$, $(\ket{0} - \ket{1})/\sqrt{2}$, $(\ket{0} + \ii \ket{1})/\sqrt{2}$, $(\ket{0} - \ii \ket{1})/\sqrt{2}$ \} to be \{ 0.975, 0.975, 0.961, 0.961, 0.961, 0.961 \}, yielding an average reduction in fidelity of about 3.5\%.

Given that the fidelity of teleporting the two states $\ket{0}$ and $\ket{1}$ depends only on the measurement imperfection on ion $b$, it would have been beneficial to have reversed the roles of the two ions in this experiment.  Doing so could have improved the average fidelity by 0.003\%.  However, this change is within the uncertainty of the results presented here; the improvement would only be noticed in a much larger sample of events.

\subsection{Discussion}
	\label{subsec:teleport_discussion}
As in the original teleportation proposal,\cite{bennett:teleportation} the successful implementation of our teleportation protocol requires the transmission of two classical bits of information: one to announce the success of the heralded quantum gate and another to determine the proper final rotation to recover the teleported state at ion $b$.  While these classical bits do not convey any information about the quantum states of either ion $a$ or $b$, in the absence of this classical information ion $b$ is left in a mixed state (Eq.~\ref{eq:teleport_ion_b_projection}), and the protocol fails.  The required classical communication also ensures that no information is transferred faster than the speed of light.

\begin{figure}
	\centering
	\includegraphics[width=0.5\columnwidth,keepaspectratio]{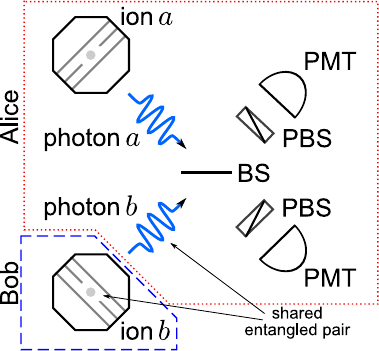}
	\caption{Comparison of the heralded teleportation protocol\protect\cite{olmschenk:teleportation} and the original proposal from Ref.~\protect\refcite{bennett:teleportation}.  The entangled qubits shared between Alice (system in the \textcolor{red}{red}, dotted box) and Bob (system in the \textcolor{blue}{blue}, dashed box) are photon $b$ and ion $b$.  The required Bell state measurement between ion $a$ (containing the information to teleport) and photon $b$ is then performed (inefficiently) by generating photon $a$ and interfering it with photon $b$ at the beamsplitter (BS).  The polarization filters (PBSs) allow only photons with the same polarization to be measured.  The coincident detection of the two photons by the photomultiplier tubes (PMTs) heralds the success of the partial Bell state measurement.  Measurement of ion $a$ and a conditional single qubit rotation on ion $b$ then completes the teleportation protocol.}
	\label{fig:bell_setup_compare}
\end{figure}

Despite the apparent differences, the heralded teleportation protocol reviewed here\cite{olmschenk:teleportation} is directly analogous to the original proposal from Ref.~\refcite{bennett:teleportation}.  In this comparison, consider ion $b$ and photon $b$ to be the entangled pair shared between Alice and Bob, as shown in Fig.~\ref{fig:bell_setup_compare}.  Thus, the three qubits used for the teleportation protocol are ion $a$, photon $b$, and ion $b$.  Photon $a$ is used in conjunction with the beamsplitter (BS) to perform an inefficient Bell-state measurement on ion $a$ and photon $b$.

Although our implementation is intrinsically probabilistic because the four two--photon Bell states are not all deterministically distinguishable,\cite{lutkenhaus:bell_no-go}  the teleportation protocol succeeds without postselection due to the two-photon coincident detection that serves as a heralding event.\cite{enk:entangle_verify}  The protocol presented here has the advantage of establishing the quantum channel between the (atomic) quantum memories using photons and entanglement swapping, allowing the atoms to be separated by a large distance from the outset.  Ultimately, the teleportation scheme demonstrated here has the potential to form the elementary constituent of a quantum repeater capable of networking quantum memories over vast distances, and may be an essential protocol for the realization of scalable quantum computation.

\section{Outlook}
	\label{sec:outlook}
The high fidelities obtained in the teleportation and quantum gate experiments is evidence of the excellent coherence properties of the photonic frequency qubit and the ``clock'' state atomic qubit.  Together, these complimentary qubits may provide a robust system for practical quantum communication and quantum computation.

\subsection{Quantum gate with infrared photons}
	\label{subsec:infrared_photons}
	
In principle, any arbitrary distance can be bridged using a quantum repeater based on the entanglement protocols reviewed in prior sections.  However, the number of nodes needed to efficiently implement the quantum repeater is approximately proportional to the inverse of the attenuation length of the photons.\cite{duan:dlcz}  In order to establish quantum channels across long-distances, it may be more practical to use photons with wavelength in the infrared region of the spectrum (rather than the 369.5 nm photons used here), as these photons experience less attenuation in fiber.  Moreover, access to additional optical frequencies may facilitate entanglement between disparate optically active systems, such as atoms and quantum dots.

The rich atomic structure of the \ybion atom results in transitions across the optical spectrum.  Two additional transitions that appear particularly amenable to the photon-mediated heralded gate described in Sec.~\ref{sec:entanglement} are the 935 nm ${}^{3}[3/2]_{1/2} \leftrightarrow {}^{2}D_{3/2}$ transition, and the 1.3 $\mu$m ${}^{2}P_{3/2} \leftrightarrow {}^{2}D_{3/2}$ transition.

\begin{figure}
	\centering
	\includegraphics[width=0.5\columnwidth,keepaspectratio]{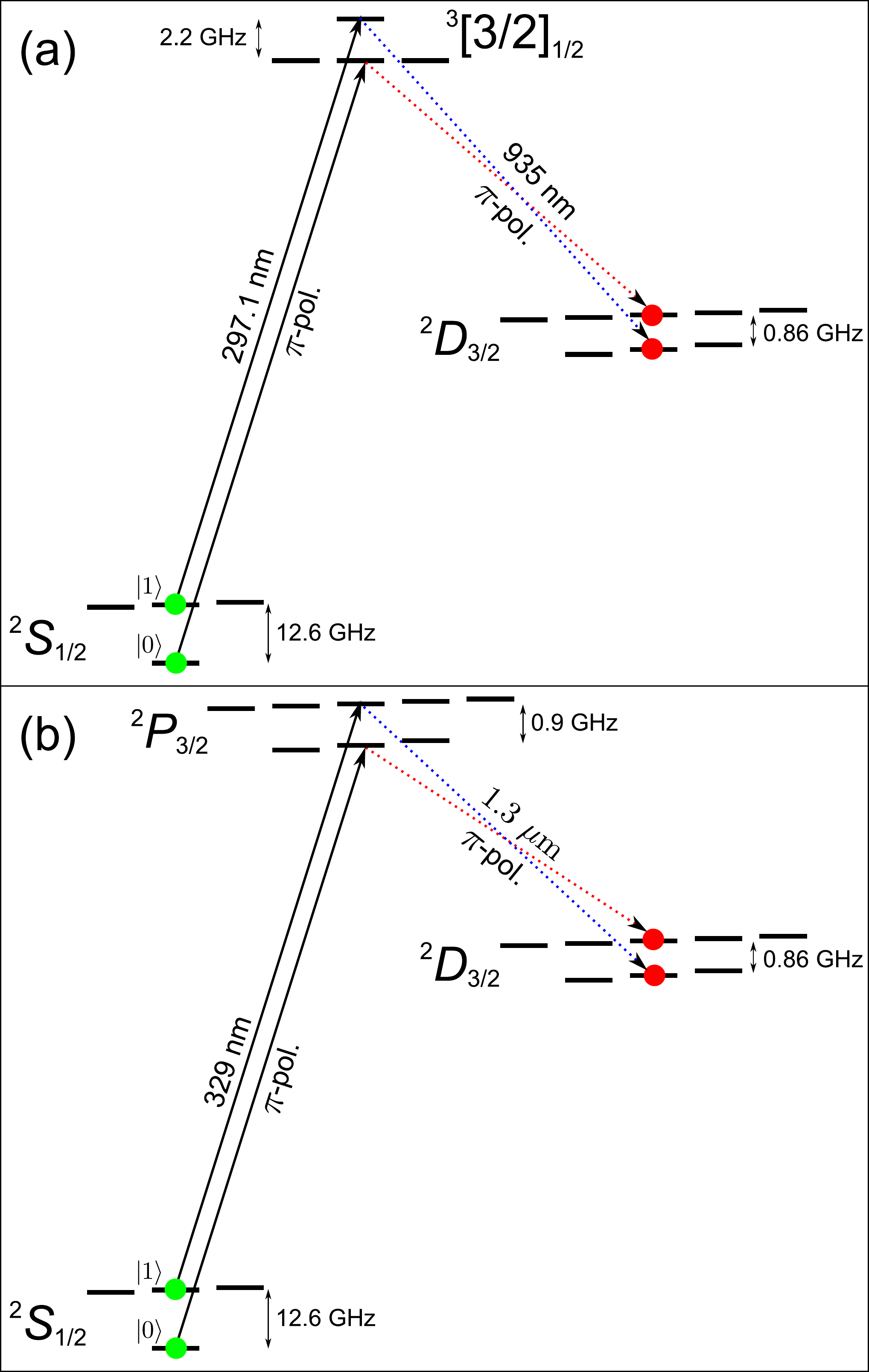}
	\caption{Quantum gate protocols utilizing infrared transitions in ${}^{171}$\ybion. (a) Procedure for generating photons at 935 nm with the frequency mode entangled with the ${}^{2}D_{3/2} \vert F=\{1,2\},m_{F}=0 \rangle$ atomic state.  As this protocol retains initial coherence in the atom, it is suitable for the implementation of a heralded quantum gate. (b) Method of generating 1.3 $\mu$m photons for the implementation of a heralded quantum gate.}
	\label{fig:infrared_entangle}
\end{figure}

Spontaneously emitted photons at 935 nm whose frequency is entangled with the internal electronic states of the atom can be generated as illustrated in Fig.~\ref{fig:infrared_entangle}(a).  In this case, the atom is initialized in the ${}^{2}S_{1/2}$ ground state and excited using ultrafast pulses at 297.1 nm.  Pulses at 297.1 nm could be generated by the third-harmonic generation of a mode-locked Ti:S laser operating at 891.4 nm.  As long as the two atomic states are mapped to separate hyperfine manifolds in the ${}^{2}D_{3/2}$ level, then selection rules can be exploited to perform state dependent fluorescence detection of the atom as before; only population in the ${}^{2}D_{3/2} \vert F=1 \rangle$ manifold will be transferred to ${}^{2}S_{1/2} \vert F=1 \rangle$ by the 935 nm light applied during detection.  Given the 52 ms natural lifetime of the ${}^{2}D_{3/2}$ level, it should be possible to obtain state detection fidelities $>98$\%.\cite{olmschenk:state-detect}

A heralded quantum gate can be performed using 935 nm photons (Fig.~\ref{fig:infrared_entangle}(a)) by first preparing the atom in a superposition of $\ket{0}$ and $\ket{1}$:
\begin{equation}
	\label{eq:935_gate_prep}
	\ket{\psi}_{a} = \alpha \ket{0} + \beta \ket{1} .
\end{equation}
A $\pi$-polarized ultrafast pulse at 297.1 nm is used to coherently transfer the population from ${}^{2}S_{1/2}$ to ${}^{3}[3/2]_{1/2}$.  Due to the selection rules involved, this will transfer $\ket{0}$ to ${}^{3}[3/2]_{1/2} \vert F=1,m_{F}=0 \rangle$ and $\ket{1}$ to ${}^{3}[3/2]_{1/2} \vert F=0,m_{F}=0 \rangle$.  The ${}^{3}[3/2]_{1/2}$ level can then decay to ${}^{2}D_{3/2}$ by spontaneously emitting a 935 nm photon.  By collecting photons at 935 nm emitted perpendicular to the quantization axis, we can use polarization filters to distinguish $\pi$- and $\sigma$-polarized photons.  A $\pi$-polarized 935 nm photon is the result of a ${}^{3}[3/2]_{1/2} \vert F=1,m_{F}=0 \rangle$ to ${}^{2}D_{3/2} \vert F=2,m_{F}=0\rangle$ or ${}^{3}[3/2]_{1/2} \vert F=0,m_{F}=0 \rangle$ to ${}^{2}D_{3/2} \vert F=1,m_{F}=0\rangle$ transition, resulting in the frequency of the emitted photon being entangled with the internal state of the atom:
\begin{eqnarray}
	\label{eq:935_gate}
	\vert \psi \rangle_{ap} = \beta \vert  F=2,m_{F}=0 \rangle \vert \nu_{r} \rangle + \alpha \vert  F=1,m_{F}=0 \rangle \vert \nu_{b} \rangle .
\end{eqnarray}
Here, $\vert F,m_{F} \rangle$ refer to states in the ${}^{2}D_{3/2}$ level, and $\Delta \nu = \nu_{b} - \nu_{r} = 3.07$ GHz is the sum of the ${}^{2}D_{3/2}$ and ${}^{3}[3/2]_{1/2}$ hyperfine splittings.  As the ion-photon entanglement process preserves the coherence initially present in the ion, the interference and detection of these photons can be used to implement a heralded quantum gate.\cite{duan:freq-qubit,maunz:heralded-gate}  Detection of the atomic state can be accomplished by fluorescence detection at 369.5 nm, as described above.  In this case, detecting fluorescence at 369.5 nm indicates the atom was originally in the ${}^{2}D_{3/2} \vert  F=1,m_{F}=0 \rangle$ state, whereas the absence of fluorescence indicates the atom is in ${}^{2}D_{3/2} \vert  F=2,m_{F}=0 \rangle$.

The disadvantage of generating 935 nm photons is that the branching ratio of the ${}^{3}[3/2]_{1/2}$ level between ${}^{2}S_{1/2}$ and ${}^{2}D_{3/2}$ has been calculated to be about 55:1,\cite{biemont:lifetimeybcalc} decreasing the probability of generating a 935 nm photon and thereby reducing the overall success probability of the entanglment and gate protocols outlined above.  Due to this reduced success probability, dark counts of the single-photon detectors could become a significant source of error.  However, there is a way to ``veto'' the contribution of dark counts in this protocol.  After a photon detection event, a microwave pulse at the ${}^{2}D_{3/2}$ hyperfine splitting of 0.86 GHz can be used to transfer the ${}^{2}D_{3/2} \vert F=1,m_{F}=0 \rangle$ atomic state to, for instance, the ${}^{2}D_{3/2} \vert F=2,m_{F}=1 \rangle$ state.  If the fluorescence detection procedure outlined previously is now performed, no 369.5 nm photons should be detected.\footnote{Here, the 14.7 GHz sideband usually used during Doppler cooling can be applied to the impinging 369.5 nm light during this detection interval, to ensure population left in the $\ket{0}$ state also produces fluorescence.}  The population transfered from ${}^{2}D_{3/2} \vert F=1 \rangle$ to ${}^{2}D_{3/2} \vert F=2 \rangle$ can be returned by a second microwave pulse, and the remainder of the protocol completed.  On the other hand, detection of 369.5 nm photons during this detection interval would indicate that the 935 nm ``detection'' was a false event, and should be discarded.

The gate protocol outlined above could also be implemented at 1.3 $\mu$m, where attenuation in optical fiber is near a minimum (Fig.~\ref{fig:infrared_entangle}(b)).  In this case, the atom is initially prepared in ${}^{2}S_{1/2}$, and then excited to ${}^{2}P_{3/2}$ by an ultrafast pulse at 329 nm.  Decay from ${}^{2}P_{3/2}$ to ${}^{2}D_{3/2}$ results in the emission of a 1.3 $\mu$m photon.  While the above protocol can be implemented in an analogous fashion at this wavelength, the branching ratio from ${}^{2}P_{3/2}$ to ${}^{2}S_{1/2}$ versus ${}^{2}D_{3/2}$ is about 475:1.\cite{biemont:lifetimeybcalc}  Thus, while the wavelength is more amenable to long-distance transmission, the decrease in the protocol success probability is even more dramatic.  In addition, ${}^{2}P_{3/2}$ can also decay to ${}^{2}D_{5/2}$, which can subsequently decay to the long-lived ${}^{2}F_{7/2}$.  Depopulating these additional metastable states would require additional optical frequencies and could limit the repetition rate of the experiment.

\subsection{Scalability}
	\label{subsec:scalability}
	
The primary impediment to scaling the current setup (as well as the infrared protocols above) to more qubits is the success probability of the heralded quantum gate.  As discussed in Sec.~\ref{sec:entanglement}, there are several avenues that can be pursued to improve the success rate.  Given that the effective solid angle of collection is about an order of magnitude smaller than the other contributing factors in Eq.~\ref{eq:prob_gate_success}, improvements here are bound to have the greatest impact.

One proposal to increase the effective solid angle is to situate reflective or refractive optics near the trapped atomic ion.\cite{streed:fresnel_optics,shu:trap_mirror}  A possible approach is to place the trapped ion at the focal point of a parabolic mirror, allowing the collection efficiency to approach unity.\cite{lindlein:4pi_parabolic_mirror,maiwald:enhanced_optical_access}  If properly segmented, a metallic mirror could also serve as the electrodes for the ion trap.  Alternatively, the spontaneous emission into free space could be replaced by the induced emission into the small mode volume of a high finesse cavity, which can reach near unit efficiency.\cite{hennrich:high_finesse_cavity,mckeever:atom-cavity,keller:ion_cavity_single_ph,russo:raman_ion_cavity}  Even though the free spectral range of the cavity would have to be 14.7 GHz to simultaneously support both frequency modes in the gate protocol, choosing a near-concentric design could still result in a small mode volume and thus in a high emission probability into a well-defined Gaussian mode.  Of course, placement of optics near the ion will require careful assessment to ensure the added structures do not destabilize the quadrupole trap.  As another option, the photonic qubit could be encoded in a temporal mode (time-bin qubit), allowing more conventional cavity structures in the implementation of the heralded quantum gate.\cite{barrett:time-bin}

While improvements that increase the success probability of the gate operation can enhance scalability, even with a low success probability, this gate can still be efficiently scaled to more complex systems.  The architecture for a quantum repeater could consist of photon-mediated operations between spatially separated nodes and local deterministic gates at each node.  This implementation would also enable deterministic long-distance quantum teleportation and remote deterministic gates.\cite{duan:prob-photon}  In this architecture, the photon-mediated operations between nodes can be attempted simultaneously.  The requirement for scaling to more nodes is therefore only that the coherence time of the atoms exceed the time needed to connect all the nodes of the quantum repeater.  Given all connections can be attempted simultaneously, then the time needed to connect all nodes is approximately given by $T_{success} \ln(N)$, where $T_{success}$ is the average time needed to connect two nodes and $N$ is the total number of nodes in the repeater ($N$ large).  With a measured coherence time of 2.5 seconds for the ${}^{171}$\ybion hyperfine qubit,\cite{olmschenk:state-detect} this means that to establish entanglement over a quantum repeater with 10 nodes, we require the average success rate to be greater than about 1 Hz.\footnote{In this estimate, we have assumed only one photon-mediated connection is being attempted between each node.  Multiplexing this operation further relaxes the required success rate.}  It should be possible to achieve this success rate with modest improvements in the photon collection efficiency and the repetition rate of the experiment.  

The stipulations on the success probability for generating large cluster states for scalable quantum computing are more stringent.  The time needed to construct an $n$ qubit 1D cluster state is approximately given by\cite{duan:robust_qip}
\begin{equation}
	\label{eq:time_for_cluster_state}
	T(n) \approx t_{a} \left( \frac{1}{P_{success}} \right)^{\log_{2}(n_{c} + 1)} + \left( \frac{t_{a}}{P_{success}} \right) \log_{2} \left( n - n_{c} \right) ,
\end{equation}
where $n_{c} \approx 4/P_{success}$ is the critical number of qubits in a single 1D cluster state chain that must be generated before multiple chains can be fused together.  Clearly, even if we assume the repetition rate of the experiment can be improved to 100 ns, at our current success probability of $2.2 \times 10^{-8}$ the time to generate a 100 qubit cluster state is prohibitively long ($>10^{36}$ years).  However, if the improvements suggested above are able to increase the gate success probability to 10\%, then a 100 qubit cluster state could be generated in less than 25 ms.

\subsection{Summary}
	\label{sec:summary}

We have reviewed the recent experimental progress towards quantum networks with distant trapped atomic ions.  We presented a detailed discussion of the recent implementation of a heralded quantum gate between remote ions based on the interference and detection of spontaneously emitted photons, and the employment of this gate to teleport a qubit between distant quantum memories.  Ultimately, this quantum gate and teleportation protocol could be essential to realizing long-distance quantum communication and scalable quantum computation.


\section*{Acknowledgments}
We would like to thank Luming Duan and Kelly Younge for their valuable contributions.  This work is supported by IARPA under ARO contract, the NSF PIF Program, and the NSF Physics Frontier Center at the Joint Quantum Institute.  D.~L.~M. acknowledges support from the Alexander von Humboldt Foundation.



\end{document}

%% file: setmathcommands_ijqi.tex
\newcommand{\ybion}{Yb${}^{+}$}

\newcommand{\ii}{\dot{\imath}}

\newcommand{\ket}[1]{\vert{#1}\rangle}
\newcommand{\bra}[1]{\langle{#1}\vert}

\newcommand{\aup}[1]{\hat{a}_{#1}^{\dagger}}
\newcommand{\adown}[1]{\hat{a}_{#1}}
\newcommand{\bup}[1]{\hat{b}_{#1}^{\dagger}}
\newcommand{\bdown}[1]{\hat{b}_{#1}}

%% file: ws-ijqi_smo_arxiv.bbl
\begin{thebibliography}{100}

\bibitem{feynman:qsim}
R.~P. Feynman,
\newblock ``Simulating physics with computers'',
\newblock {\em International Journal of Theoretical Physics} {\bf 21}, 467
  (1982).

\bibitem{deutsch:universalQC}
D.~Deutsch,
\newblock ``Quantum theory, the Church-Turing principle and the universal
  quantum computer'',
\newblock {\em Proc. Royal Society of London A} {\bf 400}, 97 (1985).

\bibitem{shor:factoring}
P.~Shor,
\newblock ``Algorithms for quantum computation: discrete logarithms and
  factoring'',
\newblock in {\em Proc. 35th Ann. Sym. Found. Comp. Sci.}, page 124, 1994.

\bibitem{bennett:bb84}
C.~H. Bennett and G.~Brassard,
\newblock ``Quantum cryptography: Public-key distribution and coin tossing'',
\newblock in {\em Proceedings of IEEE International Conference on Computers,
  Systems and Signal Processing, Bangalore, India}, page 175, IEEE Press, 1984.

\bibitem{ekert:q_crypto_bell}
A.~K. Ekert,
\newblock ``Quantum Cryptography Based on {Bell's} Theorem'',
\newblock {\em Phys. Rev. Lett.} {\bf 67}, 661 (1991).

\bibitem{wootters:no-clone}
W.~K. Wootters and W.~H. Zurek,
\newblock ``A single quantum cannot be cloned'',
\newblock {\em Nature} {\bf 299}, 802 (1982).

\bibitem{bennett:teleportation}
C.~H. Bennett et~al.,
\newblock ``Teleporting an unknown quantum state via dual classical and
  Einstein-Podolsky-Rosen channels'',
\newblock {\em Phys. Rev. Lett.} {\bf 70}, 1895 (1993).

\bibitem{bell:inequal}
J.~S. Bell,
\newblock ``On the Einstein Podolsky Rosen paradox'',
\newblock {\em Physics} {\bf 1}, 195 (1964).

\bibitem{kok:linear_op_qc}
P.~Kok et~al.,
\newblock ``Linear optical quantum computing with photonic qubits'',
\newblock {\em Rev. Mod. Phys.} {\bf 79}, 135 (2007).

\bibitem{blatt:ions_entangled}
R.~Blatt and D.~J. Wineland,
\newblock ``Entangled states of trapped atomic ions'',
\newblock {\em Nature} {\bf 453}, 1008 (2008).

\bibitem{clarke:superconducting_qubits}
J.~Clarke and F.~K. Wilhelm,
\newblock ``Superconducting quantum bits'',
\newblock {\em Nature} {\bf 453}, 1031 (2008).

\bibitem{kimble:qinternet}
H.~J. Kimble,
\newblock ``The quantum internet'',
\newblock {\em Nature} {\bf 453}, 1023 (2008).

\bibitem{bloch:ultracold_lattices}
I.~Bloch,
\newblock ``Quantum coherence and entanglement with ultracold atoms in optical
  lattices'',
\newblock {\em Nature} {\bf 453}, 1016 (2008).

\bibitem{blinov:ion-photon}
B.~B. Blinov, D.~L. Moehring, L.-M. Duan, and C.~Monroe,
\newblock ``Observation of entanglement between a single trapped atom and a
  single photon'',
\newblock {\em Nature} {\bf 428}, 153 (2004).

\bibitem{matsukevich:matter-light}
D.~N. Matsukevich and A.~Kuzmich,
\newblock ``Quantum State Transfer Between Matter and Light'',
\newblock {\em Science} {\bf 306}, 663 (2004).

\bibitem{riedmatten:atom-ensemble}
H.~{de Riedmatten} et~al.,
\newblock ``Direct Measurement of Decoherence for Entanglement between a Photon
  and Stored Atomic Excitation'',
\newblock {\em Phys. Rev. Lett.} {\bf 97}, 113603 (2006).

\bibitem{chen:atom-ensemble_entangle}
S.~Chen et~al.,
\newblock ``Demonstration of a Stable Atom-Photon Entanglement Source for
  Quantum Repeaters'',
\newblock {\em Phys. Rev. Lett.} {\bf 99}, 180505 (2007).

\bibitem{sherson:light-ensemble_teleport}
J.~F. Sherson et~al.,
\newblock ``Quantum teleportation between light and matter'',
\newblock {\em Nature} {\bf 443}, 557 (2006).

\bibitem{volz:atom-photon}
J.~Volz et~al.,
\newblock ``Observation of Entanglement of a Single Photon with a Trapped
  Atom'',
\newblock {\em Phys. Rev. Lett.} {\bf 96}, 030404 (2006).

\bibitem{wilk:atom-photon}
T.~Wilk, S.~C. Webster, A.~Kuhn, and G.~Rempe,
\newblock ``Single-Atom Single-Photon Quantum Interface'',
\newblock {\em Science} {\bf 317}, 488 (2007).

\bibitem{haffner:8_ion_w}
H.~{H\"{a}ffner} et~al.,
\newblock ``Scalable multiparticle entanglement of trapped ions'',
\newblock {\em Nature} {\bf 438}, 643 (2005).

\bibitem{kielpinski:ion_architecture}
D.~Kielpinski, C.~R. Monroe, and D.~J. Wineland,
\newblock ``Architecture for a Large-Scale Ion-Trap Quantum Computer'',
\newblock {\em Nature} {\bf 417}, 709 (2002).

\bibitem{stick:microtrap}
D.~Stick et~al.,
\newblock ``Ion Trap in a Semiconductor Chip'',
\newblock {\em Nature Physics} {\bf 2}, 36 (2006).

\bibitem{seidelin:surface_trap}
S.~Seidelin et~al.,
\newblock ``A microfabricated surface-electrode ion trap for scalable quantum
  information processing'',
\newblock {\em Phys. Rev. Lett.} {\bf 96}, 253003 (2006).

\bibitem{hensinger:t-trap}
W.~K. Hensinger et~al.,
\newblock ``T-junction ion trap array for two-dimensional ion shuttling,
  storage and manipulation'',
\newblock {\em App. Phys. Lett.} {\bf 88}, 034101 (2006).

\bibitem{blakestad:x-trap}
R.~B. Blakestad et~al.,
\newblock ``High-Fidelity Transport of Trapped-Ion Qubits through an X-Junction
  Trap Array'',
\newblock {\em Phys. Rev. Lett.} {\bf 102}, 153002 (2009).

\bibitem{turchette:heating}
Q.~A. Turchette et~al.,
\newblock ``Heating of trapped ions from the quantum ground state'',
\newblock {\em Phys. Rev. A} {\bf 61}, 063418 (2000).

\bibitem{deslauriers:needle}
L.~Deslauriers et~al.,
\newblock ``Scaling and Suppression of Anomalous Heating in Ion Traps'',
\newblock {\em Phys. Rev. Lett.} {\bf 97}, 103007 (2006).

\bibitem{labaziewicz:heating}
J.~Labaziewicz et~al.,
\newblock ``Suppression of Heating Rates in Cryogenic Surface-Electrode Ion
  Traps'',
\newblock {\em Phys. Rev. Lett.} {\bf 100}, 013001 (2008).

\bibitem{ursin:144km}
R.~Ursin et~al.,
\newblock ``Entanglement-based quantum communication over 144 km'',
\newblock {\em Nature Physics} {\bf 3}, 481 (2007).

\bibitem{briegel:quantum_repeater}
H.-J. Briegel, W.~{D\"{u}r}, J.~I. Cirac, and P.~Zoller,
\newblock ``Quantum Repeaters: The Role of Imperfect Local Operations in
  Quantum Communication'',
\newblock {\em Phys. Rev. Lett.} {\bf 81}, 5932 (1998).

\bibitem{duan:dlcz}
L.-M. Duan, M.~D. Lukin, J.~I. Cirac, and P.~Zoller,
\newblock ``Long-distance quantum communication with atomic ensembles and
  linear optics'',
\newblock {\em Nature} {\bf 414}, 413 (2001).

\bibitem{paul:nobel}
W.~Paul,
\newblock ``Electromagnetic traps for charged and neutral particles'',
\newblock {\em Rev. Mod. Phys.} {\bf 62}, 531 (1990).

\bibitem{wineland:cooling}
D.~J. Wineland, R.~E. Drullinger, and F.~L. Walls,
\newblock ``Radiation Pressure Cooling of Bound Resonant Absorbers'',
\newblock {\em Phys. Rev. Lett.} {\bf 40}, 1639 (1978).

\bibitem{neuhauser:cooling}
W.~Neuhauser, M.~Hohenstatt, P.~Toschek, and H.~Dehmelt,
\newblock ``Optical-Sideband Cooling of Visible Atom Cloud Confined in
  Parabolic Well'',
\newblock {\em Phys. Rev. Lett.} {\bf 41}, 233 (1978).

\bibitem{fisk:171s12}
P.~T.~H. Fisk, M.~J. Sellars, M.~A. Lawn, and C.~Coles,
\newblock ``Accurate Measurement of the 12.6 GHz ``Clock'' Transition in
  Trapped ${}^{171}$Yb$^+$ Ions'',
\newblock {\em IEEE Trans. Ultrasonics, Ferroelectrics, and Frequency Control}
  {\bf 44}, 344 (1997).

\bibitem{march:mass_spect}
R.~E. March,
\newblock ``Quadrupole ion trap mass spectrometry: theory, simulation, recent
  developments and applications'',
\newblock {\em Rapid Comm. in Mass Spect.} {\bf 12}, 1543 (1998).

\bibitem{fortier:fund}
T.~M. Fortier et~al.,
\newblock ``Precision atomic spectroscopy for improved limits on variation of
  the fine structure constant and local position invariance'',
\newblock {\em Phys. Rev. Lett.} {\bf 98}, 070801 (2007).

\bibitem{rosenband:AltoHg}
T.~Rosenband et~al.,
\newblock ``Frequency ratio of Al${}^{+}$ and Hg${}^{+}$ single-ion optical
  clocks; Metrology at the 17th decimal place'',
\newblock {\em Science} {\bf 319}, 1808 (2008).

\bibitem{griffiths:em}
D.~J. Griffiths,
\newblock {\em Introduction to Electrodynamics},
\newblock Prentice Hall, 3rd edition, 1999.

\bibitem{wineland:nist-jnl}
D.~J. Wineland et~al.,
\newblock ``Experimental Issues in Coherent Quantum-State Manipulation of
  Trapped Atomic Ions'',
\newblock {\em Journal of Research of the National Institute of Standards and
  Technology} {\bf 103}, 259 (1998).

\bibitem{mclachlan:mathieu}
N.~W. McLachlan,
\newblock {\em Theory and Application of Mathieu Functions},
\newblock {Dover Publications, Inc.}, 1964.

\bibitem{dehmelt:iontrap}
H.~G. Dehmelt,
\newblock ``Radiofrequency spectroscopy of stored ions I: storage'',
\newblock {\em Adv. At. Mol. Phys.} {\bf 3}, 53 (1967).

\bibitem{berkeland:micromotion}
D.~J. Berkeland, J.~D. Miller, J.~C. Bergquist, W.~M. Itano, and D.~J.
  Wineland,
\newblock ``Minimization of ion micromotion in a Paul trap'',
\newblock {\em J. Appl. Phys.} {\bf 83}, 5025 (1998).

\bibitem{madsen:planar_trap}
M.~J. Madsen, W.~K. Hensinger, D.~Stick, J.~A. Rabchuk, and C.~Monroe,
\newblock ``Planar ion trap geometry for microfabrication'',
\newblock {\em Appl. Phys. B} {\bf 78}, 639 (2004).

\bibitem{fitzpatrick:adv_calc}
P.~M. Fitzpatrick,
\newblock {\em Advanced Calculus: A Course in Mathematical Analysis},
\newblock PWS Publishing Company, 1st edition, 1996.

\bibitem{cirac:cold-ions}
J.~Cirac and P.~Zoller,
\newblock ``Quantum Computations with Cold Trapped Ions'',
\newblock {\em Phys. Rev. Lett.} {\bf 74}, 4091 (1995).

\bibitem{devoe:ba_heating}
R.~G. {DeVoe} and C.~Kurtsiefer,
\newblock ``Experimental study of anomalous heating and trap instabilities in a
  microscopic ${}^{137}$Ba ion trap'',
\newblock {\em Phys. Rev. A} {\bf 65}, 063407 (2002).

\bibitem{dubin:ion_two-photon}
F.~Dubin, D.~Rotter, M.~Mukherjee, S.~Gerber, and R.~Blatt,
\newblock ``Single-Ion Two-Photon Source'',
\newblock {\em Phys. Rev. Lett.} {\bf 99}, 183001 (2007).

\bibitem{dietrich:ba_qc}
M.~R. Dietrich et~al.,
\newblock ``Barium Ions for Quantum Computation'',
\newblock in {\em Proc. 9th Inter. Workshop on Non-Neutral Plasmas}, 2008.

\bibitem{monroe:qgate}
C.~Monroe, D.~M. Meekhof, B.~E. King, W.~M. Itano, and D.~J. Wineland,
\newblock ``Demonstration of a Fundamental Quantum Logic Gate'',
\newblock {\em Phys. Rev. Lett.} {\bf 75}, 4714 (1995).

\bibitem{benhelm:ca43}
J.~Benhelm, G.~Kirchmair, C.~F. Roos, and R.~Blatt,
\newblock ``Experimental quantum-information processing with
  ${}^{43}$Ca${}^{+}$ ions'',
\newblock {\em Phys. Rev. A} {\bf 77}, 062306 (2008).

\bibitem{home:ca_entangle}
J.~P. Home et~al.,
\newblock ``Deterministic entanglement and tomography of ion–spin qubits'',
\newblock {\em New J. Phys.} {\bf 8}, 188 (2006).

\bibitem{schulz:cooling_multitrap}
S.~A. Schulz, U.~Poschinger, F.~Ziesel, and F.~{Schmidt-Kaler},
\newblock ``Sideband cooling and coherent dynamics in a microchip
  multi-segmented ion trap'',
\newblock {\em New J. Phys.} {\bf 10}, 045007 (2008).

\bibitem{schuck:atom-sdcp}
C.~Schuck et~al.,
\newblock ``Resonant interaction of a single atom with single photons from a
  down-conversion source'',
\newblock (2009),
\newblock arXiv:0906.1719.

\bibitem{toyoda:ca_coher}
K.~Toyoda, H.~Shiibara, S.~Haze, R.~Yamazaki, and S.~Urabe,
\newblock ``Experimental study of the coherence of a terahertz-separated
  metastable-state qubit in ${}^{40}$Ca$^+$'',
\newblock {\em Phys. Rev. A} {\bf 79}, 023419 (2009).

\bibitem{lee:qubit_eom}
P.~J. Lee et~al.,
\newblock ``Atomic qubit manipulations with an electro-optic modulator'',
\newblock {\em Op. Lett.} {\bf 28}, 1582 (2003).

\bibitem{barrett:symp_be_mg}
M.~D. Barrett et~al.,
\newblock ``Sympathetic cooling of ${}^{9}$Be${}^{+}$ and ${}^{24}$Mg${}^{+}$
  for quantum logic'',
\newblock {\em Phys. Rev. A} {\bf 68}, 042302 (2003).

\bibitem{friedenauer:q-sim_magnet}
A.~Friedenauer, H.~Schmitz, J.~T. Glueckert, D.~Porras, and T.~Schaetz,
\newblock ``Simulating a quantum magnet with trapped ions'',
\newblock {\em Nature Physics} {\bf 4}, 757 (2008).

\bibitem{balzer:ybqip}
C.~Balzer et~al.,
\newblock ``Electrodynamically trapped Yb$^{+}$ ions for quantum information
  processing'',
\newblock {\em Phys. Rev. A} {\bf 73}, 041407(R) (2006).

\bibitem{olmschenk:state-detect}
S.~Olmschenk et~al.,
\newblock ``Manipulation and Detection of a Trapped Yb${}^{+}$ Hyperfine
  Qubit'',
\newblock {\em Phys. Rev. A} {\bf 76}, 052314 (2007).

\bibitem{sansonetti:nist_database}
J.~E. Sansonetti, W.~C. Martin, and S.~L. Young,
\newblock {\em Handbook of Basic Atomic Spectroscopic Data},
\newblock Number {1.1.2}, National Institute of Standards and Technology,
  {Gaithersburg, MD}, 2005,
\newblock {Available: http://physics.nist.gov/Handbook}.

\bibitem{duan:robust_qip}
L.-M. Duan and C.~Monroe,
\newblock ``{Robust Quantum Information Processing with Atoms, Photons, and
  Atomic Ensembles}'',
\newblock {\em Adv. At. Mol. Opt. Phys.} {\bf 55}, 419 (2008).

\bibitem{poyatos:strong-excite}
J.~F. Poyatos, J.~I. Cirac, R.~Blatt, and P.~Zoller,
\newblock ``Trapped ions in the strong-excitation regime: Ion interferometry
  and nonclassical states'',
\newblock {\em Phys. Rev. A} {\bf 54}, 1532 (1996).

\bibitem{garcia-ripoll:fast-gates}
J.~J. {Garc\'{i}a--Ripoll}, P.~Zoller, and J.~I. Cirac,
\newblock ``Speed Optimized Two-Qubit Gates with Laser Coherent Control
  Techniques for Ion Trap Quantum Computing'',
\newblock {\em Phys. Rev. Lett.} {\bf 91}, 157901 (2003).

\bibitem{duan:fast-gates}
L.-M. Duan,
\newblock ``Scaling Ion Trap Quantum Computation through Fast Quantum Gates'',
\newblock {\em Phys. Rev. Lett.} {\bf 93}, 100502 (2004).

\bibitem{madsen:ultrafast-rabi}
M.~J. Madsen et~al.,
\newblock ``Ultrafast Coherent Coupling of Atomic Hyperfine and Photon
  Frequency Qubits'',
\newblock {\em Phys. Rev. Lett.} {\bf 97}, 040505 (2006).

\bibitem{bell:four-level}
A.~S. Bell, P.~Gill, H.~A. Klein, and A.~P. Levick,
\newblock ``Laser cooling of trapped ytterbium ions using a four-level
  optical-excitation scheme'',
\newblock {\em Phys. Rev. A} {\bf 44}, R20 (1991).

\bibitem{lehmitz:pop-trap}
H.~Lehmitz, J.~{Hattendorf--Ledwoch}, R.~Blatt, and H.~Harde,
\newblock ``Population trapping in excited Yb ions'',
\newblock {\em Phys. Rev. Lett.} {\bf 62}, 2108 (1989).

\bibitem{bauch:pop-trap}
A.~Bauch, D.~Schnier, and C.~Tamm,
\newblock ``Collisional population trapping and optical deexcitation of
  ytterbium ions in a radiofrequency trap'',
\newblock {\em J. Mod. Opt.} {\bf 39}, 389 (1992).

\bibitem{schauer:yb_coll_pop}
M.~M. Schauer et~al.,
\newblock ``Collisional population transfer in trapped Yb$^+$ ions'',
\newblock {\em Phys. Rev. A} {\bf 79}, 062705 (2009).

\bibitem{olmschenk:yb_lifetime}
S.~Olmschenk et~al.,
\newblock ``Precision measurement of the lifetime of the $6p$ ${}^{2}P_{1/2}$
  level of Yb${}^{+}$'',
\newblock (2009),
\newblock arXiv:0906.0586.

\bibitem{yu:lifetime}
N.~Yu and L.~Maleki,
\newblock ``Lifetime measurements of the $4f^{14}5d$ metastable states in
  single ytterbium ions'',
\newblock {\em Phys. Rev. A} {\bf 61}, 022507 (2000).

\bibitem{taylor:d52}
P.~Taylor et~al.,
\newblock ``Investigation of the ${}^{2}S_{1/2}-{}^{2}D_{5/2}$ clock transition
  in a single ytterbium ion'',
\newblock {\em Phys. Rev. A} {\bf 56}, 2699 (1997).

\bibitem{berends:beam-laser-YbII}
R.~W. Berends, E.~H. Pinnington, B.~Guo, and Q.~Ji,
\newblock ``Beam-laser lifetime measurements for four resonance levels of Yb
  II'',
\newblock {\em J. Phys. B} {\bf 26}, L701 (1993).

\bibitem{roberts:171f72}
M.~Roberts, P.~Taylor, G.~P. Barwood, W.~R.~C. Rowley, and P.~Gill,
\newblock ``Observation of the ${}^{2}S_{1/2}$--${}^{2}F_{7/2}$ electric
  octupole transition in a single ${}^{171}$Yb${}^{+}$ ion'',
\newblock {\em Phys. Rev. A} {\bf 62}, 020501(R) (2000).

\bibitem{biemont:lifetimeybcalc}
E.~{Bi\'{e}mont}, J.-F. Dutrieu, I.~Martin, and P.~Quinet,
\newblock ``Lifetime calculations in Yb II'',
\newblock {\em J. Phys. B: At. Mol. Opt. Phys.} {\bf 31}, 3321 (1998).

\bibitem{berkeland:coherent-pop}
D.~J. Berkeland and M.~G. Boshier,
\newblock ``Destabilization of dark states and optical spectroscopy in
  Zeeman-degenerate atomic systems'',
\newblock {\em Phys. Rev. A} {\bf 65}, 033413 (2002).

\bibitem{hong:HOM}
C.~K. Hong, Z.~Y. Ou, and L.~Mandel,
\newblock ``Measurement of Subpicosecond Time Intervals between Two Photons by
  Interference'',
\newblock {\em Phys. Rev. Lett.} {\bf 59}, 2044 (1987).

\bibitem{shih:HOM}
Y.~H. Shih and C.~O. Alley,
\newblock ``New Type of Einstein-Podolsky-Rosen-Bohm Experiment Using Pairs of
  Light Quanta Produced by Optical Parametric Down Conversion'',
\newblock {\em Phys. Rev. Lett.} {\bf 61}, 2921 (1988).

\bibitem{legero:photon-interfere}
T.~Legero, T.~Wilk, A.~Kuhn, and G.~Rempe,
\newblock ``Time-resolved two-photon quantum interference'',
\newblock {\em Appl. Phys. B} {\bf 77}, 797 (2003).

\bibitem{jackson:em}
J.~D. Jackson,
\newblock {\em Classical Electrodynamics},
\newblock John Wiley \& Sons, Inc., 3rd edition, 1999.

\bibitem{enk:single-particle_entanglment}
S.~J. {van Enk},
\newblock ``Single-particle entanglement'',
\newblock {\em Phys. Rev. A} {\bf 72}, 064306 (2005).

\bibitem{drezet:comment_single-particle_entanglement}
A.~Drezet,
\newblock ``Comment on 'Single-particle entanglement''',
\newblock {\em Phys. Rev. A} {\bf 74}, 026301 (2006).

\bibitem{enk:reply_single-particle_entanglment}
S.~J. {van Enk},
\newblock ``Reply to 'Comment on 'Single-particle entanglement'''',
\newblock {\em Phys. Rev. A} {\bf 74}, 026302 (2006).

\bibitem{yuan:single-photon}
Z.~Yuan et~al.,
\newblock ``Electrically Driven Single-Photon Source'',
\newblock {\em Science} {\bf 295}, 102 (2002).

\bibitem{maunz:interference}
P.~Maunz et~al.,
\newblock ``Quantum interference of photon pairs from two remote trapped atomic
  ions'',
\newblock {\em Nature Physics} {\bf 3}, 538 (2007).

\bibitem{benhelm:ms-gate}
J.~Benhelm, G.~Kirchmair, C.~F. Roos, and R.~Blatt,
\newblock ``Towards fault-tolerant quantum computing with trapped ions'',
\newblock {\em Nature Physics} {\bf 4}, 463 (2008).

\bibitem{raussendorf:one-way_qc}
R.~Raussendorf and H.~J. Briegel,
\newblock ``A One-Way Quantum Computer'',
\newblock {\em Phys. Rev. Lett.} {\bf 86}, 5188 (2001).

\bibitem{duan:prob_gate}
L.-M. Duan and R.~Raussendorf,
\newblock ``Efficient Quantum Computation with Probabilistic Quantum Gates'',
\newblock {\em Phys. Rev. Lett.} {\bf 95}, 080503 (2005).

\bibitem{walther:photon_one-way_qc}
P.~Walther et~al.,
\newblock ``Experimental one-way quantum computing'',
\newblock {\em Nature} {\bf 434}, 169 (2005).

\bibitem{lu:photon_graph_states}
C.-Y. Lu et~al.,
\newblock ``Experimental entanglement of six photons in graph states'',
\newblock {\em Nature Physics} {\bf 3}, 91 (2007).

\bibitem{bodiya:linear_op_graph}
T.~P. Bodiya and L.-M. Duan,
\newblock ``Scalable Generation of Graph-State Entanglement Through Realistic
  Linear Optics'',
\newblock {\em Phys. Rev. Lett.} {\bf 97}, 143601 (2006).

\bibitem{maunz:heralded-gate}
P.~Maunz et~al.,
\newblock ``Heralded Quantum Gate between Remote Quantum Memories'',
\newblock {\em Phys. Rev. Lett.} {\bf 102}, 250502 (2009).

\bibitem{duan:freq-qubit}
L.-M. Duan et~al.,
\newblock ``Probabilistic quantum gates between remote atoms through
  interference of optical frequency qubits'',
\newblock {\em Phys. Rev. A} {\bf 73}, 062324 (2006).

\bibitem{braunstein:bs_bell_measure}
S.~L. Braunstein and A.~Mann,
\newblock ``Measurement of the Bell operator and quantum teleportation'',
\newblock {\em Phys. Rev. A} {\bf 51}, R1727 (1995).

\bibitem{nielsen:qcqi}
M.~A. Nielsen and I.~L. Chuang,
\newblock {\em Quantum Computation and Quantum Information},
\newblock Cambridge University Press, 2000.

\bibitem{moehring:ion-ion}
D.~L. Moehring et~al.,
\newblock ``Entanglement of single-atom quantum bits at a distance'',
\newblock {\em Nature} {\bf 449}, 68 (2007).

\bibitem{matsukevich:bell_ion}
D.~N. Matsukevich, P.~Maunz, D.~L. Moehring, S.~Olmschenk, and C.~Monroe,
\newblock ``Bell Inequality Violation with Two Remote Atomic Qubits'',
\newblock {\em Phys. Rev. Lett.} {\bf 100}, 150404 (2008).

\bibitem{james:qubit_measure}
D.~F.~V. James, P.~G. Kwiat, W.~J. Munro, and A.~G. White,
\newblock ``Measurement of qubits'',
\newblock {\em Phys. Rev. A} {\bf 64}, 052312 (2001).

\bibitem{gottesman:qc_teleportation}
D.~Gottesman and I.~L. Chuang,
\newblock ``Demonstrating the viability of universal quantum computation using
  teleportation and single-qubit operations'',
\newblock {\em Nature} {\bf 402}, 390 (1999).

\bibitem{bouwmeester:photon_teleport}
D.~Bouwmeester et~al.,
\newblock ``Experimental quantum teleportation'',
\newblock {\em Nature} {\bf 390}, 575 (1997).

\bibitem{boschi:photon_teleport}
D.~Boschi, S.~Branca, F.~{De Martini}, L.~Hardy, and S.~Popescu,
\newblock ``Experimental Realization of Teleporting an Unknown Pure Quantum
  State via Dual Classical and Einstein-Podolsky-Rosen Channels'',
\newblock {\em Phys. Rev. Lett.} {\bf 80}, 1121 (1998).

\bibitem{furusawa:cont_var_teleport}
A.~Furusawa et~al.,
\newblock ``Unconditional Quantum Teleportation'',
\newblock {\em Science} {\bf 282}, 706 (1998).

\bibitem{chen:light-ensemble_teleport}
Y.-A. Chen et~al.,
\newblock ``Memory-built-in quantum teleportation with photonic and atomic
  qubits'',
\newblock {\em Nature Physics} {\bf 4}, 103 (2008).

\bibitem{riebe:ion_teleport}
M.~Riebe et~al.,
\newblock ``Deterministic quantum teleportation with atoms'',
\newblock {\em Nature} {\bf 429}, 734 (2004).

\bibitem{barrett:ion_teleport}
M.~D. Barrett et~al.,
\newblock ``Deterministic quantum teleportation of atomic qubits'',
\newblock {\em Nature} {\bf 429}, 737 (2004).

\bibitem{riebe:ion_teleport2}
M.~Riebe et~al.,
\newblock ``Quantum teleportation with atoms: quantum process tomography'',
\newblock {\em New J. Phys.} {\bf 9}, 211 (2007).

\bibitem{olmschenk:teleportation}
S.~Olmschenk et~al.,
\newblock ``Quantum Teleportation between Distant Matter Qubits'',
\newblock {\em Science} {\bf 323}, 486 (2009).

\bibitem{altepeter:tomography}
J.~B. Altepeter, E.~R. Jeffrey, and P.~G. Kwiat,
\newblock ``Photonic State Tomography'',
\newblock {\em Adv. At. Mol. Opt. Phys.} {\bf 52}, 105 (2006).

\bibitem{obrien:process_tomography}
J.~L. {O'Brien} et~al.,
\newblock ``Quantum Process Tomography of a Controlled-NOT Gate'',
\newblock {\em Phys. Rev. Lett.} {\bf 93}, 080502 (2004).

\bibitem{horodecki:teleport_channel}
M.~Horodecki, P.~Horodecki, and R.~Horodecki,
\newblock ``{General teleportation channel, singlet fraction, and
  quasidistillation}'',
\newblock {\em Phys. Rev. A} {\bf 60}, 1888 (1999).

\bibitem{lutkenhaus:bell_no-go}
N.~{L\"{u}tkenhaus}, J.~Calsamiglia, and K.-A. Suominen,
\newblock ``Bell measurements for teleportation'',
\newblock {\em Phys. Rev. A} {\bf 59}, 3295 (1999).

\bibitem{enk:entangle_verify}
S.~J. {van Enk} and N.~{L\"{u}tkenhaus }and H.~J.~Kimble,
\newblock ``Experimental procedures for entanglement verification'',
\newblock {\em Phys. Rev. A} {\bf 75}, 052318 (2007).

\bibitem{streed:fresnel_optics}
E.~W. Streed, B.~G. Norton, J.~J. Chapman, and D.~Kielpinski,
\newblock ``{Scalable, efficient ion-photon coupling with phase Fresnel lenses
  for large-scale quantum computing}'',
\newblock (2008),
\newblock arXiv:0805.2437.

\bibitem{shu:trap_mirror}
G.~Shu, M.~R. Dietrich, N.~Kurz, and B.~B. Blinov,
\newblock ``Trapped Ion Imaging with a High Numerical Aperture Spherical
  Mirror'',
\newblock (2009),
\newblock arXiv:0901.4742.

\bibitem{lindlein:4pi_parabolic_mirror}
N.~Lindlein et~al.,
\newblock ``A New 4$\pi$ Geometry Optimized for Focusing on an Atom with a
  Dipole-Like Radiation Pattern'',
\newblock {\em Laser Physics} {\bf 17}, 927 (2007).

\bibitem{maiwald:enhanced_optical_access}
R.~Maiwald et~al.,
\newblock ``Ion traps with enhanced optical and physical access'',
\newblock (2009),
\newblock arXiv:0810.2647.

\bibitem{hennrich:high_finesse_cavity}
M.~Hennrich, T.~Legero, A.~Kuhn, and G.~Rempe,
\newblock ``Vacuum-Stimulated Raman Scattering Based on Adiabatic Passage in a
  High-Finesse Optical Cavity'',
\newblock {\em Phys. Rev. Lett.} {\bf 85}, 4872 (2000).

\bibitem{mckeever:atom-cavity}
J.~McKeever et~al.,
\newblock ``Deterministic Generation of Single Photons from One Atom Trapped in
  a Cavity'',
\newblock {\em Science} {\bf 303}, 1992 (2004).

\bibitem{keller:ion_cavity_single_ph}
M.~Keller, B.~Lange, K.~Hayasaka, W.~Lange, and H.~Walther,
\newblock ``A calcium ion in a cavity as a controlled single-photon source'',
\newblock {\em New J. Phys.} {\bf 6}, 95 (2004).

\bibitem{russo:raman_ion_cavity}
C.~Russo et~al.,
\newblock ``Raman spectroscopy of a single ion coupled to a high-finesse
  cavity'',
\newblock {\em Appl. Phys. B} {\bf 95}, 205 (2009).

\bibitem{barrett:time-bin}
S.~D. Barrett and P.~Kok,
\newblock ``Efficient high-fidelity quantum computation using matter qubits and
  linear optics'',
\newblock {\em Phys. Rev. A} {\bf 71}, 060310(R) (2005).

\bibitem{duan:prob-photon}
L.-M. Duan, B.~B. Blinov, D.~L. Moehring, and C.~Monroe,
\newblock ``Scalable Trapped Ion Quantum Computation with a Probabilistic
  Ion-Photon Mapping'',
\newblock {\em Quant. Inf. Comp.} {\bf 4}, 165 (2004).

\end{thebibliography}
